\documentclass{statsoc}

\usepackage{geometry}

\usepackage{etoolbox}

\makeatletter
\patchcmd{\@makecaption}
  {\parbox}
  {\advance\@tempdima-\fontdimen2} 
  {}{}
\makeatother 
\usepackage{lmodern}			
\usepackage[T1]{fontenc}		
\usepackage[utf8]{inputenc}		
\usepackage{lastpage}			
\usepackage{color}		        
\usepackage{xcolor}
\usepackage{pmboxdraw}
\usepackage{graphicx}			
\usepackage{microtype} 			
\usepackage[subentrycounter,seeautonumberlist,nonumberlist=true]{glossaries}
\usepackage{lipsum}			
\usepackage[portuguese,onelanguage]{algorithm2e}	
 \usepackage{amsbsy}			
 \usepackage{amscd}			
 \usepackage{amsfonts}			
 \usepackage{amsmath}			
 \usepackage{amssymb}			
 \usepackage{amstext}			
\usepackage{hyperref}			
\usepackage{cleveref}			
\usepackage{dsfont}			
\usepackage{listings}           	
 \usepackage{lscape}             	
 \usepackage{pdfpages}           	
\usepackage{textcomp}                   

\usepackage{rotating}
\usepackage{booktabs}
\usepackage{longtable}
\usepackage{array}
\usepackage{multirow}
\usepackage{wrapfig}
\usepackage{float}
\usepackage{colortbl}
\usepackage{pdflscape}
\usepackage{tabu}
\usepackage{threeparttable}
\usepackage{threeparttablex}
\usepackage[normalem]{ulem}
\usepackage{makecell}
\usepackage{blindtext}

\usepackage{natbib}

\newcommand{\Yijt}{Y_{ij}(t)}
\newcommand{\Yij}[1]{Y_{ij}(#1)}
\newcommand{\Wijmct}{W_{ijcm}(t)}
\newcommand{\tc}{\alpha}
\newcommand{\byij}{\mathbf{y}_{ij}}

\newcommand{\bsY}{\mathbf{Y}}

\newcommand{\myeqref}[1]{Equation \eqref{#1}}
\newcommand{\transp}{\intercal}

\newcommand{\group}{substation}
\newcommand{\groups}{substations}
\newcommand{\mjc}{m_{jc}}
\newcommand{\mj}[1]{m_{j#1}}
\newcommand{\basisfunc}{\phi}
\newcommand{\vecbasisfunc}{\boldsymbol \phi}

\newcommand{\prob}[1]{\mathbb{P}\Bigl( #1 \Bigr)}

\newcommand{\mysum}[2]{\sum_{#1 =1}^{#2}}

\newcommand{\erro}{\varepsilon}
\newcommand{\berro}{\boldsymbol \varepsilon}
\newcommand{\erroij}{\erro_{ij}}
\newcommand{\erroijmc}{\erro_{ijmc}}
\newcommand{\erroijmct}{\erro_{ijmc}(t)}
\newcommand{\erroijt}{\erro_{ij}(t)}
\newcommand{\erroijdd}{\erro_{ij}(\cdot,\cdot)}

\newcommand{\vecerroij}{\boldsymbol \varepsilon_{ij}}

\newcommand{\by}{\mathbf{y}}
\newcommand{\meancurve}{\mu}
\newcommand{\bmc}{\boldsymbol \meancurve}

\newcommand{\covmatrix}{\Sigma}
\newcommand{\bcovmatrix}{\boldsymbol \Sigma}
\newcommand{\wcovmatrix}{\Psi}
\newcommand{\bbeta}{\boldsymbol \beta}

\newcommand{\bphi}{\boldsymbol \phi}

\newcommand{\covar}{D}
\newcommand{\bcovar}{\mathbf{\covar}}
\newcommand{\covarpar}{\gamma}
\newcommand{\veccovarpar}{\boldsymbol \covarpar}
\newcommand{\bcovarpar}{\veccovarpar}
\newcommand{\varfunc}{\eta}

\newcommand{\bvarpar}{\boldsymbol \sigma}
\newcommand{\corfunc}{\rho}
\newcommand{\corpar}{\omega}
\newcommand{\bcorpar}{\boldsymbol \corpar}

\newcommand{\parsVec}{\boldsymbol \Theta}
\newcommand{\bpi}{\boldsymbol \pi}
\newcommand{\indic}[1]{\mathcal{I} \bigl( #1\bigr)}
\newcommand{\X}{X}
\newcommand{\bX}{\mathbf{X}}

\newcommand{\latvar}{Z}
\newcommand{\slatvar}{z}
\newcommand{\idlat}{b}
\newcommand{\bigidlat}{B}
\newcommand{\latpar}{\pi}
\newcommand{\pjbr}{p_{jb}^{(r)}}

\title[Aggregated functional data model]{Aggregated functional data model applied on clustering and disaggregation of UK electrical load profiles}
\author[Gabriel Franco {\it et al.}]{Gabriel Franco}
\address{University of Campinas (UNICAMP),
Campinas,
Brazil.}
\email{gabrielfranco89@gmail.com}
\author{Camila P. E. Souza}
\address{The University of Western Ontario,
London,
Canada.}
\author{Nancy L. Garcia}
\address{University of Campinas (UNICAMP),
Campinas,
Brazil.}

\begin{document}

\maketitle

\begin{abstract}
   Understanding electrical energy demand at the consumer level plays an important role in planning the distribution of electrical networks and offering of off-peak tariffs, but observing individual consumption patterns is still expensive. On the other hand, aggregated load curves are normally available at the substation level. The proposed methodology separates substation aggregated loads into estimated mean consumption curves, called typical curves, including information given by explanatory variables. In addition, a model-based clustering approach for substations is proposed based on the similarity of their consumers’ typical curves and covariance structures. 
   The methodology is applied to a real substation load monitoring dataset from the United Kingdom and tested in eight simulated scenarios.
\end{abstract}

\section{Introduction}

In 2019, electricity accounted for 17\% of the United Kingdom's final energy consumption. This proportion has been relatively stable in recent years. Moreover, when stratified by sector, residential consumers account for 30\% of electricity demand (see Table 1.1 and Chart 5.4 in \cite{dukes2020}), and they have a significant influence of peak demand in the early evening and the peak is more pronounced in winter \citep{hamidi2009demand}. For example, in Brazil, typical curves for residential consumers have a spike in electricity consumption from 6 pm to 8 pm, due partially to the use of electric showers after the workday \citep{lenzi2017analysis}. It is 
also well known that in the UK, there is large power surge for 3 to 5 minutes at the end of the most popular TV shows or sporting events, exactly the time that takes to make a cup of tea. One alternative and efficient strategy to not only reduce the chance of overload but also to maximize the use of existing equipment is to provide cheaper off-peak tariffs, such as Economy 7, with off-peak rate usually running from midnight to 7am, while the more expensive daytime rate covers the rest of the day. 



 Understanding individual customer consumption behaviour is essential to comprehend electrical energy demand and consequently to take action to reduce substation load, such as educational programs and off-peak tariff policies, or even to consider bigger projects like new power plants and network distribution redesign. Solutions such as the aggregated data model proposed in this work provide estimated typical curves for each customer type based only on aggregated data and enhance comprehension of the covariance structure to assess data uncertainty. On the other hand, it is not expected that these typical curves will be the same for all times, locations and consumers. As said before, there are explanatory, such as temperature, TV programming, and location, that can improve the inference and clustering of these typical curves. 
 
 Increasing network distribution and the rise of smart grids have drawn attention to load profile monitoring~\citep{wang2015load}. Multiple articles have been published in the literature proposing  clustering techniques to segment customers and reduce variability~\citep{prahastono2007review, li2015development1,bouveyron2018functional}. Efforts are also underway to achieve short-term load forecasting using machine learning ~\citep{sousa2014short}  and deep learning methods~\citep{shi2017deep}. Although load profile modeling is an important task to comprehend electrical demand variability, it does not provide information on the customer level like smart meters~\citep{d2014smart,gouveia2016unraveling,de2017switching}, appliances monitoring~\citep{hart1992nonintrusive,arghira2012prediction,adjei2020} and disaggregation methods~\citep{schirmer2019evaluation}.
 


Several widely used machine learning and regression methods were used to study energy disaggregation such as artificial neural networks~\citep{lin2015advanced,hosseini2017non}, random forests~\citep{bilski2017generalized,schirmer2019integration}, Support Vector Machines~\citep{basu2014nonintrusive,schirmer2020energy}, wavelet component analysis~\citep{zhu2014load} and $K$-Nearest-Neighbours~\citep{kim2014electrical}. Reviews and comparisons of multiple statistical methods for energy disaggregation are available in \cite{schirmer2019evaluation}.


The family of  aggregated data models considered in this work was first proposed  by~\cite{dias2009non}. Using the observed electrical load from energy transformers and their  market information, the authors composed a non-parametric model to estimate the typical consumption curve of customers in the city of Campinas, Brazil, using basis function expansion and the sample covariance matrix as the model covariance structure.
More sophisticated structures, under the Bayesian paradigm, were proposed to study the transformer load curves \citep{dias2013hierarchical,dias2015aggregated}.
Later, considering the market (distribution of customers per type) as random, the  aggregated data model could identify errors in  energy customer classification~\citep{lenzi2017analysis}.

In this work, a generalization of the aggregated data model described above is proposed. Our novel approach performs the disaggregation task by assuming a Gaussian process with mean functional response as an aggregated linear combination of the market, typical customer curves, and explanatory variables. Additional functional variables are incorporated in the typical curve model to comprehend, for example, the impact of temperature on customer load profile. A model-based clustering approach is also proposed to group energy substations with similar disaggregated curves using a mixture of Gaussian processes~\citep{shi2005hierarchical,shi2011gaussian,tresp2001mixtures} estimated by the Expectation-Maximization algorithm~\citep{dempster1977maximum,mclachlan2007algorithm}. Finally, structured covariance functionals are proposed to model load variability and correlation decay over time. To show the strength of our method, we analyzed a dataset from electrical load profiles from several substations across the UK.  

This paper is organized as follows. Section 2 presents the UK electrical dataset which is analysed in Section 4. The proposed methodology is presented in Section 3. Section 5 provides some discussion and conclusion.  Further analysis of artificial datasets is available as supplementary material.  The code is currently available online as an R package at a GitHub repository (https://github.com/gabrielfranco89/aggrmodel). 

\section{UK electrical data}
\label{sec:motivation}


The dataset analyzed in this work is a subset of the data first introduced by \cite{li2015development1} and \cite{li2015development2}. The initial data contain information on electrical load profiles observed every 10 minutes across 407 electrical energy substations in the northwest portion of the United Kingdom. Observations were taken from October 28, 2012, to March 30, 2013, for a total of 154 days.

Each substation supplies energy to up to eight types of customers. This eight-customer division dates from the 1990s and is organized as two domestic types, unrestricted and ``Economy 7''; two non-domestic types, also unrestricted and ``Economy 7''; and four non-domestic classes of maximum-demand customers according to their peak-load factor. This distribution has proven to be inefficient because a small delicatessen and a supermarket can be assigned to the same customer type~\citep{wilks2010demand}. The variability of non-domestic groups makes the aggregated data model unsuitable because there is no typical curve that could, for example, represent both a supermarket and a small delicatessen. Hence, to apply the proposed model, a subset of the data, consisting of substations with only two types of domestic customers, was considered, resulting in a dataset with 12 substations and the following two types of customers: unrestricted (C1) and ``Economy 7'' (C2) domestic customers, with the latter referring to a program with cheaper electrical tariffs during the off-peak period. 



Only working days were considered in the dataset to remove the weekend effect on electrical energy consumption because it is possibly different than the domestic routine between Monday and Friday. Also, to avoid variability during the Christmas and New Year holidays, observations from January 3 to March 30, 2013, were used instead.

\begin{figure}[t]
  \centering
  \includegraphics[width=.8\textwidth]{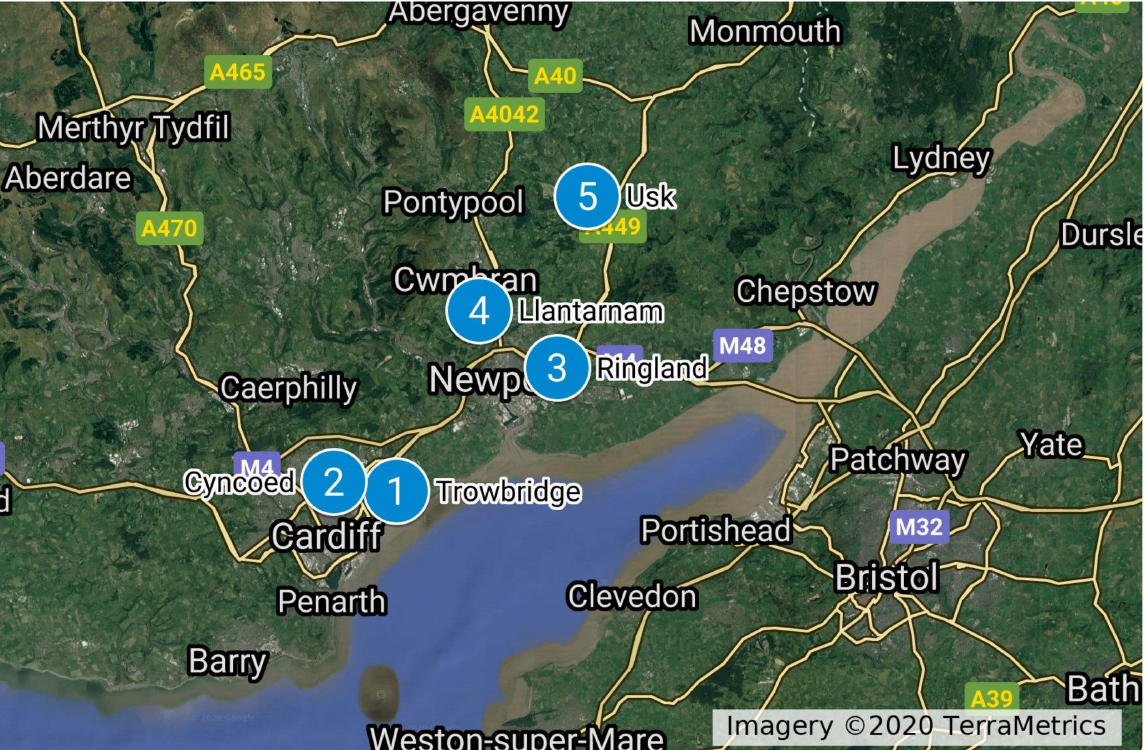}
  \caption{Geographic location of substation primaries in United Kingdom: (1) Trowbridge, (2) Cyncoed, (3) Ringland, (4) Llantarnam and (5) Usk.}
  \label{fig:map}
\end{figure}

Temperature measurements were obtained through the API of the World Weather Online  Web site (worldweatheronline.com), using the substation primary, generally representing a community or a district in Wales, as the location reference (see Table \ref{tab:gavin-mkt} and Figure \ref{fig:map}). The downloaded historical temperature data, however, contain observations only every three hours. Hence, to achieve the same observation frequency of 10 minutes as in the electrical load dataset, temperature data were interpolated via a cubic B-spline fit. 

\begin{table}
\caption{\label{tab:gavin-mkt}Primaries, substation names, substation IDs and number of customers of types C1 and C2.}
\centering
\fbox{
\begin{tabular}{rlcc}
Primary & Substation & C1 & C2\\
\hline
  Trowbridge & S1 & 228 & 3\\
  & S2  & 146 & 5\\
  & S3  & 151 & 5\\
  \hline
 Cyncoed & S4  & 21 & 88\\
  & S5 & 218 & 7\\
  \hline
  Ringland Newport & S6  & 155 & 17\\
  & S7 & 194 & 12\\
  \hline
 Llantarnam Primary & S8  & 173 & 9\\
  & S9  & 163 & 12\\
  & S10  & 158 & 2\\
 & S11 & 244 & 10\\
 \hline
 Usk & S12  & 46 & 23\\
\end{tabular}}
\end{table}

\begin{figure}[thp]
  \centering
\includegraphics[width=.8\linewidth]{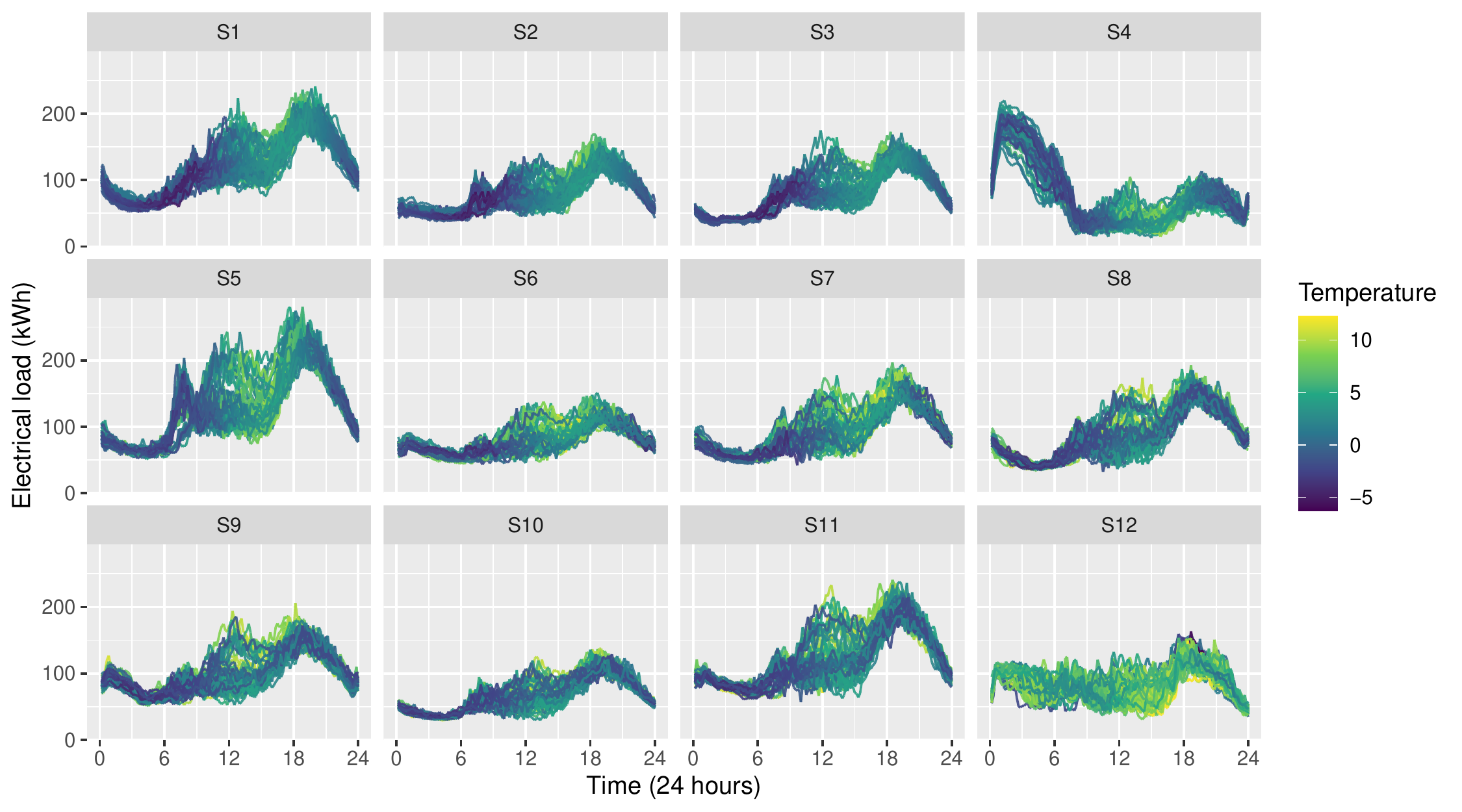}(a)
\includegraphics[width=.8\linewidth]{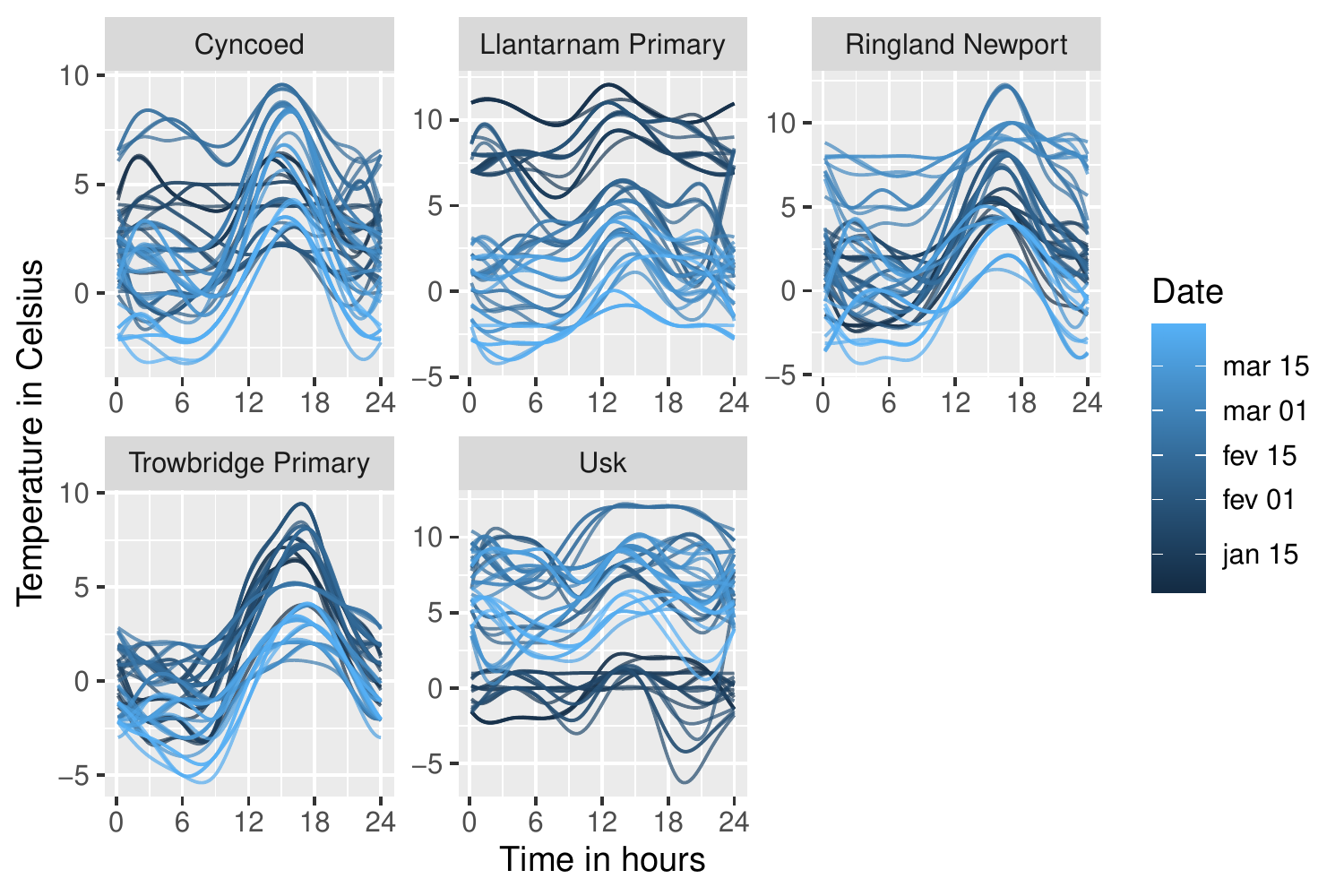}(b) 
\caption{(a) Electrical load profile data in kWh observed every 10 minutes from 12 substations color-coded by the current air temperature in Celsius and (b) the observed temperature in the five primaries over 61 winter days in the United Kingdom.}
  \label{fig:gavin1}
\end{figure}

Figure~\ref{fig:gavin1}a provides a visualization of the electrical load profiles corresponding to the 61 days from January 3 to March 30, 2013, for each one of the 12 substations, coloured according to the temperature scale located above the panel. Figure~\ref{fig:gavin1}b shows the observed temperature at the five substation primaries for the same 61 days as in Figure~\ref{fig:gavin1}a. The associated market of each substation, that is, the number of C1 and C2 residences, is displayed in Table~\ref{tab:gavin-mkt}, which shows that the great majority of the customers are unrestricted domestic customers (C1), dominating more than 90\% of the market in 10 of 12 substations, whereas substation S4 is the only one with a majority of  ``Economy 7'' domestic customers  (C2), representing 80.73\% of its market.

Except for S4 and S12, all substations presented a similar pattern. Early morning showed the lowest energy consumption until approximately 9 AM, with apparently homogeneous variance during this period. The period between 10 AM and 4 PM showed the largest variability, probably because this is the period when people tend to leave their houses to work, but some stay at a home office, for example. From 5 PM onward, the variability apparently stabilized again, even at the load peak at 7 PM. However, substations S4 and S12 not only did not follow this pattern, but also were distinct one from the other. Their peaks, at late night in substation S12 and before sunrise in S4, probably occur because of lower tariffs at night, encouraging energy consumption outside daytime. Exceptionally, substation S12 has one-third of its market consisting of customers of type C2, which results in relatively higher loads in early morning compared to substations with a great majority of C1 customers, and more variability before 9 AM.



Figures~\ref{fig:gavin1}a and \ref{fig:gavin1}b show that it may be useful to extend the aggregated model to include temperature as a functional covariate to better explain the variability of electrical load profiles. Indeed, most substations experience temperature fluctuations during the daytime, whereas night periods tend to be more stable. Nonetheless, substation S12 is an exception for the temperature pattern as well; it is the only one that shows higher temperature values both day and night.


The particularity of substation S12 may be explained by its geographic location within the Usk primary, shown in Figure \ref{fig:map}. 
S12 is in the town of Monmouth, in the countryside of Wales, with a population of less than three thousand, and is the smallest of all the primaries.
To the southwest is Llantarnam, a community in the suburb of Cwmbran, with a population size slightly larger than four thousand, where substations S8 to S11 are located, all consisting mostly of customers of type C1. Not far away is Ringland, in the city of Newport, where substations S6 and S7 are located, with populations approximately double that of Llantarnam. Closer to the capital of Wales, there are two primaries: Trowbridge and Cyncoed. Both are in communities with a population greater than ten thousand (16,194 and 11,148, respectively) located in the urban area of Cardiff Central. In fact, Cyncoed, the only substation with a majority of C2 customers, has some of the highest property prices in the country. All cited demographic data are available in the 2011 census of the United Kingdom  \cite{uk2011census}.

These different characteristics among substations raise the question if we should assume typical customer curves to be the same for every substation. For example, can it be expected that an unrestricted domestic residence in an urban area like London will have the same load profile as a house located in the countryside? The answer is probably no and this is the motivation to introduce a latent variable to cluster substations based on their disaggregated typical curves and covariance structures.

\section{Methods}

\subsection{Simple aggregated data model}



Let us first introduce the aggregated data model in its simplest form, as proposed in~\cite{dias2009non}.
The observed data consist of aggregated energy consumption curves for $J$ substations observed over $I$ days. Each substation constitutes a distinct market with $C$ types of consumers, -- e.g., residential, industrial and business. Each aggregated curve is the sum of all individual consumer curves served by that substation.
Suppose that $\Wijmct$, the unobserved energy consumption of customer $m$ of type $c$ at time $t$ from substation $j$ on day $i$, can be represented as
\begin{equation}
  \Wijmct = \tc_c(t) + \erroijmct,
  \label{eq:wmodel}
\end{equation}

\noindent with $\tc_c(\cdot)$ being the typical curve of a customer of type $c$ and $\erroijmct$ a Gaussian Process (GP)~\citep{shi2011gaussian} with zero mean and covariance structure $\wcovmatrix_c(\cdot,\cdot)$ to be detailed in Section~\ref{sec:covstructures}.

Let $\Yijt$ be the observable aggregated energy consumption at substation $j$, day $i$ and time $t$.  $\Yijt$  can then be represented as the sum of individual customer  curves, that is,
\begin{eqnarray}
  \label{eq:yij}
  \Yijt &=&
      \sum_{c=1}^C
      \sum_{m=1}^{\mjc}
      \Wijmct  \nonumber \\
  &=&
      \sum_{c=1}^C
      \sum_{m=1}^{\mjc}
      \alpha_c(t)
      +
      \erroijmct  \nonumber \\
  &=&
      \sum_{c=1}^C
      \mjc
       \tc_c(t)
      +
      \erroijt, 
\end{eqnarray}
with $m_{jc}$ being the fixed and known number of customers of type $c$ in substation $j$, $\erroijdd \sim GP \big( 0, \covmatrix_j(\cdot,\cdot)\big)$, where, 
assuming independence among individual customers, the covariance structure $\bcovmatrix_j(\cdot,\cdot)$ can be written as $ \covmatrix_j(s,t) = \sum_{c=1}^C \mjc \wcovmatrix_c(s,t)$.

The mean component $\tc_c(\cdot)$ in \myeqref{eq:wmodel} represents the typical curve of customers of type $c$ and can be modelled using a basis function expansion as $ \tc_c(t) = \mysum{k}{K} \basisfunc_k(t) \beta_{ck} = \bphi(t) \bbeta_c$,
where $\bbeta_c \in \mathbb{R}^K$ is the vector of expansion parameters or coefficients and $\basisfunc_k(\cdot)$ the $k$-th basis function, which can be B-Splines, Fourier transforms, wavelets or a polynomial basis~\citep{ramsay2005functional}. As in  previous studies using aggregated data analysis \citep{dias2009non,dias2013hierarchical,dias2015aggregated,lenzi2017analysis}, cubic B-Splines basis are used with the assumption that the number of basis functions $K$ is known. Selection of knot placement and number of basis functions it is an important factor and several studies have dealt with this research topic \citep{devore2003best, dias2007consistent, kohn2000wavelet,luo1997hybrid,dias1998density}. 

\subsection{Full aggregated data model}
\label{sec:method-fm}

In the electrical energy consumption setup, suppose that the typical curve depends not only on the time $t$, but also on functional covariates such as the air temperature on day $i$, $v_i(t)$. Additionally, one can incorporate $P$ explanatory variables related  to the substations, namely $\covar_{ij1}, \dots, \covar_{ijP}$ to the full aggregated data model in \eqref{eq:yij},  getting 
\begin{eqnarray*}
  \Yijt &=&
            \left(
            \sum_{c=1}^C
            \mjc
            \tc_{c} \big( t, v_i(t) \big)
            \right)
            +
          \covar_{ij1} \covarpar_1 +
          \dots +
          \covar_{ijP} \covarpar_P +
          \erroijt. 
\end{eqnarray*}

Again, we will assume that $\alpha_c:\mathbb{R}^+ \times \mathbb{R}$ can be expanded as a tensor product of basis functions as 
\begin{eqnarray}
  \tc_{c} \big( t, v_i(t) \big)
  =&
      \mysum{k}{K} \mysum{l}{L}
      \basisfunc_k \big( t\big)
      \varphi_l \big( v_i(t)\big)
      \,
      \beta_{lkc}, \label{eq:surf-tensor}
\end{eqnarray}
where $\basisfunc(\cdot)$ and $\varphi(\cdot)$ are basis functions and $\beta_{lkc}$ are expansion parameters. 

Therefore, we can write the model as
\begin{eqnarray}
  \label{eq:surfmodel}
  \Yijt    &=&
          \left(\,
            \mysum{c}{C}\mysum{k}{K} \mysum{l}{L}
            \mjc
          \basisfunc_k \big( t\big)
          \varphi_l \big( v_i(t)\big)
          \,
          \beta_{lkc}
          \, \right)
          +
          \bcovar_{ij} \veccovarpar
          +
            \erroijt.
\end{eqnarray}



\noindent In vector representation, $\Yijt = \mysum{c}{C} \mjc \vecbasisfunc_i(t) \bbeta_c + \bcovar_{ij} \veccovarpar + \erroijt$
with $\veccovarpar \in \mathbb{R}^P$ being the parameters corresponding to the \group ~explanatory variables in $\mathbf{D}_{ij}$,
\begin{align*}
    \vecbasisfunc_i(t) &= \Bigl( \phi_1(t)  \varphi_1( v_i(t)), \phi_1(t)  \varphi_2( v_i(t)),\ldots, \phi_1(t)  \varphi_L( v_i(t)),\\
    & \phi_2(t)  \varphi_1( v_i(t)),\ldots, \phi_2(t)  \varphi_L( v_i(t)),\ldots, \phi_K(t)  \varphi_L( v_i(t))\Bigr) \text{\: and}\\
    \bbeta_c &= \big(\beta_{11c},\beta_{12c},\ldots,\beta_{1Lc},\beta_{21c},\ldots \beta_{2Lc},\ldots, \beta_{KLc} \big). 
\end{align*}


Note that the simple model is nested inside the full aggregated data model, because it represents the case when temperature and explanatory variables have no effect on the typical curve.

\subsection{Covariance structures} \label{sec:covstructures}

Let $\erroijmc(\cdot)$ be the Gaussian Process introduced in \myeqref{eq:wmodel} with zero mean and covariance structure defined by the functional $\wcovmatrix_c(\cdot,\cdot)$.  This presentation will use they following decomposition~\citep{dias2013hierarchical}:
\begin{eqnarray}
  \label{eq:covstructure}
  \wcovmatrix_c(s,t)
  =
      \text{Cov} \big(
      \erroijmc(s),
      \erroijmct
      \big) = \varfunc_c(s) \, \corfunc_c(s,t) \, \varfunc_c(t), \nonumber
\end{eqnarray}

\noindent where $\varfunc_c(\cdot)$ and $\corfunc_c(\cdot,\cdot)$ are variance and correlation functionals, respectively. To guarantee the positive definiteness of the Gaussian Process covariance structure,  $\corfunc_c(\cdot,\cdot)$ must be a proper positively defined correlation functional. The following sections describe the different nested forms of $\varfunc_c(\cdot)$ and $\corfunc_c(\cdot,\cdot)$.

\subsubsection{Variance functionals}
\label{sec:varfunc}

The variance functional $\varfunc_c(\cdot)$ describes the variability of customers of type $c$ over time.~ 
The identifiability of the model is guaranteed only if $\eta_c(\cdot)$ is positive, otherwise any function multiplied by $-1$ is also an optimal solution. Hence, the results of~\cite{ramsay2005functional} can be used, and the variance function $\eta_c(\cdot)$ can be written as:
\begin{equation}
  \label{eq:completefunc_exp}
  \varfunc_c(\cdot)
  =
  \exp
  \left\{
  \mysum{k}{K^\prime}
  \basisfunc^\varfunc_k(\cdot)
  \beta_{kc}^\star
  \right\}.  
\end{equation}

Furthermore, nested functional variances can be created based on a different parametrization of the expansion coefficients of \myeqref{eq:completefunc_exp}~\citep{dias2013hierarchical}. If
\begin{eqnarray}
  \label{eq:nest}
  \sigma_c^\star = \frac{1}{K^\prime} \sum_{k=1}^{K^\prime} \beta_{kc}^\star \;\; \mbox{and} \;\; \beta_{kc}^{\varfunc} = \beta_{kc}^\star - \sigma_c^\star, \nonumber
\end{eqnarray}

\noindent then 
\begin{equation}
  \label{eq:completefunc2}
  \varfunc_c(\cdot)
  =
  \exp
  \left\{
  \sigma_c^\star
  +
  \mysum{k}{K^\prime}
  \basisfunc^\varfunc_k(\cdot)
  \beta_{kc}^\varfunc
  \right\},
\end{equation}

\noindent with $\sum_{k=1}^{K^\prime} \beta_{kc}^\varfunc = 0$. Now if $\beta_{kc}^\varfunc = 0, \forall k$, then there is a homogeneous variance $\sigma_c = e^{\sigma_c^\star}$ over time for each customer type and if $\sigma_c = \sigma, \forall c$ we have an uniform homogeneity for all types of customer.  Hence, the three forms of nested variance functionals are


\begin{enumerate}
\item
  Homogeneous uniform: $\varfunc_c(t) = \sigma, \: \forall c$;
\item
  Homogeneous: $\varfunc_c(t) = \sigma_c$;
\item
  Complete: $  \varfunc_c(\cdot)
  =
  \sigma_c
  \exp
  \left\{
  \mysum{k}{K^\prime}
  \basisfunc^\varfunc_k(\cdot)
  \beta_{kc}^\varfunc
  \right\}$.
\end{enumerate}



\subsubsection{Correlation functional}
\label{sec:corfunc}

The correlation functional $\corfunc_c(s,t)$ quantifies the relationship between the energy consumption of a customer of type $c$ at two points in time  $s$ and $t$ in the time interval $[0,T]$. It is assumed that this relationship is defined by an exponential decay proportional to the absolute difference $|t-s|$ and with parameter $\omega_c>0, \forall c$, that is,
\begin{equation}
    \corfunc_c(s,t) = \exp \left\{-2\frac{1}{\corpar_c} \frac{|t-s|}{T}\right\}. \nonumber
\end{equation}

\subsection{Aggregated model likelihood and estimation}
\label{sec:model-lk}

The full aggregated data model in \myeqref{eq:surfmodel} includes an error $\erroij(\cdot)$, which is a Gaussian Process with zero mean and covariance $\bcovmatrix_j(\cdot,\cdot)$. Therefore, we can write $\Yij{\cdot} \sim GP \big( \mu_{ij}(\cdot), \, \covmatrix_j(\cdot,\cdot) \big)$ with
\begin{align}
  \mu_{ij}(t)
  =&     
     \:
     \mysum{c}{C}
     \mjc \,
     \vecbasisfunc_i(t) \bbeta_c
      +  
      \bcovar_j \veccovarpar \label{eq:muj} \qquad \text{ and }\\
  \covmatrix_j(s,t)
  =&\:
  \mysum{c}{C}
  \,
  \mjc
  \:
  \varfunc_c(s) \,
  \corfunc_c(s,t) \,
  \varfunc_c(t). \nonumber   \label{eq:aggrcov}
\end{align}

\noindent Let $\by$ be a sample of $N$ daily observations from $J$ substations over $I$ days, say
\begin{equation}
  \label{eq:sample}
  \by
  =
  \Bigl\{
  \byij \, : \:
  \byij
  =
  \big(
    y_{ij}(t_1), \dots, y_{ij}(t_N)
    \big)
    \, \text{with } \, i=1,\dots,I \text{ and }
    j=1,\dots,J
  \Bigr\}.
\end{equation}

\noindent Note that $\by$ can be made up of substations observed on different days at different time frequencies. However, to simplify the notation it is assumed that all data are observed on the same days and at the same time frequency. Let $\parsVec$ be the set containing all model parameters. Assuming independence among days and substations, the log-likelihood of the aggregated data model can be written as
\begin{align}
  \ell
  \big(
    \parsVec | \by
  \big)
  \equiv&
     \sum_{i=1}^I
     \sum_{j=1}^J
     \,
     \log f \big( \byij ; \parsVec \big) \nonumber \\
  =&
     -\frac{1}{2}\sum_{i=1}^I
     \sum_{j=1}^J
     \big[ \log |\bcovmatrix_j| +
     (\bmc_{ij} - \byij)^\transp
     \bcovmatrix_j^{-1}
     (\bmc_{ij} - \byij)
    \big ] + C,
     \label{eq:aggrmodellikelihood}
\end{align}

\noindent where 
 \begin{align*}
   \boldsymbol{\mu}_{ij}
   &=
   \big\{
     \mu_{ij}(t_1),
     \dots,
     \mu_{ij}(t_N)
   \big\} \qquad \text{and} \\
   \bcovmatrix_j
   &=
   \big\{
     \bcovmatrix_j \in \mathbb{R}^{N \times N} 
     \text{ with elements } \covmatrix_j(s,t) : 
     s = t_1, \dots, t_N; \,
     t = t_1, \dots, t_N; \,
   \big\}.
 \end{align*}
To facilitate parameter estimation, the mean function $\mu_{ij}(\cdot)$ in Equation~(\ref{eq:muj}) is written as 
\begin{equation}
    \mu_{ij}(t) = \bX_{ij}(t) \bbeta, 
    \label{eq:muasXbeta} \nonumber
\end{equation}
where $\X_{ij}(\cdot)$ is a matrix composed of the basis functions multiplied by its respective market $\mjc$ and the covariates $\bcovar_j$, that is, 
\begin{align}
  \bX_{ij}(t)
  =&
     \begin{pmatrix}
       \mj{1} \vecbasisfunc_i(t)
       &    \mj{2} \vecbasisfunc_i(t)
       & \cdots
       &    \mj{C} \vecbasisfunc_i(t)
       & \bcovar_j
     \end{pmatrix}_{1 \times (KLC+P)}, \label{eq:xj} \nonumber
\end{align}
     
\noindent and $\bbeta = (\bbeta_1 \;\; \bbeta_2 \;\; \cdots \;\; \bbeta_C \;\; \bcovarpar)_{1\times(KLC+P)}$ is made up of the parameters of the basis expansion and the coefficients of the explanatory variables. 
With this vector representation, we can write the aggregated data model as
\begin{equation}
  \label{eq:yijvec}
  \mathbf{Y}_{ij} = \bX_{ij} \bbeta + \vecerroij, \nonumber
\end{equation}

\noindent where $\bX_{ij}^T = [\bX_{ij}(t_1)  \cdots \bX_{ij}(t_N)]_{(KLC+P)\times N}$ and $\vecerroij =(\erroij(t_1),\ldots, \erroij(t_N))^T$.

Furthermore, the model can be represented across all $J$ substations over $I$ days using a single vector $\mathbf{Y}$, that is,
\begin{equation}
  \label{eq:byijmodel}
  \mathbf{Y} = \bX \bbeta + \berro, \nonumber
\end{equation}

\noindent where 
\begin{align*}
 &\mathbf{Y}_{(NIJ) \times 1} = (\bsY_{11},\bsY_{21}, \ldots, \bsY_{I1},\bsY_{12},\ldots, \bsY_{IJ})^T,\\
 &\bX_{NIJ \times (KLC+P)} = (\bX_{11},\ldots,\bX_{I1},\bX_{12},\ldots,\bX_{IJ}) \text{ and} \\
 &\berro_{(NIJ)\times 1} = (\berro_{11}, \berro_{21},\ldots, \berro_{I1},\berro_{12},\ldots,\berro_{IJ})^T.
\end{align*}



Let $\bcovmatrix \in \mathbb{R}^{NIJ \times NIJ}$ be a sparse block diagonal covariance matrix composed of the matrices $\bcovmatrix_1, \dots, \bcovmatrix_J$. 
Hence, the log-likelihood of the aggregated data model in \myeqref{eq:aggrmodellikelihood} can be written as
\begin{align}
  \ell
  \big(
  \parsVec ;
  \by
  \big)
  =&
     -\frac{1}{2} \log |\bcovmatrix| 
     +
     -\frac{1}{2}
     \big(
     \bX \bbeta - \by
     \big)^\transp
     \bcovmatrix^{-1}
     \big(
     \bX \bbeta - \by
     \big).
    \label{eq:vecaggrmodellk}
\end{align}

Equation~(\ref{eq:vecaggrmodellk}) configures a Gaussian process regression likelihood~\citep{shi2011gaussian,ramsay2005functional}. Let $\parsVec_{\bcovmatrix} = \big( \bvarpar,\bcorpar,\bbeta^\varfunc \big)$ be the parameters describing the covariance matrix $\bcovmatrix$. The estimator of  $\bbeta$ is obtained using weighted least squares and $\parsVec_{\bcovmatrix}$ is estimated using the \textit{BFGS} Quasi-Newton numerical optimization method. The estimation steps are described as follows. 

Fix a precision value $\xi>0$. Given a sample $\by$, at run $r=0$ get initial values for $\bbeta^{(0)}$. At run $r>0$, do
  \begin{enumerate}
  \item[1.] Fix $\bbeta^{(r-1)}$ to obtain $\parsVec_{\bcovmatrix}^{(r)}$ by optimizing the log-likelihood in \eqref{eq:vecaggrmodellk}.
  \item[2.] Fix $\parsVec_{\bcovmatrix}^{(r)}$ to obtain $\bbeta^{(r)}$ via
    \begin{equation}
      \label{eq:betahat}
      \bbeta^{(r)}
      =
      \Bigl(
      \bX^\transp
      \big(\bcovmatrix^{(r)}\big)^{-1}
      \bX
      \Bigr)^{-1}
      \Bigl(
      \bX^\transp
      \big(\bcovmatrix^{(r)}\big)^{-1}
      \by
      \Bigr).
    \end{equation}
  \item[3.] If $$\Big{|}\ell\big( \parsVec^{(r)};\by \big) - \ell\big( \parsVec^{(r-1)};\by \big) \Big{|} < \xi,$$ then stop. If not, add one unit to run $(r)$ and repeat.
  \end{enumerate}

The precision value $\xi>0$, also called the convergence criterion, is typically set to $10^{-6}$.
Because the least squares estimator for $\bbeta$ is unbiased and its expected value does not depend on $\bcovmatrix$, the initial values for $\bbeta$ can be obtained by fitting a linear model with no covariance structure for the aggregated data model. 

To improve computational performance, \eqref{eq:betahat} can be written in terms of the covariance matrices for each substation in the block diagonal matrix $\bcovmatrix$, that is,
\begin{equation}
      \label{eq:betahat_opt}
      \bbeta^{(r)}
      =
      \left(
      \sum_{i=1}^I
      \sum_{j=1}^J
      \bX_{ij}^\transp
      \big(\bcovmatrix_j^{(r)}\big)^{-1}
      \bX_{ij}
      \right)^{-1}
      \left(
      \sum_{i=1}^I  \sum_{j=1}^J
      \bX_{ij}^\transp
      \big(\bcovmatrix_j^{(r)}\big)^{-1}
      \byij
      \right). \nonumber
    \end{equation}

\paragraph{Conditions for identifiability.}
\label{sec:ident}

To ensure the existence of the inverse of the left-hand size of $\bbeta^{(r)}$ in \eqref{eq:betahat} the number of substations in sample $\by$ must be greater than the number of subject types. In other words, $J>C$. Also, to avoid multicolinearity, markets must be linearly independent, that is, there must be no $M \in \mathbb{R}$ such that $\mathbf{m}_{j} = M\mathbf{m}_{j^\prime}$, for any $j \neq j^\prime$.

\subsection{Model-based clustering analysis}
\label{sec:clustering}

Assume that substations can belong to $B$ distinct clusters depending on the similarity of their consumers typical curves. Let $\latvar_j$ be the latent variable that identifies to which cluster substation $j$ belongs, with $\pi_{b}$ being the probability of substation $j$ belonging to cluster $b$. In other words, for each substation $j = 1,\dots,J$, let $\latvar_j$ be a random multinomial variable such that $\mathbb{P}(\latvar_j = \idlat) = \latpar_{\idlat},$ for $ \idlat = 1,2,\dots,\bigidlat $ and $\mysum{\idlat}{\bigidlat} \latpar_{\idlat} = 1$.
It is assumed that given $\latvar_j = b$, the typical curve of a consumer of type $c$ is given by $\alpha_{cb}(\cdot)$ and the aggregated load is a Gaussian process with mean function $\mu_{jb}(\cdot)$ and covariance function $\Sigma_{jb}(\cdot,\cdot)$, that is,
\begin{equation}
  \Yij{\cdot} | \latvar_j = b
  \sim
  GP
  \big(
  \mu_{jb}(\cdot) ,\,
  \covmatrix_{jb}(\cdot,\cdot) 
  \big), \nonumber
\end{equation}

\noindent where $\mu_{jb}(t) = \sum_{c=1}^C m_{jc} \alpha_{cb}(t)$ and therefore the introduction of the latent variable $\latvar_j$ leads to a mixture of Gaussian process regression \citep{shi2005hierarchical}.

\subsubsection{Clustering model likelihood and inference}
\label{sec:cluster-lk}

Let $\by$ be the vector of observed aggregated energy consumption over $I$ days at $J$ substations, as in \myeqref{eq:sample}, let $\mathbf{\slatvar} = \big( \slatvar_1, \dots, \slatvar_J\big)$ be the vector of latent variables and $\bpi = \big(\pi_{1} \dots, \pi_{B}  \big)$ its associated parameters. ~
Consider $\by_{\cdot j}=\big( \by_{1j},\ldots,\by_{Ij} \big)^\transp$, the observed data log-likelihood can be written as
\begin{align}
    \ell
    \big(
  \parsVec, \bpi | \by 
  \big)  
  =&
\sum_{j=1}^J 
     \log 
     \Big(
     \sum_{b=1}^B
     \pi_b
     f \big( \by_{\cdot j} | \slatvar_j , \parsVec, \bpi \big)
     \Big).     
     \label{eq:obs-lk}
\end{align}
The direct maximization of \myeqref{eq:obs-lk} to obtain parameter estimates is difficult due to the presence of the logarithm of a summation. Hence, we develop an Expectation-Maximization (EM) algorithm~\citep{dempster1977maximum,mclachlan2007algorithm}, which performs an iterative maximization of \myeqref{eq:obs-lk} using the joint distribution of $\byij$ and $\mathbf{\slatvar}_j$, with the so-called complete data log-likelihood given by:
\begin{align}
  \ell
  \big(
  \parsVec, \bpi | \by, \mathbf{\slatvar} 
  \big)
  = &
      \sum_{j=1}^J
      \sum_{b=1}^B \indic{\slatvar_j = b}
      \Big(
      \log f \big( \by_{\cdot j} | \slatvar_j ; \, \parsVec \big) + 
      \log \prob{\latvar_j = \slatvar_j | \bpi}
      \Big) \nonumber \\
      =&
     \mysum{j}{J}
     \mysum{b}{B}
    \indic{\slatvar_j = b} \times \nonumber \\
  \qquad&
     \Big(
          \log \pi_{b}
        -\frac{1}{2} 
        \sum_{i=1}^I 
        \big[
        \log \big |\bcovmatrix_{jb}| 
        +
       (\bmc_{jb} - \byij)^\transp
       \bcovmatrix_{jb}^{-1}
       (\bmc_{jb} - \byij)
       \big]
     \Big) + C,
      \label{eq:clusterlk}
\end{align}
\noindent where, similarly to Equation (\ref{eq:muasXbeta}), $\bmc_{jb}$ can be written as $\bmc_{jb}= \mathbf{X}_j\bbeta_{b}$.

The E-step of the EM algorithm calculates the expected value of $\ell \big(\parsVec, \bpi | \by,\mathbf{\slatvar}\big)$ in \eqref{eq:clusterlk} with respect to the conditional distribution of $\mathbf{\latvar}$ given the observed data and current parameter estimates $\parsVec^{(r)}$ and $\bpi^{(r)}$ at iteration $r$ to obtain  
\begin{align}
  &Q
  \big(
    \parsVec, \bpi
    |\,
    \parsVec^{(r)}, 
    \bpi^{(r)}
  \big)
  \equiv \:
     \mathbb{E}_{
       \mathbf{\latvar}    
       |
       \by,\parsVec^{(r)}, \bpi^{(r)}
     }
     \Big[
        \ell
        \big(
          \parsVec, \bpi
          |
          \by,
          \mathbf{\slatvar}
         \big)
     \Big] \nonumber \\
  & \qquad =
    \mysum{j}{J}
     \mysum{b}{B}
    \prob{\latvar_j = b | \by_{1j},\ldots,\by_{Ij}; \parsVec^{(r)}, \bpi^{(r)}} \times \nonumber \\
  &\qquad \qquad
     \Big( \log \pi_{b}  
             -\frac{I}{2} \log |\bcovmatrix_{jb}| 
          -\frac{1}{2}  \mysum{i}{I} 
             (\bmc_{jb} - \byij)^\transp
             \bcovmatrix_{jb}^{-1}
             (\bmc_{jb} - \byij) 
     \Big) + C.
    \label{eq:qfunction}
\end{align}


The probability $\mathbb{P}(\cdot)$ in \myeqref{eq:qfunction} can be computed using Bayes Theorem and written as
\begin{eqnarray}
    &\prob{\latvar_j = b | \by_{\cdot j}; \parsVec^{(r)}, \bpi^{(r)}}  
    = \frac
      {\left[
      \prod_{i=1}^I
      f\big( \by_{ij}
      | \slatvar_j=b ; \parsVec^{(r)}_b \big) \right]
      \times 
       \pi^{(r)}_{b}}
      {\mysum{b^\prime}{B} {
      \left[
      \prod_{i=1}^I
      f\big( \by_{ij} | \slatvar_j=b^\prime ; \parsVec^{(r)}_{b'}\big) 
      \right]
      \times 
      \pi^{(r)}_{b'} }}, \nonumber
    \label{eq:prob}
\end{eqnarray}

\noindent where the product of densities is possible because independence among days $i = 1, 2, \dots, I$ is assumed.


In the M-step we maximize the function $Q(\cdot)$ in \myeqref{eq:qfunction} with respect to the parameters $\parsVec = \big\{ \bbeta, \parsVec_{\bcovmatrix} \big\}$ and $\bpi$, where $\parsVec_{\bcovmatrix}$ contains the parameters $\bbeta^{\eta}$ and $\boldsymbol \omega$ of the covariance matrix $\bcovmatrix_{jb}$ described in Section~\ref{sec:model-lk}. The E-step probabilities $\mathbb{P}(\latvar_j = b | \by_{\cdot j}; \parsVec^{(r)}, \bpi^{(r)})$ are treated as fixed in the M-step since they depend only on the previous parameter estimates. Let 
\begin{equation}
    \pjbr \equiv \prob{\latvar_j = b | \by_{\cdot j}; \parsVec^{(r)}, \bpi^{(r)}} \nonumber
\end{equation}
\noindent and let $Q(\cdot)$  be written as a sum of two terms: one that depends only on $\bpi$ and another term that depends only on $\parsVec$, that is,
\begin{equation}
    Q
  \big(
  \parsVec, \bpi
  |\,
  \parsVec^{(r)}, \bpi^{(r)}
  \big)
  =
  Q_1
  \big(
  \bpi
  |\,
  \parsVec^{(r)}, \bpi^{(r)}
  \big)
  +
    Q_2
  \big(
  \parsVec
  |\,
  \parsVec^{(r)}, \bpi^{(r)}
  \big)
  \label{eq:qfunc-parted} \nonumber
\end{equation}

\noindent where
\begin{align}
   Q_1
  \big(
  \bpi
  |\,
  \parsVec^{(r)}, \bpi^{(r)}
  \big)
  &=
  \sum_{j=1}^J
  \sum_{b=1}^B
  \pjbr
  \log \pi_{b} \text{ and}
  \label{eq:q1}\\
  Q_2
  \big(
  \parsVec
  |\,
  \parsVec^{(r)}, \bpi^{(r)}
  \big) \nonumber
  &=
  -\frac{1}{2}
  \sum_{j=1}^J
  \sum_{b=1}^B
  \pjbr
  \big(
  \log |\bcovmatrix_{jb}|
  + \nonumber \\
  &\qquad \sum_{i=1}^I
     (\bX_j \bbeta_{b} - \byij)^\transp
     \bcovmatrix_{jb}^{-1}
     (\bX_j \bbeta_{b} - \byij) 
  \big).
  \label{eq:q2}
\end{align}

Because $Q_1$ does not depend on $\parsVec$, $\pi_b^{(r+1)}$ can be obtained by maximizing \myeqref{eq:q1} with respect to $\pi_b$, subject to $\mysum{b}{B} \pi_{b} = 1$. Therefore, using Lagrange multipliers it can be shown that
\begin{equation}
    \pi_b^{(r+1)} 
    = 
    \frac{1}{J}
    \sum_{j=1}^J
    \pjbr, 
    \label{eq:piq}
\end{equation}
\noindent for $b=1,\dots,B$.

To obtain $\bbeta_b^{(r+1)}$ and $\parsVec_{\bcovmatrix}^{(r+1)}$, we use the so-called Expectation/Conditional Maximization (ECM) algorithm~\citep{meng1993maximum, mclachlan2007algorithm}, where  $\parsVec_{\bcovmatrix}$ is set equal to $\parsVec_{\bcovmatrix}^{(r)}$ and  $Q_2$ in (\ref{eq:q2}) is maximized with respect to $\bbeta_b$ to obtain
\begin{equation}
    \bbeta_b^{(r+1)}
    =
    \left(
    I
    \mysum{j}{J}
    \bX_j^\transp
  \big(\bcovmatrix_{jb}^\star 
  \big)^{-1}
  \bX_j
  \right)^{-1}
  \left(
  \mysum{i}{I}
  \mysum{j}{J}
  \bX_j^\transp
  \big(\bcovmatrix_{jb}^\star \big)^{-1}
  \byij
  \right),
   \label{eq:betaq}
\end{equation}

\noindent for $b=1,\dots,B$ and $\bcovmatrix_{jb}^\star
    =
    \pjbr \times 
  \bcovmatrix_{jb}^{-1}$.
  
Next,  $\bbeta_b$ is set to its updated value $\bbeta_b^{(r+1)}$ and  $Q_2$ is maximized with respect to $\parsVec_{\bcovmatrix}$ through numerical optimization algorithms to obtain $\parsVec_{\bcovmatrix}^{(r+1)}$. E and M steps are then iterated until convergence is reached, that is, when $\big{|}\ell \big( \parsVec^{(r)},\bpi^{(r)};\by\big) - \ell \big( \parsVec^{(r-1)},\bpi^{(r-1)};\by\big) \big{|} < \xi $ for $\xi>0$.

\subsubsection{Initial values and number of clusters}
\label{sec:init}

Obtaining initial values for all parameters might be a challenge if no previous information is available to guide the initialization. In this work, the following approach is proposed for clustering and parameter initialization. 

The first step is to fix the number of clusters $B$ and the number of trials $G$. For each trial $g \in G$, each \group ~ is randomly assigned to a cluster, where the number of substations in each cluster must be greater than the number of customer types to preserve model identifiability. For each trial $g$, the clusters are split into datasets with their respective substations, and  a simple aggregated data model is fitted to each one. Then the model with the smallest squared error among the $G$ trials is selected to provide an initial $\bbeta_b$. The initial $\bpi$ is the proportion of \groups ~ in each cluster and the winning trial can also be used to provide initial covariance parameters.

The total number of clusters $B$ is highly dependent on  previous user information. As a first step, one might use the suggested approach of multiple fits with different numbers of clusters to select the configuration with the smallest squared error, which implies in high computing cost, or one might use an approximation of Bayes factors to select the best number of clusters $B$~\citep{schwarz1978estimating}. The latter approach is detailed in Section~\ref{sec:modelcheck} and is the approach used in this work. 

It is also possible to assume that $B$ is a random variable and obtain its estimated value through its posterior probability using approaches like the reversible jump algorithm~\citep{green1995reversible}, but with intensive computation. 

\subsubsection{Identifiability condition}
\label{sec:ident2}

As in Section~\ref{sec:ident}, there are necessary conditions for model fitting. Because there are at least  $B$ times the number of parameters, the procedure requires $J > CB$, that is, the number of \groups ~must be greater than the number of estimated typical curves. Furthermore, substation markets must not be proportional to ensure full rank matrices in least squares computations.

\subsection{Model check}
\label{sec:modelcheck}

This section will examine how to assess the uncertainty of the estimated mean curves and their covariance parameters.
Inferences on the disaggregated mean curves can be performed by taking the closed form of the parameters in \eqref{eq:betahat} and \eqref{eq:betaq}, because they are functions of the Gaussian process $Y_{ij}(\cdot)$~\citep{shi2011gaussian,tresp2001mixtures}. In fact, it can be said that
\begin{equation}
  \label{eq:betadist}
  \hat{\bbeta}
  \sim
  Normal
  \big(
  \bbeta,
  \mathbf{A}
  \bcovmatrix
  \mathbf{A}^\transp
  \big), 
\end{equation}

\noindent with $\bbeta \in \mathbb{R}^{CK}$ as the true expansion parameters and $\mathbf{A} \in \mathbb{R}^{CK \times NIJ}$ defined as $\mathbf{A} =(\bX^\transp \bcovmatrix^{-1}\bX \big)^{-1} \bX^\transp \bcovmatrix^{-1}$.

Using the distribution of $\hat{\bbeta}$ given by \eqref{eq:betadist}, confidence intervals can be determined based on the standard errors in the diagonal of $  \mathbf{A}\bcovmatrix \mathbf{A}^\transp$.
On the other hand, the covariance parameters $\boldsymbol \Theta_{\bcovmatrix}$ are obtained by numerical optimization using the Quasi-Newton methods available in the R language \citep{fletcher2013practical,rlang}, and the parameter standard errors can be obtained from the observed Hessian matrix $\mathbf{H}$, that is,
\begin{equation}
  \label{eq:secov}
  SE
  \big(
  \boldsymbol \Theta_{cov}
  \big)
  \equiv
  \sqrt{diag
    \big(
    \mathbf{H}^{-1}
  \big)}. \nonumber
\end{equation}

\noindent 

The proposed covariance structures make the aggregated data models a family of nested models, where the uniformly homogeneous one is a particular case of the homogeneous model which is a particular case of the complete model. Two model fits can be compared using the likelihood ratio test. Let $\mathcal{M}_1$ and $\mathcal{M}_2$ be the two  aggregated data models to be compared, with $\mathcal{M}_1$ nested in $\mathcal{M}_2$. Denote by $\ell(\mathcal{M}_1)$ and $\ell(\mathcal{M}_2)$ the log-likelihood of models $\mathcal{M}_1$ and $\mathcal{M}_2$, respectively. Then the likelihood ratio statistic $L$ is defined by
\begin{equation}
    L = -2
    \big(
    \ell(\mathcal{M}_2)
    -
    \ell(\mathcal{M}_1)
    \big),
\end{equation}
\noindent where the test statistic is asymptotically $\chi^2$ distributed with degrees of freedom equal to the difference in the number of  parameters between models. 

When comparing clustering models, if they have the same number of clusters but different covariance structures, then the same approach can be used to compare them. However, to compare models with distinct numbers of clusters,  the \textit{Bayesian information criterion} (BIC)  comparison is recommended~\citep{shi2008curve}. Let  $\ell\big(\parsVec,\bpi;\by \big)$ be the observed data log-likelihood and let $\hat{\parsVec}$ and $\hat{\bpi}$ be the maximum likelihood estimates, then the BIC  is given by
\begin{equation}
    BIC
    =
    -2
    \ell
    \big(
    \hat{\parsVec},
    \hat{\bpi} ;
    \by
    \big)
    +
    H \log (IJN), \nonumber
    \label{eq:bic}
\end{equation}
\noindent where $H$ is the total number of parameters, $I$  the total number of days, $J$ the number of \groups and $N$ the number of observed point in  time. Simulation studies for Gaussian process mixtures have shown that models with the smallest BIC tend to have the  correct number of clusters~\citep{shi2008curve}. 


Finally, if a model has a good fit to the observed data, the residual curves can be expected to oscillate randomly around the zero line.

\section{Analysis of UK electrical substation data}
\label{chap:results}

In this section we apply the proposed clustering and full aggregated data models to the electrical load profiles from twelve energy substations in the United Kingdom presented in Section \ref{sec:motivation}. In what follows, Section~\ref{sec:gavin-homog} presents the results of the simple aggregated model fit, followed by the full aggregated model in Section~\ref{sec:gavin-fm}. Finally, Section~\ref{sec:gavin-cluster} describes the clustering analysis results.

\subsection{Simple homogeneous aggregated data model}
\label{sec:gavin-homog}

\begin{figure}[t!]
  \centering
\includegraphics[width=.8\linewidth]{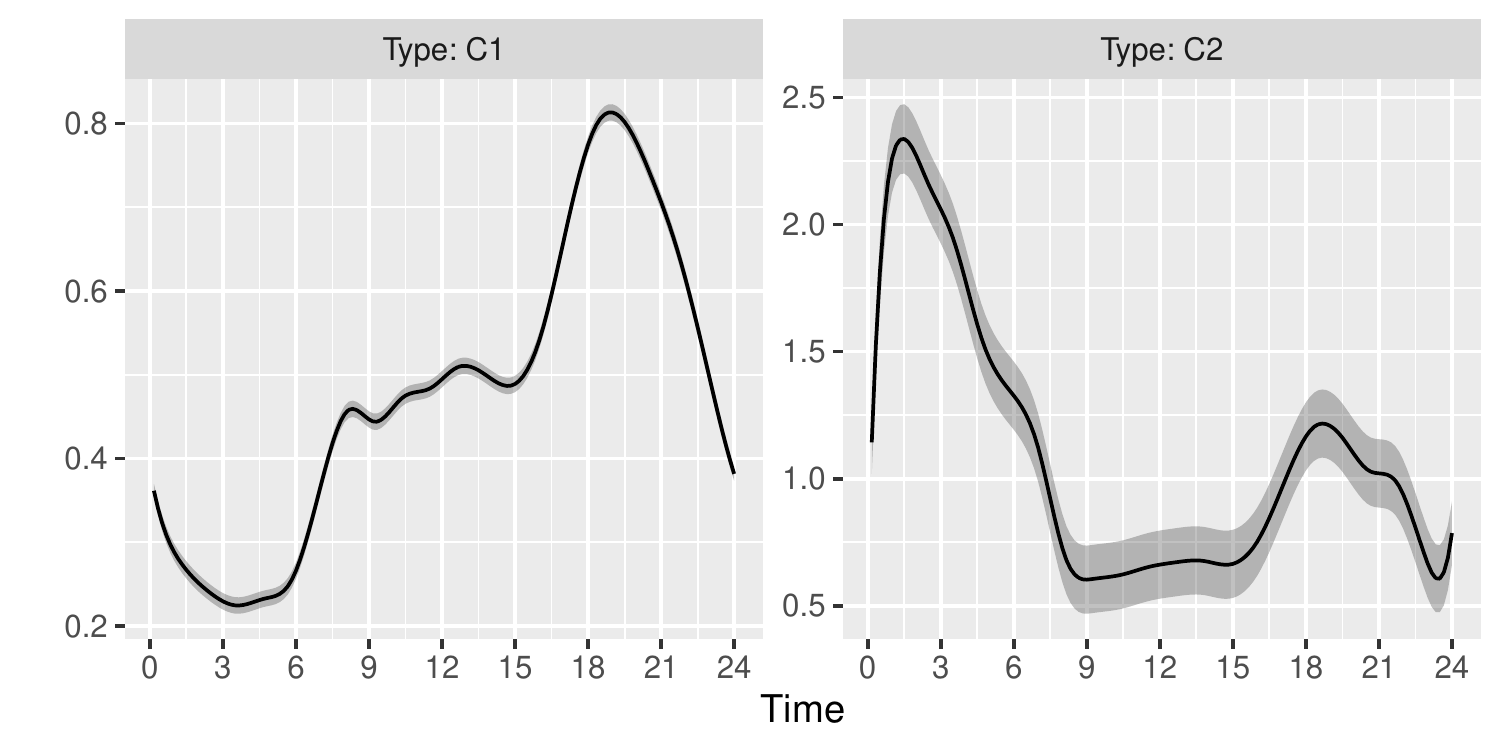} 

 \caption{Estimated typical curves in kWh and their confidence band (in gray) for unrestricted (C1) and ``Economy 7'' (C2) domestic customers using an homogeneous aggregated model.}
  \label{fig:gavin-mc1}
\end{figure}

The simple homogeneous aggregated data model described in Section 2.1 assumes the same homogeneous dispersion and decay parameters for all customer types. This might be a naive approach, but its results can be used as initial values for the full model, drastically improving its computational performance.

Figure~\ref{fig:gavin-mc1} displays the estimated typical curves considering 24 cubic B-Splines functions expansion. The unrestricted domestic (C1) customers  consume less energy than ``Economy 7'' domestic (C2) customers. The typical consumption curve of C1 customers shows modest values early in the morning, rising to a higher baseline in the traditional work period between 9 AM and 4 PM to finally reach their peak at 7 PM and complete the cycle by slowly returning to low consumption late at night. On the right panel, the typical curve for C2 customers is almost a mirror image of C1: the curve has its peak right after midnight and is constantly decreasing until it reaches its lowest values at 9 AM, when the cheaper tariffs cease. Later, there is a local peak around 7 PM, higher than C1, but still considerably lower than the early morning peak. 

Because both these customer types are domestic customers, certain behaviours can be conjectured to justify their typical curves. For example, unrestricted customers seem to have the habit of getting up in the morning and turning on electrical appliances that increase their load values, such as tea kettles, microwaves and hairdryers, for example. The work period presents many possibilities: most people leave their houses to go to work, decreasing home energy consumption, but some household members may stay at home to work in a home office. At night, when people arrive from their jobs, the appliances that are now turned on have higher energy consumption, such as washing machines and dryers that were not used in the morning. In contrast, the C2 typical curve has its major peak right after midnight, maybe due to appliances with higher energy cost making use of cheaper tariffs.  From 9am onward both typical curves increase their loads up to a plateau until approximately 4pm, where the consumption rapidly increase to the local peak. Furthermore, the confidence band for the C2 typical curve is larger than for C1 because the C1 class contains most of the market share (around 90\% or more) in most substations  (see Table \ref{tab:gavin-mkt}),  and consequently the amount of information available to estimate the C1 typical curve is greater than for C2. 

The estimated covariance parameters for the  homogeneous aggregated data model are displayed in Table~\ref{tab:gavin-pars1}. The dispersion parameter for C2 is considerably greater than for C1, with larger confidence bands in Figure~\ref{fig:gavin-mc1}. The small decay parameter for C1 indicates that correlation between energy consumption at two distinct points in time decays faster for C1 than for C2. This means that, given the same time window, energy consumption in C2 has a stronger dependence on values in its time neighbourhood than C1. Furthermore, the confidence intervals for the covariance parameters reveal no evidence in favour of the homogeneous uniform model because the intervals for each customer type do not overlap.

\begin{table}
\caption{\label{tab:gavin-pars1}Estimated covariance parameters of the simple homogeneous aggregated data model for the UK electrical energy dataset.}
\centering
\fbox{
\begin{tabular}{llrr}
Parameter & Type & Value & 95\% Confidence Interval\\
\midrule
 & C1 & 0.6608 & (0.6452, 0.6764)\\

\multirow{-2}{*}{\raggedright\arraybackslash $\sigma_c$} & C2 & 5.6094 & (5.4494, 5.7693)\\
\cmidrule{1-4}
 & C1 & 0.0404 & (0.0384, 0.0425)\\

\multirow{-2}{*}{\raggedright\arraybackslash $\omega_c$} & C2 & 0.8205 & (0.7721, 0.8689)\\
\end{tabular}}
\end{table}

\begin{figure}[pth]
  \centering
\includegraphics[width=.8\linewidth]{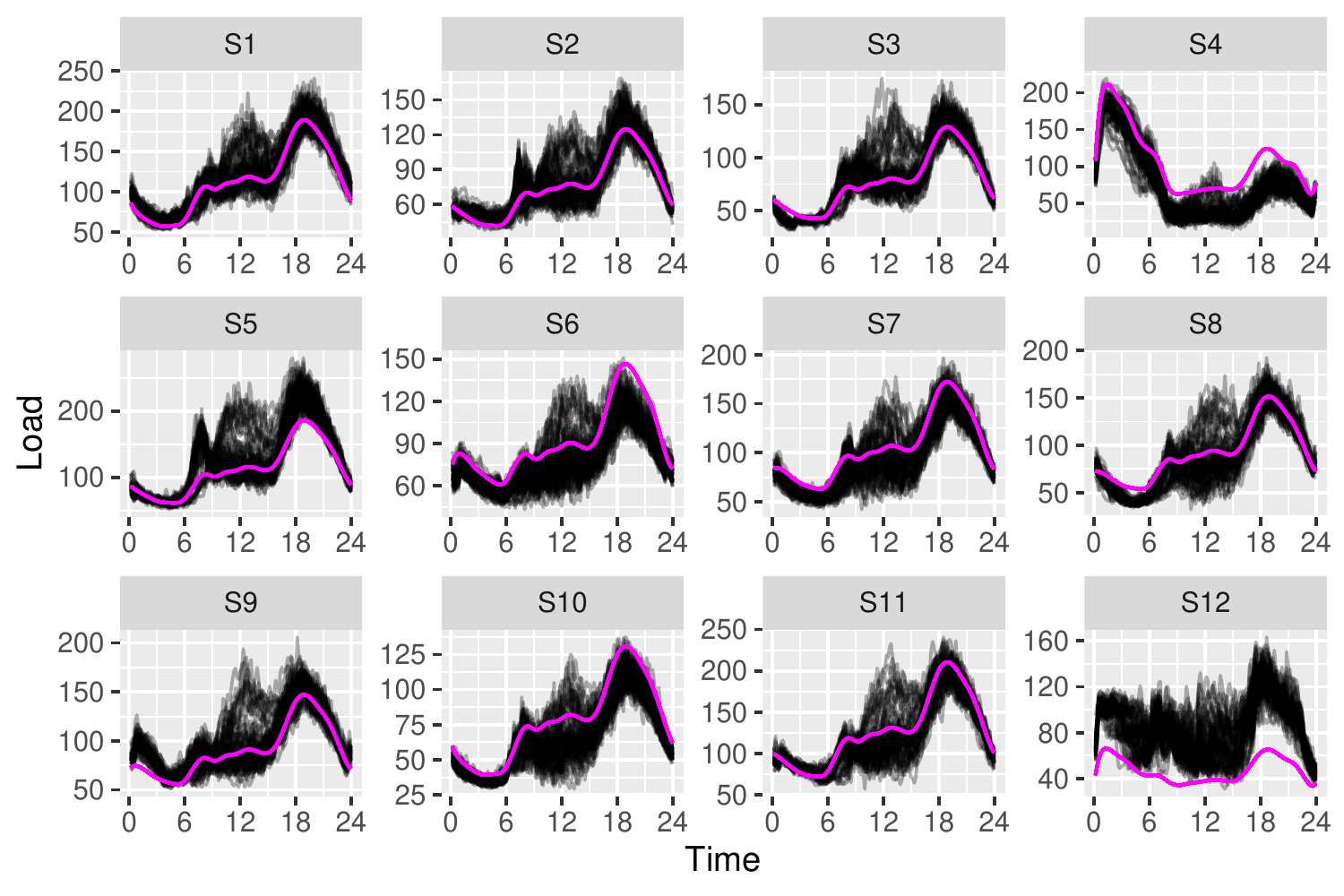}(a) 
\includegraphics[width=.8\linewidth]{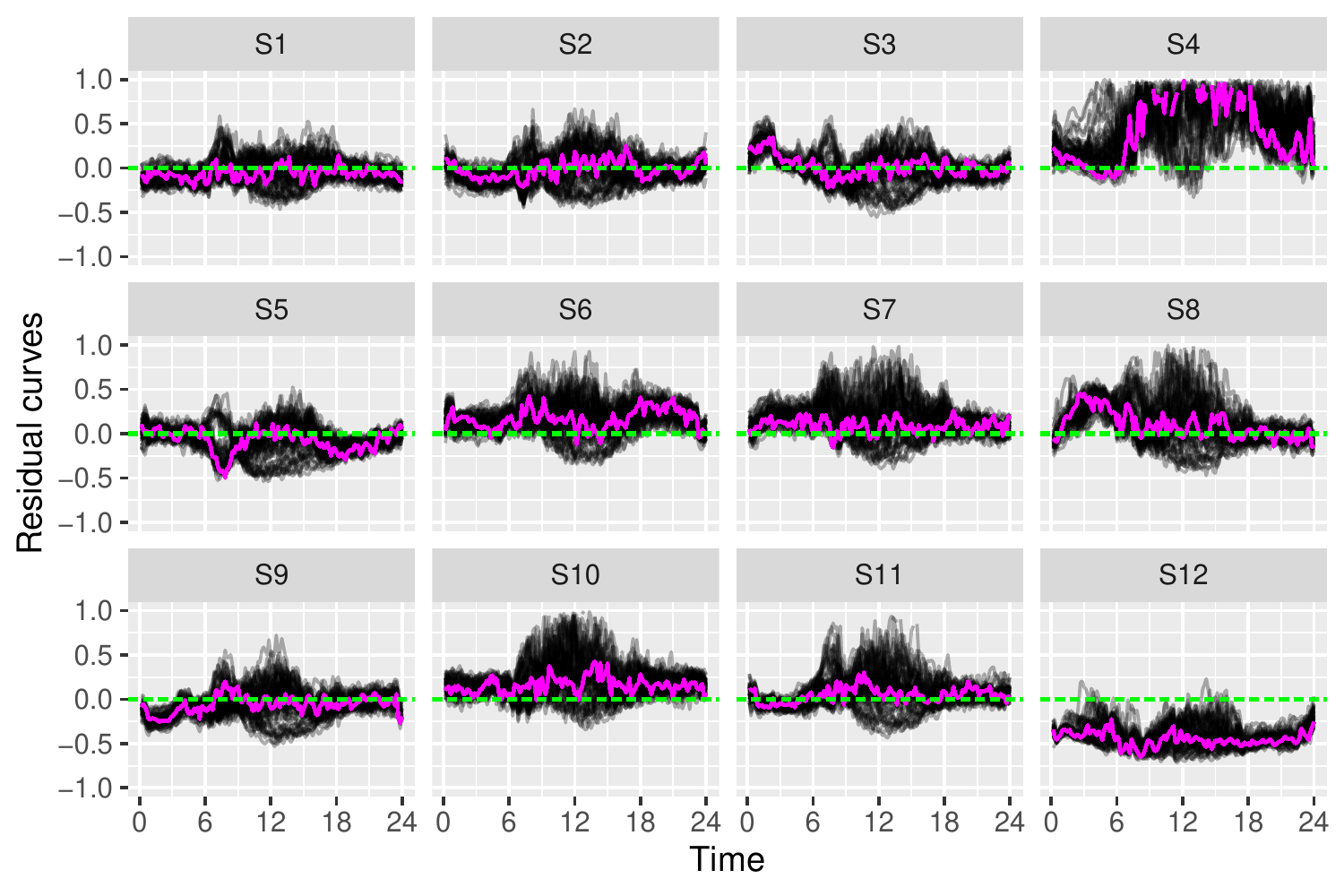}(b) 
  \caption{Simple homogeneous aggregated data model fit: (a) observed aggregated curves in kWh (in gray) and fitted aggregated curves (in magenta) and (b) Relative residual curves (in gray), median residual curve (in magenta) and zero reference line (in green) for each substation.}
    \label{fig:gavin-res1}
\end{figure}

To evaluate whether the model is suitable for the available data, the fitted aggregated curve is plotted along with the observed data in Figure~\ref{fig:gavin-res1}a. Apparently, the homogeneous model can capture the main features of the data, but fails to fit the aggregated load in some substations such as S4 and S12: S4 has overestimated fitted curves, whereas S12 has underestimated curves. This suggests that it might be interesting to add other explanatory variables, or dummy variables indicating these two substations in the full model approach because there appears to be a vertical shift of the fitted curves. Other small discrepancies are visible in other substations, but in general they follow the main features of the observed data.

Figure~\ref{fig:gavin-res1}b shows the relative residual curves defined in Section~\ref{sec:modelcheck} for  each substation with a reference line at zero and their median curve in green.  Ideally, the median should almost coincide with the zero-reference line, but note that there are curves positioned above or below the zero reference. Specifically, substations S4 and S12 are clearly under- and overestimated respectively. Furthermore, the homogeneous dispersion hypothesis does not hold for these data because the dispersion of the residual curves varies over time, which is another piece of evidence in favour of the complete aggregated  data model with variance functionals to capture this feature.

Important insights can be extracted from the homogeneous model fit before proceeding to the next level of the aggregated model, that is, the full model approach with additional covariates and a complete covariance structure. The bias in the fitted values for substations S4 and S12 suggests that indicator variables specific to these substations could be used as explanatory variables in the full model. In addition, the air temperature information is incorporated as a functional covariate to build the typical surface and potentially reduce the residual curves dispersion in the work period between 9am and 5pm.


\subsection{Full aggregated data model}
\label{sec:gavin-fm}

The full aggregated data model enables a functional covariate to be incorporated to produce typical surface responses for each customer type, as well as explanatory variables to better explain the aggregated data variability. For the tensorial product expansion in \myeqref{eq:surf-tensor},  $K=24$ and $L=6$ are used to estimate the typical surface. In addition, two explanatory variables are considered as indicators of substations S4 and S12, as mentioned in Section~\ref{sec:gavin-homog}. The variance functionals used in the complete covariance structure are expanded as in \myeqref{eq:completefunc_exp} with $K^\prime=6$.






\begin{table}
 \caption{\label{tab:gavin-temp-tab} Summary statistics of air temperature in degrees Celsius over the 61 observed days in the dataset for each substation primary (Q = Quartile).}
\centering
\fbox{
\begin{tabular}{lrrrrr}
Primary & Minimum & 1st Q & Median & 3rd Q & Maximum\\
\midrule
Cyncoed & -3.24 & 0.95 & 2.54 & 4.53 & 9.58\\
Llantarnam Primary & -4.22 & 0.89 & 3.30 & 7.90 & 12.06\\
Ringland Newport & -4.35 & 0.49 & 2.57 & 5.37 & 12.29\\
Trowbridge Primary & -5.41 & -0.78 & 1.26 & 3.29 & 9.44\\
Usk & -6.27 & 1.00 & 5.74 & 8.07 & 12.22\\
\end{tabular}}
\end{table}

As mentioned in Section~\ref{sec:motivation}, temperature data were extracted for each primary every three hours, interpolated by cubic B-Splines and displayed in Figure~\ref{fig:gavin1}b.  Table~\ref{tab:gavin-temp-tab} shows the summary statistics of the observed temperatures for each primary. Simulation studies (see Supplementary Material) showed  satisfactory estimated typical surfaces for temperature intervals frequently observed in the data, but higher dispersion in the estimate for rarely observed temperatures. In the case of the UK data, except for Trowbridge, temperature data are concentrated approximately between 1\textdegree C and 4 \textdegree C, and therefore the estimated typical surfaces may be well estimated within this range, but present some difficulties outside it. 


\subsubsection{Full aggregated data model}
\label{sec:gavin-fmfit}

\begin{figure}[thp]
\centering

\includegraphics[width=.8\linewidth]{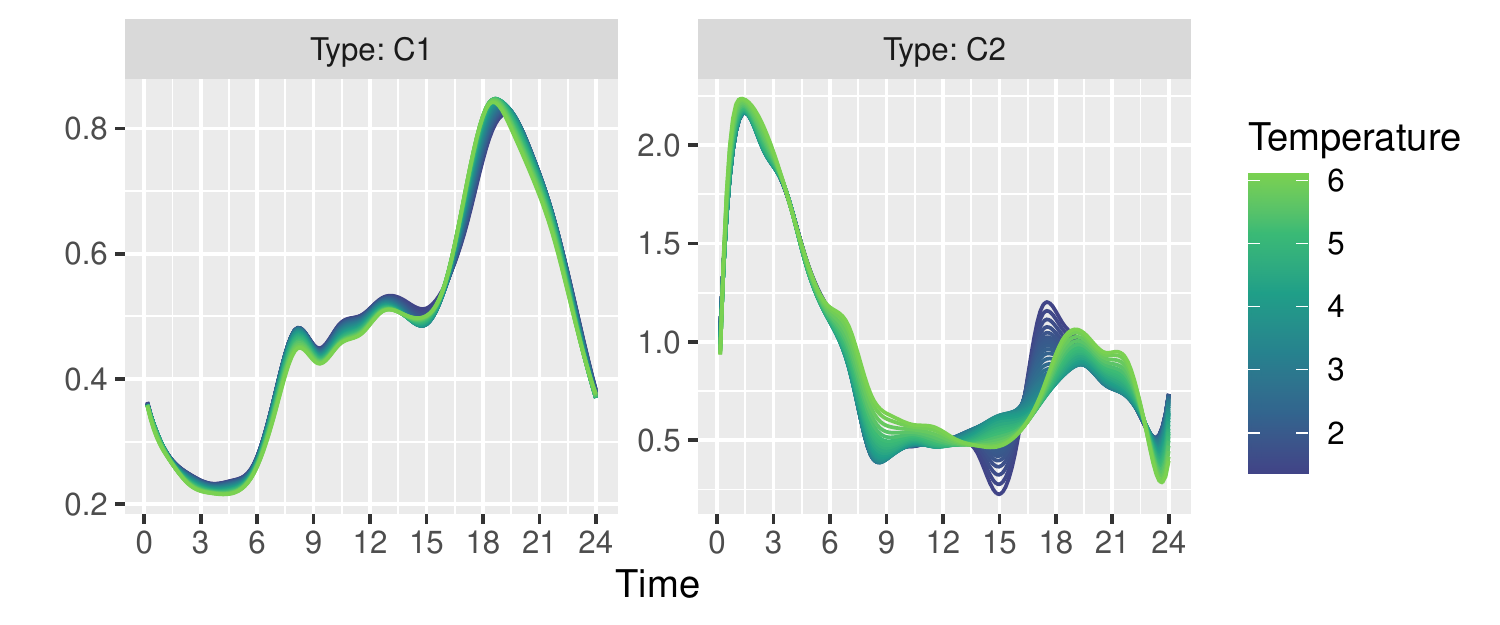}(a)


\includegraphics[width=.65\linewidth]{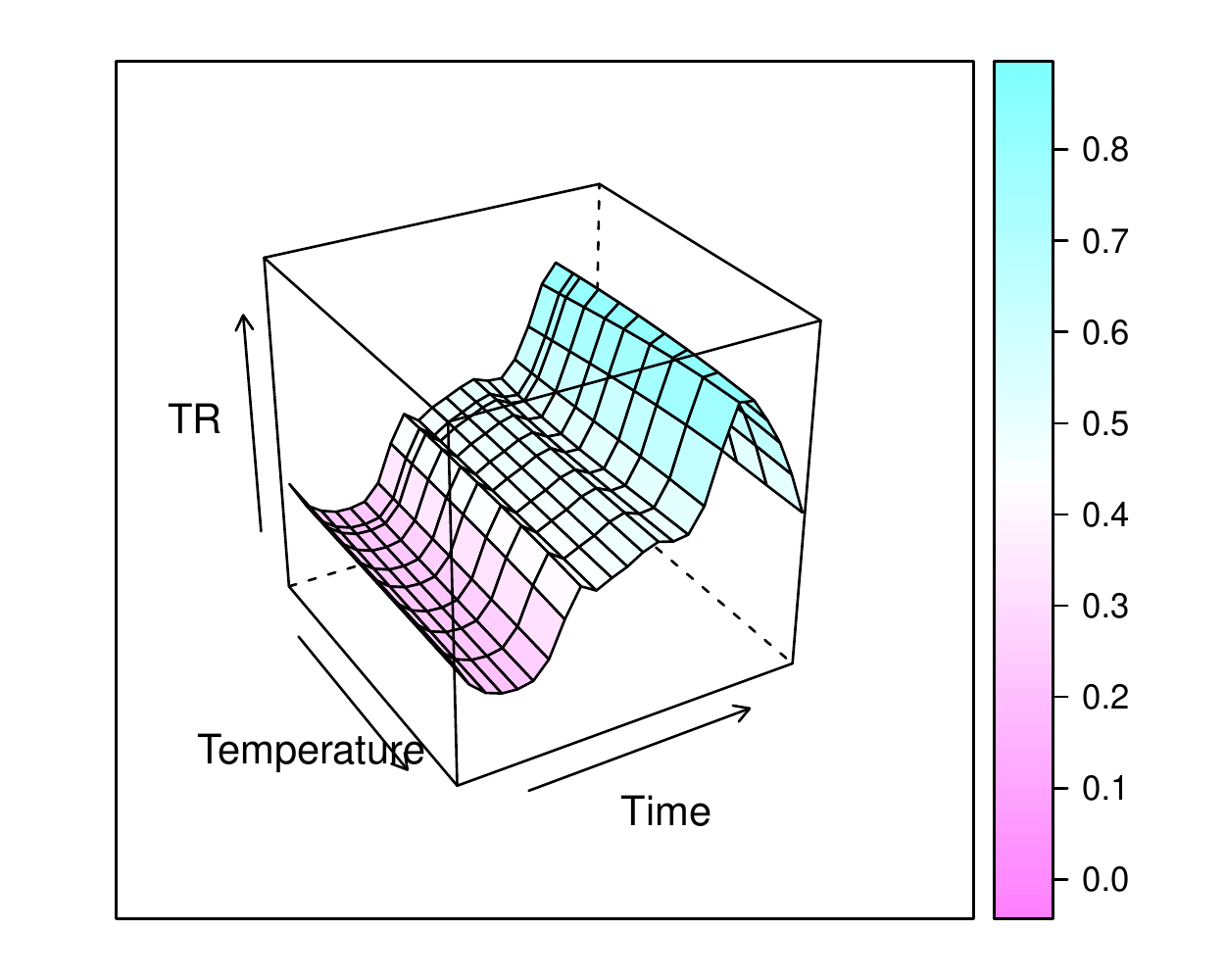}(b)


\includegraphics[width=.65\linewidth]{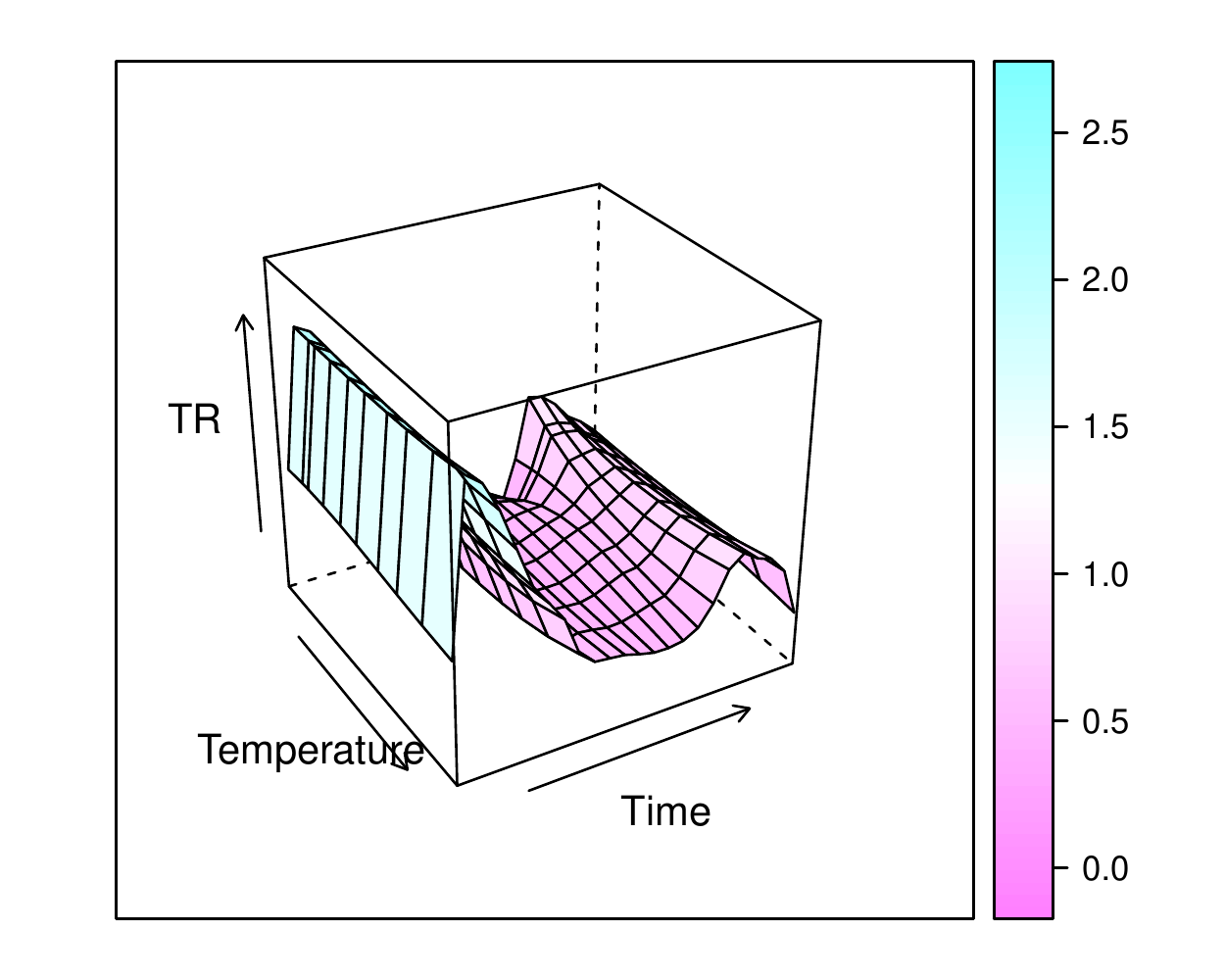}(c)


  
  \caption{(a) Estimated typical curves in kWh for customers of type C1 and C2 coloured according to temperatures between 1.21\textdegree C and 5.89\textdegree, (b) estimated typical surface response  in kWh for customers of type C1 between 1.21\textdegree C and 5.89\textdegree C and (c) estimated typical surface response in kWh for customers of type C2 between 1.21\textdegree C and 5.89\textdegree C. TR denotes the for typical response in kWh.}
    \label{fig:gavin-fm-mc}
  \end{figure}


Figure~\ref{fig:gavin-fm-mc}a shows the estimated typical curves for certain fixed temperature values, and Figures~\ref{fig:gavin-fm-mc}b and \ref{fig:gavin-fm-mc}c  show the estimated typical surfaces for C1 and C2 customer types, respectively, for temperatures between 1.21\textdegree C and 5.89\textdegree C. The selected temperature range contains 60\% of the observed values in the dataset. Hence, that is the interval where the typical surfaces are well estimated avoiding discrepant values that do not contribute to the analysis as shown in Section 1.3 of our Supplementary Material. On the time axis, the estimated typical surfaces have similar characteristics to the curves estimated by the simple aggregated model shown in Figure~\ref{fig:gavin-mc1}. On the temperature axis, unrestricted domestic C1 customers present robust behaviour for different temperatures, but C2 customers are subject to greater variation of energy consumption between 12 PM and 8 PM at different temperatures. In the latter case, extreme temperatures must be considered with caution because for values outside the selected range, the typical curves are unstable and may present negative or extremely high values.



\begin{table}
\caption{\label{tab:gavin-pars-fm}  Estimated coefficients of explanatory variables and estimated covariance parameters followed by their 95\% confidence intervals using the full aggregated data model.}
\centering
\fbox{
\begin{tabular}{llr}
Parameter  & Value & 95\% Confidence Interval\\
\midrule
S12 & 37.3968 & (33.5605, 41.2331)\\
\cmidrule{1-3}
S4  & -8.1977 & (-19.9304, 3.5350)\\
\cmidrule{1-3}
 $\omega_{C1}$ &0.0333 & (0.0313, 0.0353)\\

$\omega_{C2}$ & 0.6127 & (0.5803, 0.6450)\\
\end{tabular}}
\end{table}

The first two lines of Table~\ref{tab:gavin-pars-fm} show the estimated effect values for the dummy explanatory variables corresponding to substations S4 and S12. Note that the estimated effect of substation S12 is a shift of 37.40, which is a considerable value because the aggregated observations in this location are mainly around 50 kWh and 120 kWh. Substation S4 results in an estimated effect of -8.20, but its 95\% confidence interval contains zero, revealing that it may have no effect on the aggregated load data. The remaining lines of Table~\ref{tab:gavin-pars-fm} present the estimated covariance decay parameters for C1 and C2 customers ($\omega_{C1}$  and $\omega_{C2}$), which are much like to the ones obtained in Table~\ref{tab:gavin-pars1}. The correlation of neighbouring observations is stronger in type C2, with a decay parameter estimated at 0.61 versus 0.03 for type C1. 

\begin{figure}[t]\centering

\includegraphics[width=.8\linewidth]{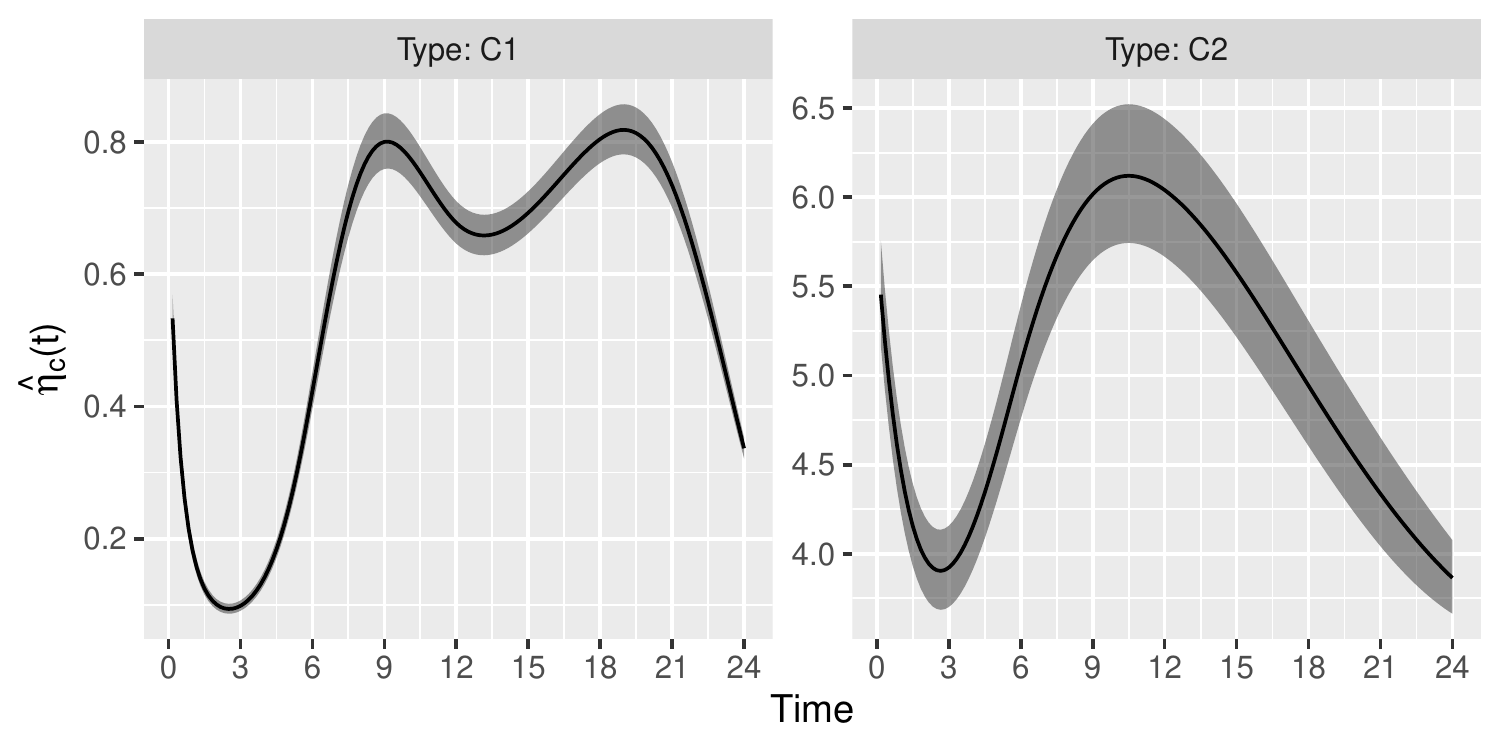} 
\caption{Estimated variance functionals for C1 and C2 customers along with their confidence bands using the full aggregated data model.}
\label{fig:gavin-varfunc}
\end{figure}

Figure~\ref{fig:gavin-varfunc} shows the estimated variance functionals for C1 and C2 customers along with their confidence bands built using their standard error as described in Section~\ref{sec:modelcheck}. The left panel reveals the higher values of dispersion at 9 AM, when people tend to leave their houses, and at 8 PM, the peak of the estimated typical curve. The lowest values are observed in early morning, a period with few or no activities in residences. On the right panel, note the peak around 10 AM and the lowest values around 10 PM and 3 AM. Interestingly, except for midnight, the lower tariff period has lower dispersion values.

\begin{figure}[t]\centering

\includegraphics[width=.8\linewidth]{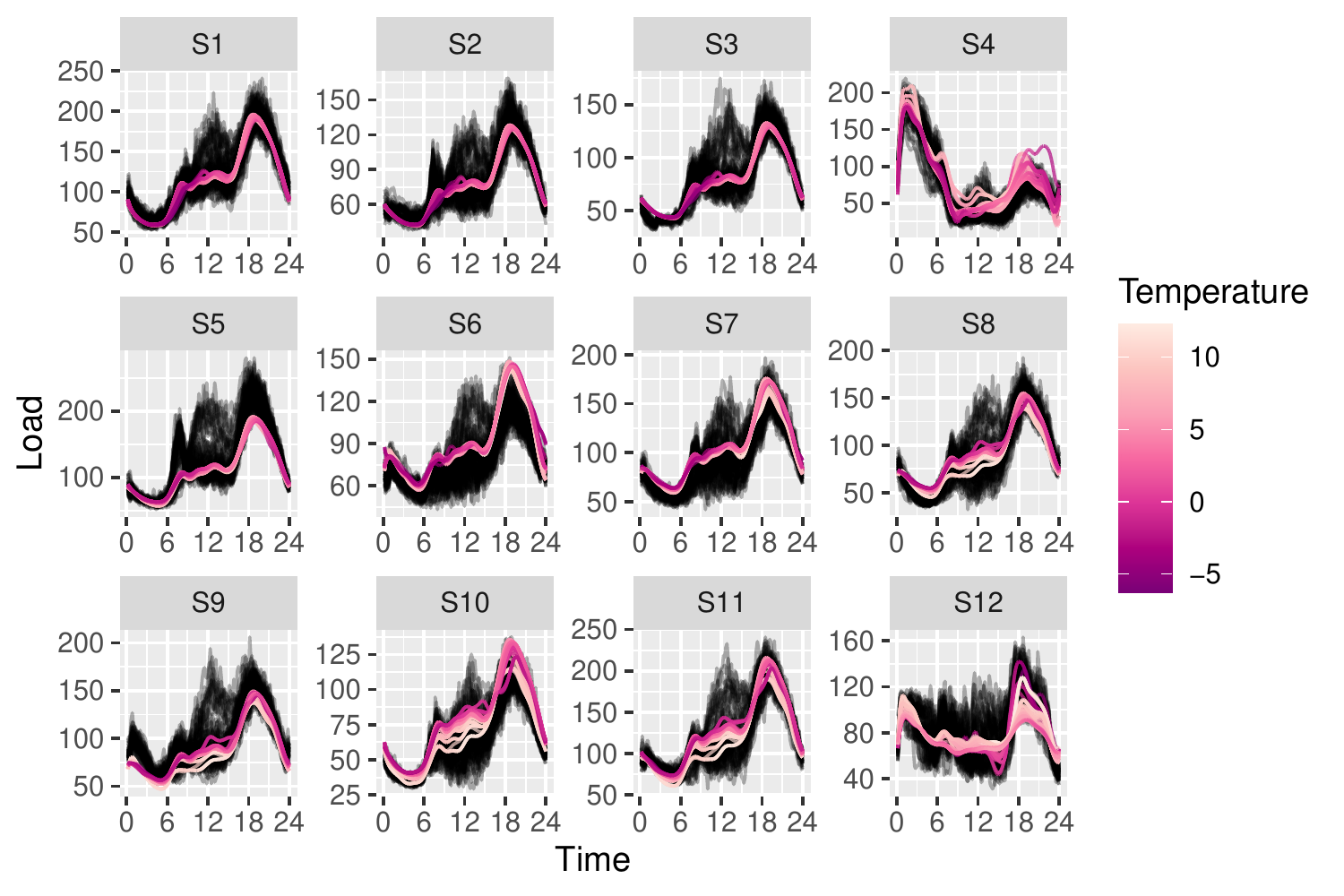}(a)


\includegraphics[width=.8\linewidth]{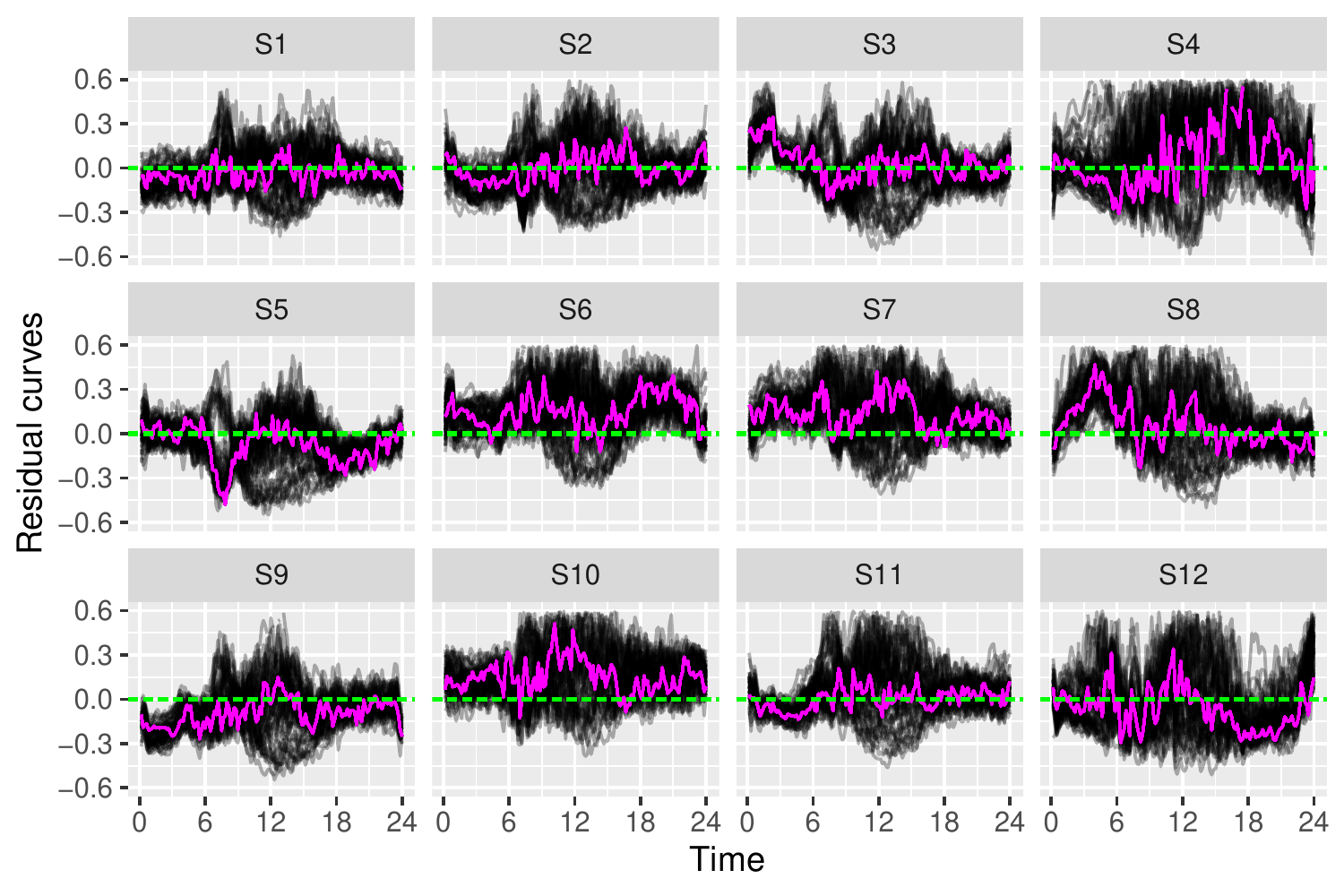}(b)

\caption{Full aggregated data model typical curves results: (a) Observed aggregated load data (in gray) in kWh over estimated aggregated curves (in tones of magenta) and (b) relative error curves (in gray), median residual curves (in green) and zero reference line (in magenta) for the 12 substations using the full aggregated data model.}
\label{fig:gavin-res2}
\end{figure}

Figure~\ref{fig:gavin-res2}a shows the fit for the full aggregated data model along with the observed aggregated data. The over- and underestimation problem for substations S4 and S12 is solved by adding dummy variables, and including temperature captures a portion of data variability, but it is still difficult to explain the work period variability, which may be associated with other factors.  Figure~\ref{fig:gavin-res2}b presents the associated relative residual curves. Greater variability can be observed between 9 AM and 5 PM, but the residuals for S4 and S12 are closer to the zero-reference line than in Figure~\ref{fig:gavin-res1}b.

\subsubsection{Comparison with simple homogeneous aggregated data model}

This section compares the proposed full aggregated data model with the simple homogeneous aggregated data model described in Section~\ref{sec:gavin-homog} to assess whether there are advantages to incorporating explanatory variables and temperature in terms of model fitting. 

Table \ref{tab:mod-res-comparison} shows the functional mean squared relative error (fMSRE) between fitted and observed values for each substation and fitted model given by:
\begin{equation}
\mbox{fMSRE}_j = \frac{1}{I}\sum_{i=1}^I \Big\{ \frac{T}{N} \sum_{t=t_1}^{t_N} \big( \hat{y}_{ij}(t) - y_{ij}(t) \big)^2 \Big\}. \nonumber
\end{equation}

\noindent Except for substations S8 and S9, the full aggregated data model has smaller fMSRE for most substations. The larger differences are observed in S4 and S12, both substations that were not well fitted by the homogeneous aggregated model, as shown in Figure~\ref{fig:gavin-res1}a. Considering all substations, the average fMSRE of the full model is better in terms of fMSRE, with a value of 0.208, whereas the homogeneous model has an average fMSRE of 0.266.

\begin{table}
\caption{\label{tab:mod-res-comparison}Functional mean squared relative error (fMSRE) of fitted and observed data for each substation and fitted model.}
\centering
\fbox{
\begin{tabular}{lrr}
Substation & Model & fMSRE\\
\midrule
 & Homogeneous & 0.1350\\

\multirow{-2}{*}{\raggedright\arraybackslash S1} & Full & 0.1295\\
\cmidrule{1-3}
 & Homogeneous & 0.1527\\

\multirow{-2}{*}{\raggedright\arraybackslash S2} & Full & 0.1518\\
\cmidrule{1-3}
 & Homogeneous & 0.1636\\

\multirow{-2}{*}{\raggedright\arraybackslash S3} & Full & 0.1629\\
\cmidrule{1-3}
 & Homogeneous & 0.8097\\

\multirow{-2}{*}{\raggedright\arraybackslash S4} & Full & 0.3811\\
\cmidrule{1-3}
 & Homogeneous & 0.1713\\

\multirow{-2}{*}{\raggedright\arraybackslash S5} & Full & 0.1690\\
\cmidrule{1-3}
 & Homogeneous & 0.2390\\

\multirow{-2}{*}{\raggedright\arraybackslash S6} & Full & 0.2241\\
\cmidrule{1-3}
 & Homogeneous & 0.2279\\

\multirow{-2}{*}{\raggedright\arraybackslash S7} & Full & 0.2221\\
\cmidrule{1-3}
 & Homogeneous & 0.2485\\

\multirow{-2}{*}{\raggedright\arraybackslash S8} & Full & 0.2387\\
\cmidrule{1-3}
 & Homogeneous & 0.1609\\

\multirow{-2}{*}{\raggedright\arraybackslash S9} & Full & 0.1680\\
\cmidrule{1-3}
 & Homogeneous & 0.2763\\

\multirow{-2}{*}{\raggedright\arraybackslash S10} & Full & 0.2699\\
\cmidrule{1-3}
 & Homogeneous & 0.1733\\

\multirow{-2}{*}{\raggedright\arraybackslash S11} & Full & 0.1693\\
\cmidrule{1-3}
 & Homogeneous & 0.4389\\

\multirow{-2}{*}{\raggedright\arraybackslash S12} & Full & 0.2154\\
\end{tabular}}
\end{table}

Because the homogeneous aggregated data model is nested inside the full aggregated data model, the likelihood ratio test can be performed to verify whether the model fits are statistically different. Hence, by designating the full model as $\mathcal{M}_1$ and the simple model as $\mathcal{M}_2$, the test statistic $L$ can be computed as
\begin{align}
    L &= -2 
    \big(
    \ell(\mathcal{M}_1)
    -
    \ell(\mathcal{M}_2)
    \big) 
    = -2
    \big(
    -344,770.3
    -
    (-358,204.1)
    \big) \nonumber\\
    & = 26,867.67. \nonumber
\end{align}

\noindent Under the null hypothesis, the test statistic has approximately a chi-square distribution with 254 degrees of freedom obtained from the difference of the number of model parameters. Hence,  the difference between the models is statistically significant with $p$-value very close to zero. 

Therefore, the full aggregated data model is a better fit, improving the explanation of the aggregated load variability by adding the temperature component and dummy variables. The assumption of a complete covariance structure could capture the variability over time by means of the estimated variance functionals. Although the estimated surfaces might be used with caution in temperature ranges with few observations, in general they are useful to assess electrical energy consumption under different weather conditions. In addition, many other functional variables can be included in the model either as a higher-dimensional surface of additive linear or non-linear terms or as other explanatory variables, scalar or functional, to explain the aggregated data variability.


 %

\subsection{Clustering analysis}
\label{sec:gavin-cluster}

The clustering aggregated data model groups substations with similar typical curves and covariance structure for  domestic customers of type unrestricted and ``Economy 7''. The model assumes that the aggregated observed data are a mixture of B aggregated models with distinct mean curves, with B as the total number of clusters. This section describes the fitting of a mixture of aggregated models with homogeneous covariance structure considering two and three clusters to obtain the best substation clustering that explains observed aggregated data variability using the Bayesian Information Criterion. There will be no explanatory variables or temperature components in this model. Models with four or more clusters do not meet the condition of identifiability to obtain typical curves and covariance parameter estimates because there are only 12 substations.


\subsubsection{Two clusters}
\label{sec:gavin-2cl}



\begin{figure}[t]
  \centering

\includegraphics[width=.8\linewidth]{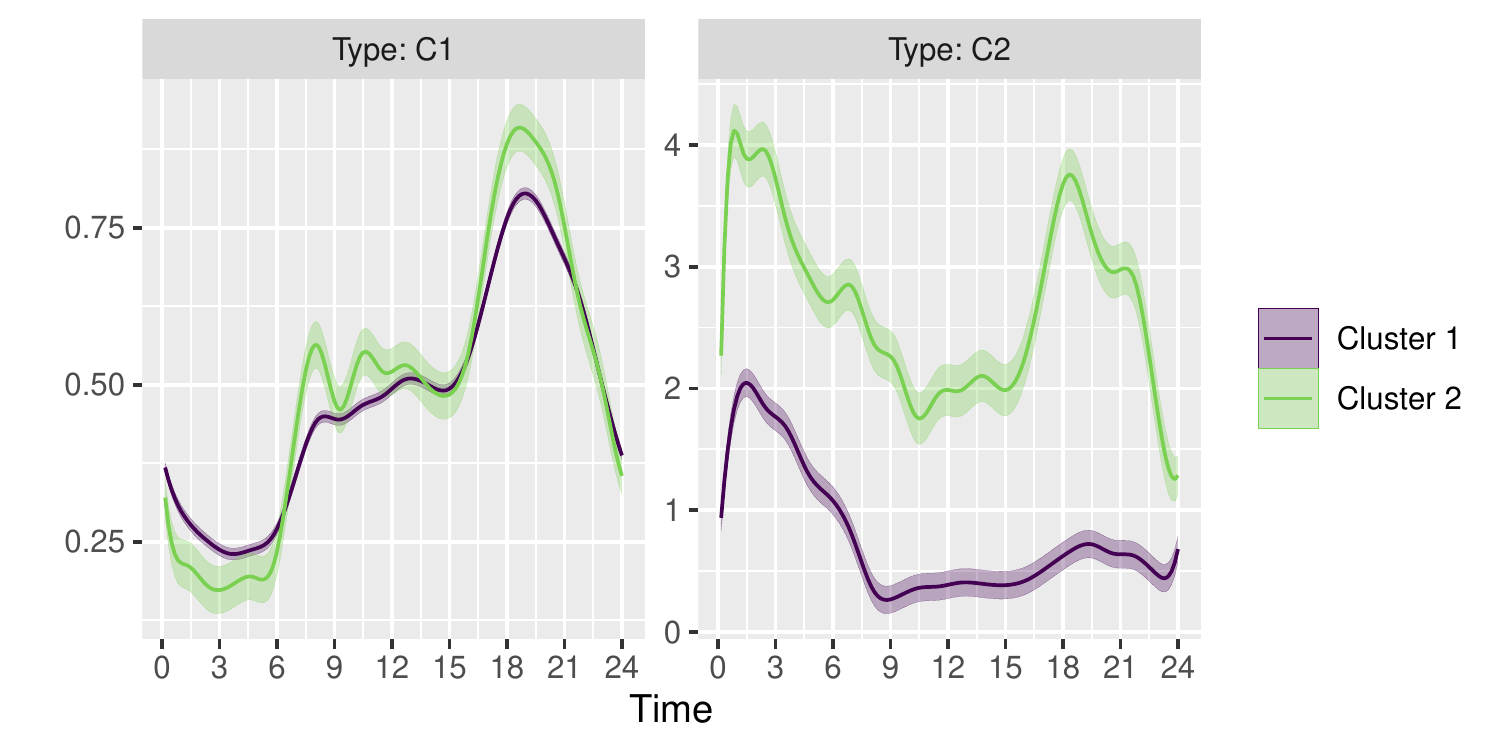}

  \caption{Estimated typical curves in kWh and their confidence band for unrestricted (C1) and ``Economy 7'' (C2) domestic customers under two clusters aggregated data model fit.}
  \label{fig:gavin-cl2-mc}
\end{figure}

Figure~\ref{fig:gavin-cl2-mc} shows the estimated typical curves for customers of type C1 and C2 in Clusters 1 and 2. Type C1 curves share characteristics in both clusters like the increasing load around 9am, the plateau in the middle of the day, and the highest consumption at 8pm. Customers of type C2 have peaks at 2am in both  clusters, but with different magnitudes, with Cluster 1 being the smaller one. The clustering aggregated model reveals new features for C2 customers, such as the different 8 PM peak, that could not be identified with the aggregated data models in Section~\ref{sec:gavin-fm}.


\begin{table}
\caption{\label{tab:gavin-cl-pi}Estimated probability $\hat{p}_{jb}$ of substation $j$ belonging to cluster $b$, under the two cluster fit.}
\centering
\fbox{
\begin{tabular}{lrrrrrrrrrrrr}
\multicolumn{1}{c}{ } & \multicolumn{3}{c}{Trowbridge} & \multicolumn{2}{c}{Cyncoed} & \multicolumn{2}{c}{Ringland} & \multicolumn{4}{c}{Llantarnam} & \multicolumn{1}{c}{Usk} \\
\cmidrule(l{3pt}r{3pt}){2-4} \cmidrule(l{3pt}r{3pt}){5-6} \cmidrule(l{3pt}r{3pt}){7-8} \cmidrule(l{3pt}r{3pt}){9-12} \cmidrule(l{3pt}r{3pt}){13-13}
  & S1 & S2 & S3 & S4 & S5 & S6 & S7 & S8 & S9 & S10 & S11 & S12\\
\midrule
$\hat{p}_{j1}$ & \cellcolor[HTML]{CECECE}{1} & \cellcolor[HTML]{CECECE}{1} & \cellcolor[HTML]{CECECE}{1} & \cellcolor[HTML]{CECECE}{1} & \cellcolor[HTML]{FFFFFF}{0} & \cellcolor[HTML]{CECECE}{1} & \cellcolor[HTML]{CECECE}{1} & \cellcolor[HTML]{CECECE}{1} & \cellcolor[HTML]{CECECE}{1} & \cellcolor[HTML]{CECECE}{1} & \cellcolor[HTML]{CECECE}{1} & \cellcolor[HTML]{FFFFFF}{0}\\
$\hat{p}_{j2}$ & \cellcolor[HTML]{FFFFFF}{0} & \cellcolor[HTML]{FFFFFF}{0} & \cellcolor[HTML]{FFFFFF}{0} & \cellcolor[HTML]{FFFFFF}{0} & \cellcolor[HTML]{CECECE}{1} & \cellcolor[HTML]{FFFFFF}{0} & \cellcolor[HTML]{FFFFFF}{0} & \cellcolor[HTML]{FFFFFF}{0} & \cellcolor[HTML]{FFFFFF}{0} & \cellcolor[HTML]{FFFFFF}{0} & \cellcolor[HTML]{FFFFFF}{0} & \cellcolor[HTML]{CECECE}{1}\\
\end{tabular}}
\end{table}

Table~\ref{tab:gavin-cl-pi} shows the estimated probability $\hat{p}_{jb}$  of substation $j$ being allocated to cluster $b$. Substations S5 and S12 are grouped in Cluster 2, and Cluster 1 gathers the remaining substations into a large cluster with 10 elements. Interestingly, S12 is the substation located far to the north, as shown in Figure~\ref{fig:map}, and one of the few substations that does not show an extreme dominance of C1 customer type; on the other hand, substation S5 has one of the markets dominated by C1 customers.


\begin{figure}[t]
  \centering

\includegraphics[width=.8\linewidth]{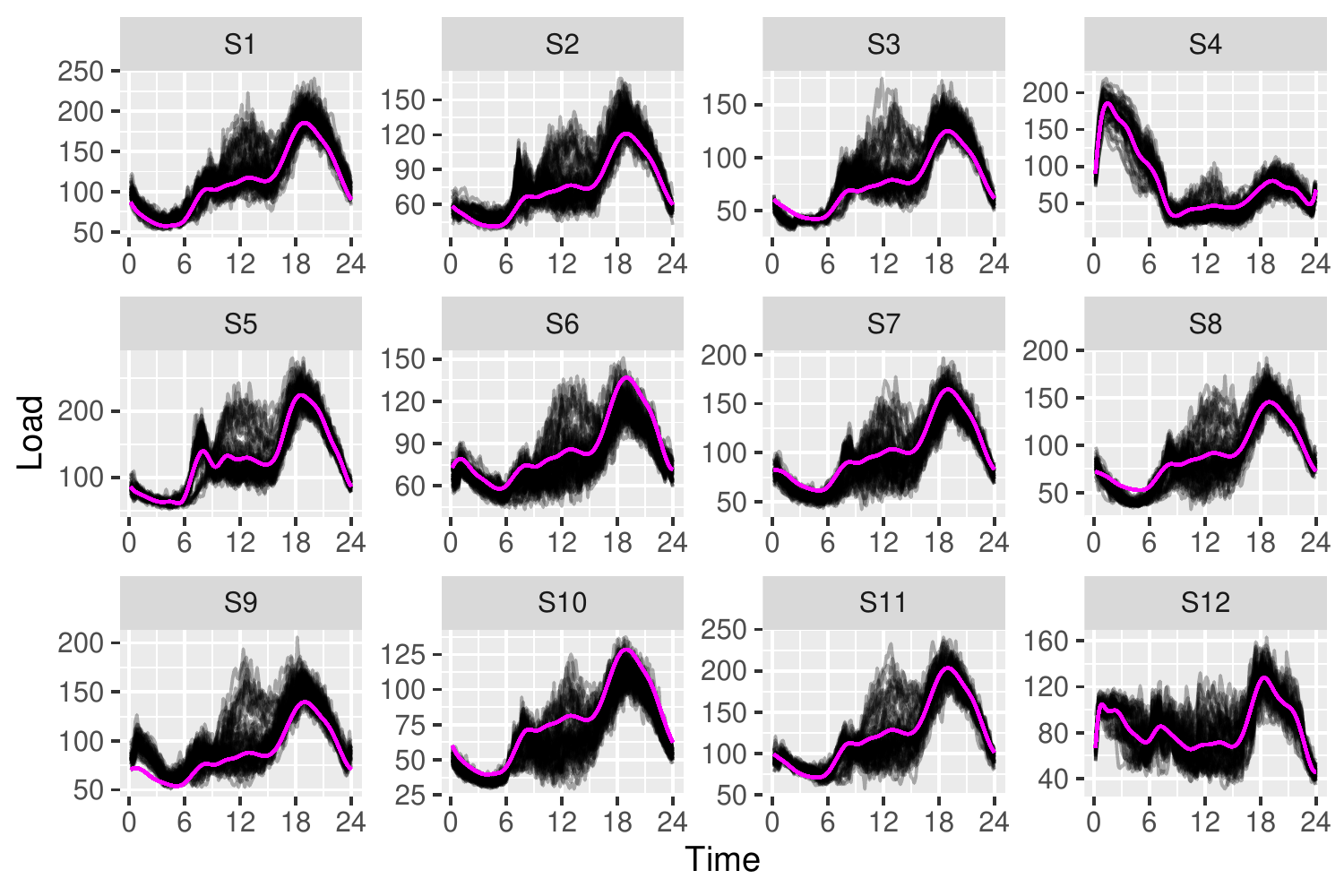}(a)


\includegraphics[width=.8\linewidth]{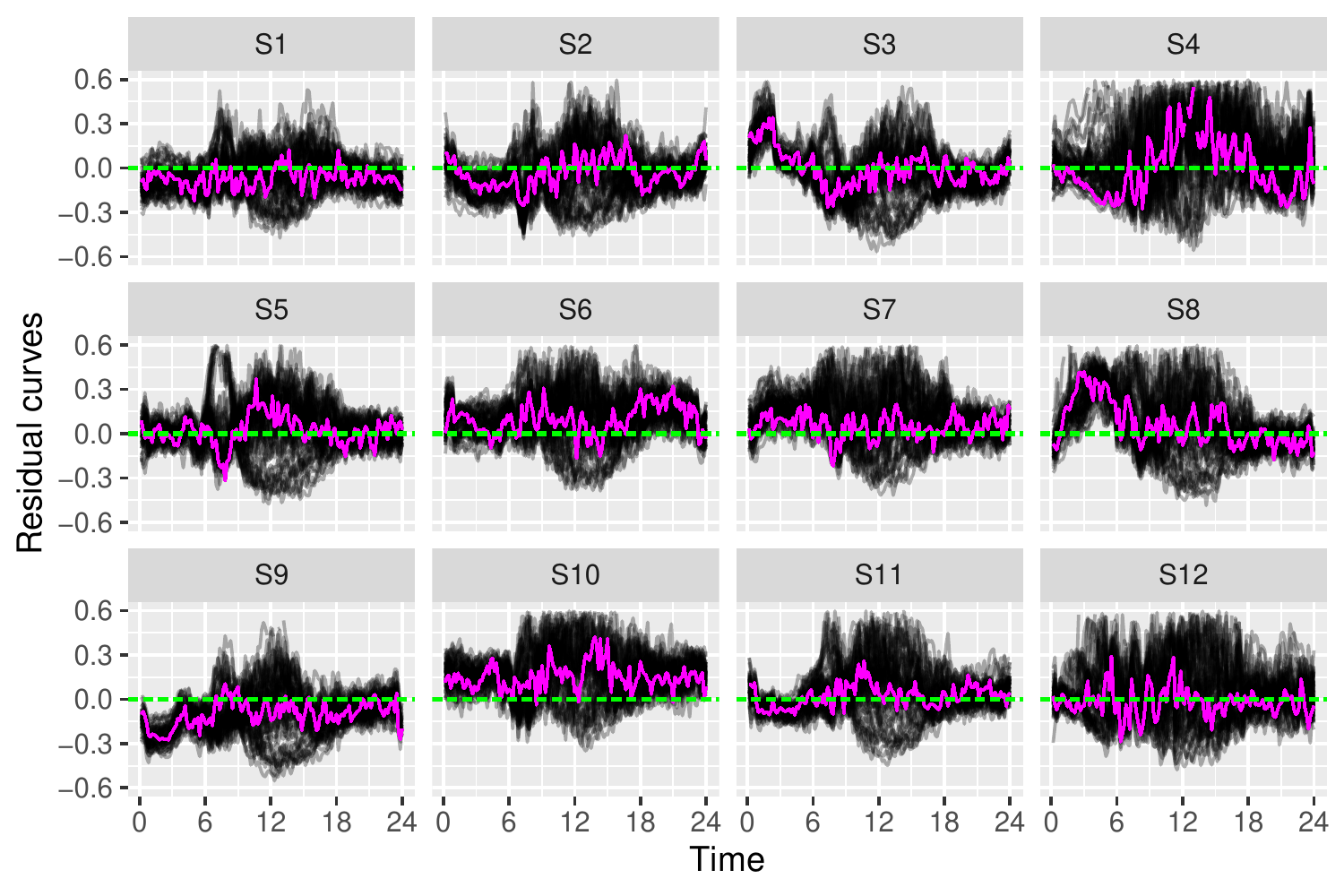}(b)

    \caption{Results of the clustering aggregated data model with two clusters: (a) Observed aggregated load data (in gray) in kWh over estimated aggregated curves (in magenta) and (b) relative error curves (in gray), median residual curves (in green) and zero reference line (in magenta) for the 12 substations.}
      \label{fig:gavin-cl2-res}
\end{figure}

Figure~\ref{fig:gavin-cl2-res}a shows the fitted values plotted along with observed aggregated load data. Note that the model can explain most of the aggregated data variability. In contrast to the full homogeneous aggregated data model, the impact of the clustering approach is visible on substation S12. The fact that the model enables this substation to have an estimated typical curve different than most of the others shows that this clustering approach is sufficient to explain the electrical load variability without a dummy explanatory variable to shift estimated fitted values. In substation S5, this impact is not evident because C2 has only 3.11\% of market share. Furthermore, the relative residual curves in Figure~\ref{fig:gavin-cl2-res}b shows that the clustering aggregated model has median residual curve oscillating around the zero reference line and indicating that it is well adjusted to the observed data. However, the difference in variability over time in the residual curves suggests that the complete covariance structure with variance functionals might be more suitable. 

\begin{table}
\caption{\label{tab:gavin-pars-cl2}Estimated dispersion ($\sigma_{cb}$) and decay ($\omega_{cb}$) parameters for customer type $c$ in cluster $b$ of the clustering aggregated model considering two clusters.}
\centering
\fbox{
\begin{tabular}{lrr}
Parameter & Value & 95\% Confidence Interval\\
\midrule
$\sigma_{11}$ & 0.7016 & (0.6856, 0.7175)\\
$\sigma_{21}$ & 4.3629 & (4.2214, 4.5045)\\
$\omega_{11}$ & 0.0491 & (0.0466, 0.0515)\\
$\omega_{21}$ & 1.0033 & (0.9406, 1.0661)\\
\addlinespace
$\sigma_{12}$ & 1.5410 & (1.4834, 1.5985)\\
$\sigma_{22}$ & 1.5375 & (1.461, 1.6139)\\
$\omega_{12}$ & 0.1588 & (0.146, 0.1716)\\
$\omega_{22}$ & 0.0277 & (0.0244, 0.0310)\\
\end{tabular}}
\end{table}

The estimated covariance parameters for both clusters are displayed in  Table~\ref{tab:gavin-pars-cl2}. When compared to the homogeneous model of Section~\ref{sec:gavin-homog}, the estimated parameters of Cluster 1 are closer to those presented in Table~\ref{tab:gavin-pars1}. Still in Cluster 1, the results present a large estimated dispersion parameter for C2 customer types, possibly related to the small number of customers in the market and the difficulty of representing the variability of the period between 9 AM and 5 PM by a single typical curve.

In summary, the clustering aggregated data model with two clusters provides satisfactory fitted curves (see Figure~\ref{fig:gavin-cl2-res}) and typical curves that capture different characteristics for each cluster, especially customers of type C2. Even with no explanatory variables or additional temperature component it was possible to explain most of the variability of the load profiles.

\subsubsection{Three clusters}
\label{sec:gavin-3cl}

\begin{figure}[t]
  \centering
\includegraphics[width=.8\linewidth]{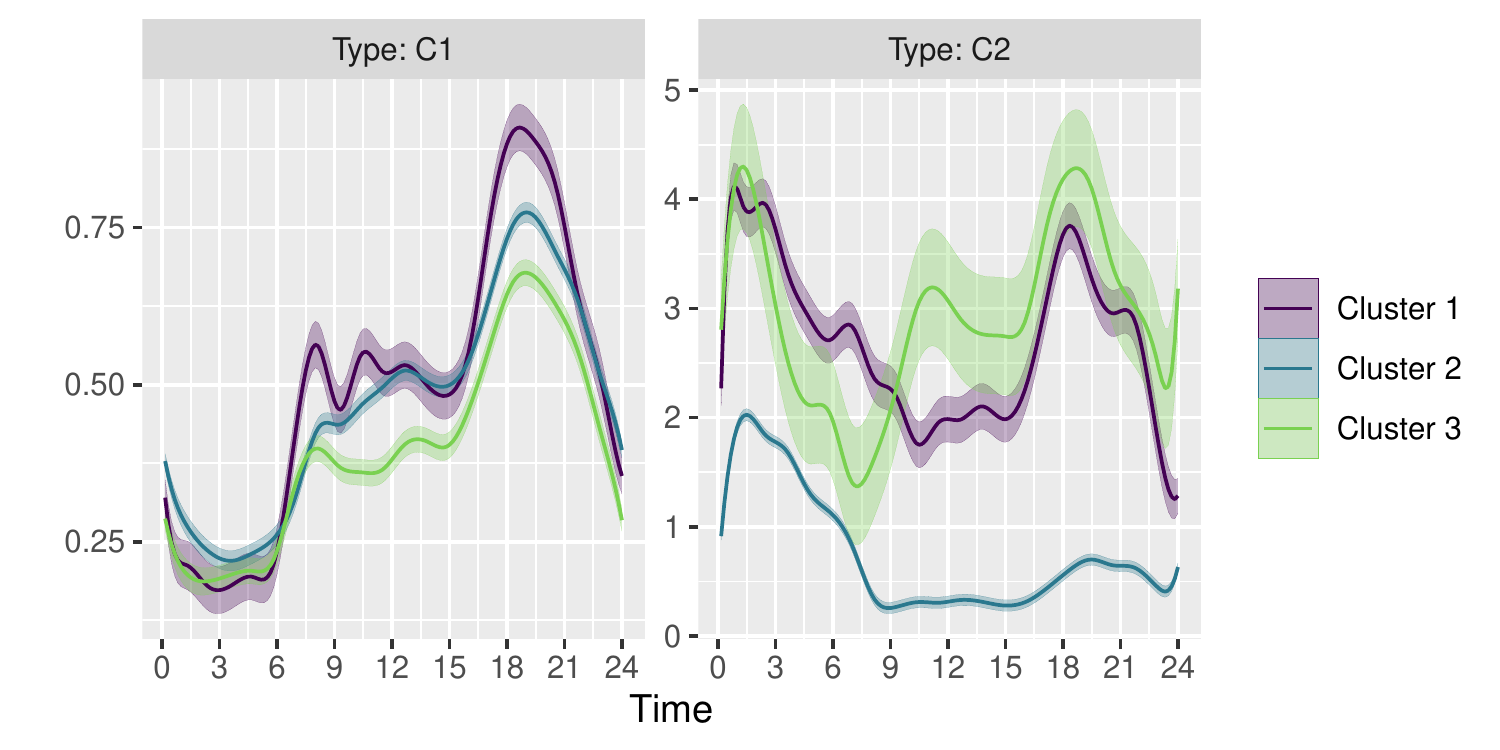} 
  \caption{Estimated typical curves in kWh and their confidence band for unrestricted (C1) and ``Economy 7'' (C2) domestic customers under three clusters aggregated data model fit}
  \label{fig:gavin-cl3-mc}
\end{figure}

The next step was to consider three clusters to fit the clustering aggregated data model assuming homogeneous covariance structure. Figure \ref{fig:gavin-cl3-mc} shows the estimated typical curves for C1 and C2 for the three clusters. The unrestricted customers C1 have once more similar curves in all clusters, with small observable differences during the work period between 9 AM and 5 PM and at the 8 PM peak at night. In contrast, the ``Economy 7'' customers C2 have distinct estimated load profiles among clusters. The estimated C2 typical curve for Cluster 2 has  the lowest energy consumption and its only peak in the early morning, whereas Clusters 1 and 3 share some characteristics like the double peak right after midnight and at 8 PM, but minor differences in the morning and during the work period.

The estimated cluster assignment probabilities are shown in Table~\ref{tab:gavin-cl3-pi}, where each substation is allocated with high probability to its cluster. Again, S5 and S12 form one cluster whereas the the large cluster in Section~\ref{sec:gavin-2cl} with 10 substations was divided into two clusters: one cluster composed by substations from Llantarnam primary and two from Trowbridge, and another cluster with Ringland substations plus S1 and S4. The clustering results show estimated typical curves that share some characteristics, but representing different morning and work period behaviors as seen in Figure~\ref{fig:gavin-cl3-mc}.

\begin{table}
\caption{\label{tab:gavin-cl3-pi}Estimated probability $\hat{p}_{jb}$ of substation $j$ belonging to cluster $b$, under the three cluster fit.}
\centering
\fbox{
\begin{tabular}{lrrrrrrrrrrrr}
\multicolumn{1}{c}{ } & \multicolumn{3}{c}{Trowbridge} & \multicolumn{2}{c}{Cyncoed} & \multicolumn{2}{c}{Ringland} & \multicolumn{4}{c}{Llantarnam} & \multicolumn{1}{c}{Usk} \\
\cmidrule(l{3pt}r{3pt}){2-4} \cmidrule(l{3pt}r{3pt}){5-6} \cmidrule(l{3pt}r{3pt}){7-8} \cmidrule(l{3pt}r{3pt}){9-12} \cmidrule(l{3pt}r{3pt}){13-13}
  & S1 & S2 & S3 & S4 & S5 & S6 & S7 & S8 & S9 & S10 & S11 & S12\\
\midrule
$\hat{p}_{j1}$ & \cellcolor[HTML]{FFFFFF}{0} & \cellcolor[HTML]{FFFFFF}{0} & \cellcolor[HTML]{FFFFFF}{0} & \cellcolor[HTML]{FFFFFF}{0} & \cellcolor[HTML]{CECECE}{1} & \cellcolor[HTML]{FFFFFF}{0} & \cellcolor[HTML]{FFFFFF}{0} & \cellcolor[HTML]{FFFFFF}{0} & \cellcolor[HTML]{FFFFFF}{0} & \cellcolor[HTML]{FFFFFF}{0} & \cellcolor[HTML]{FFFFFF}{0} & \cellcolor[HTML]{CECECE}{1}\\
$\hat{p}_{j2}$ & \cellcolor[HTML]{CECECE}{1} & \cellcolor[HTML]{FFFFFF}{0} & \cellcolor[HTML]{FFFFFF}{0} & \cellcolor[HTML]{CECECE}{1} & \cellcolor[HTML]{FFFFFF}{0} & \cellcolor[HTML]{CECECE}{1} & \cellcolor[HTML]{CECECE}{1} & \cellcolor[HTML]{FFFFFF}{0} & \cellcolor[HTML]{FFFFFF}{0} & \cellcolor[HTML]{FFFFFF}{0} & \cellcolor[HTML]{FFFFFF}{0} & \cellcolor[HTML]{FFFFFF}{0}\\
$\hat{p}_{j3}$ & \cellcolor[HTML]{FFFFFF}{0} & \cellcolor[HTML]{CECECE}{1} & \cellcolor[HTML]{CECECE}{1} & \cellcolor[HTML]{FFFFFF}{0} & \cellcolor[HTML]{FFFFFF}{0} & \cellcolor[HTML]{FFFFFF}{0} & \cellcolor[HTML]{FFFFFF}{0} & \cellcolor[HTML]{CECECE}{1} & \cellcolor[HTML]{CECECE}{1} & \cellcolor[HTML]{CECECE}{1} & \cellcolor[HTML]{CECECE}{1} & \cellcolor[HTML]{FFFFFF}{0}\\
\end{tabular}}
\end{table}

\begin{figure}[t]
  \centering

\includegraphics[width=.8\linewidth]{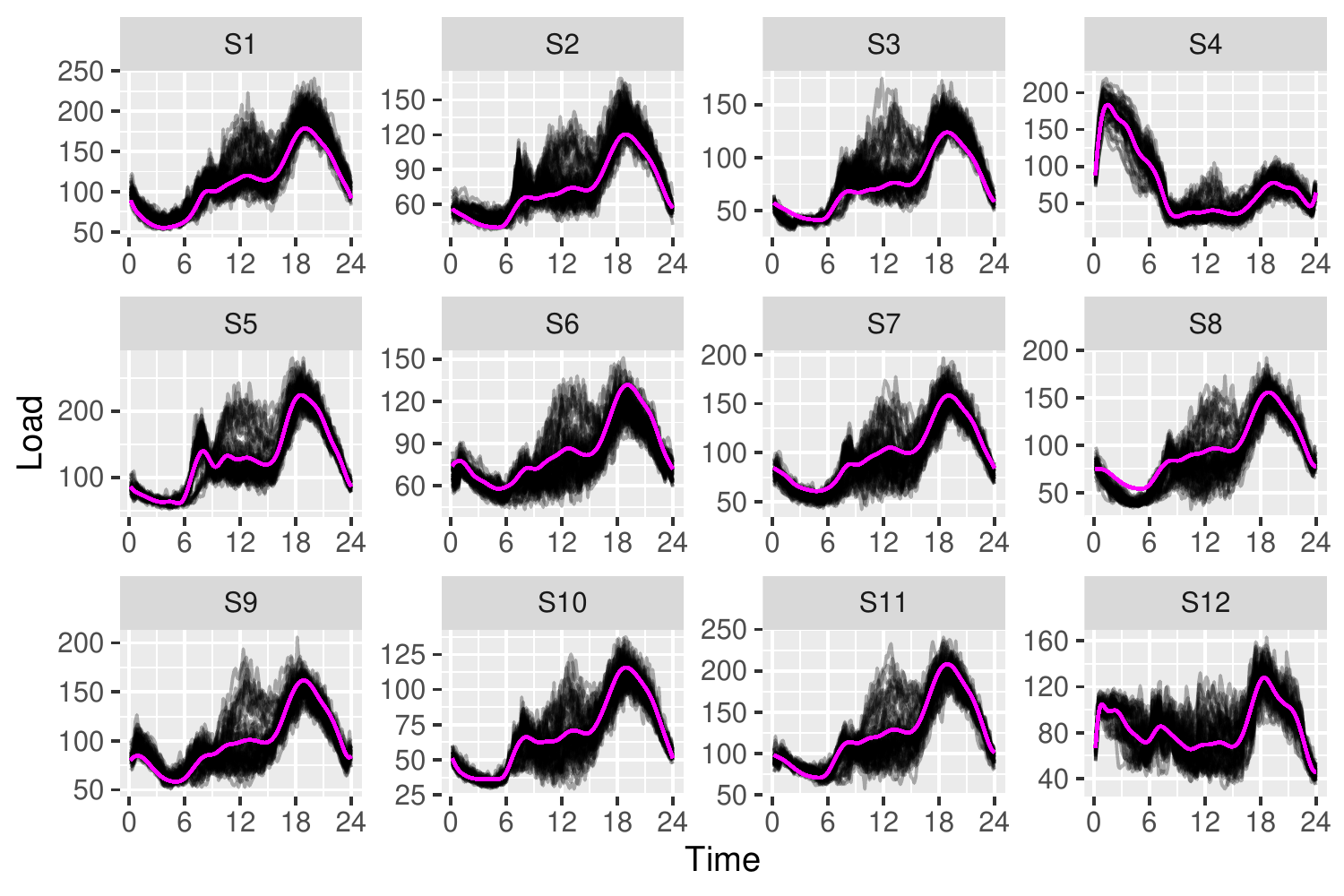}(a)


\includegraphics[width=.8\linewidth]{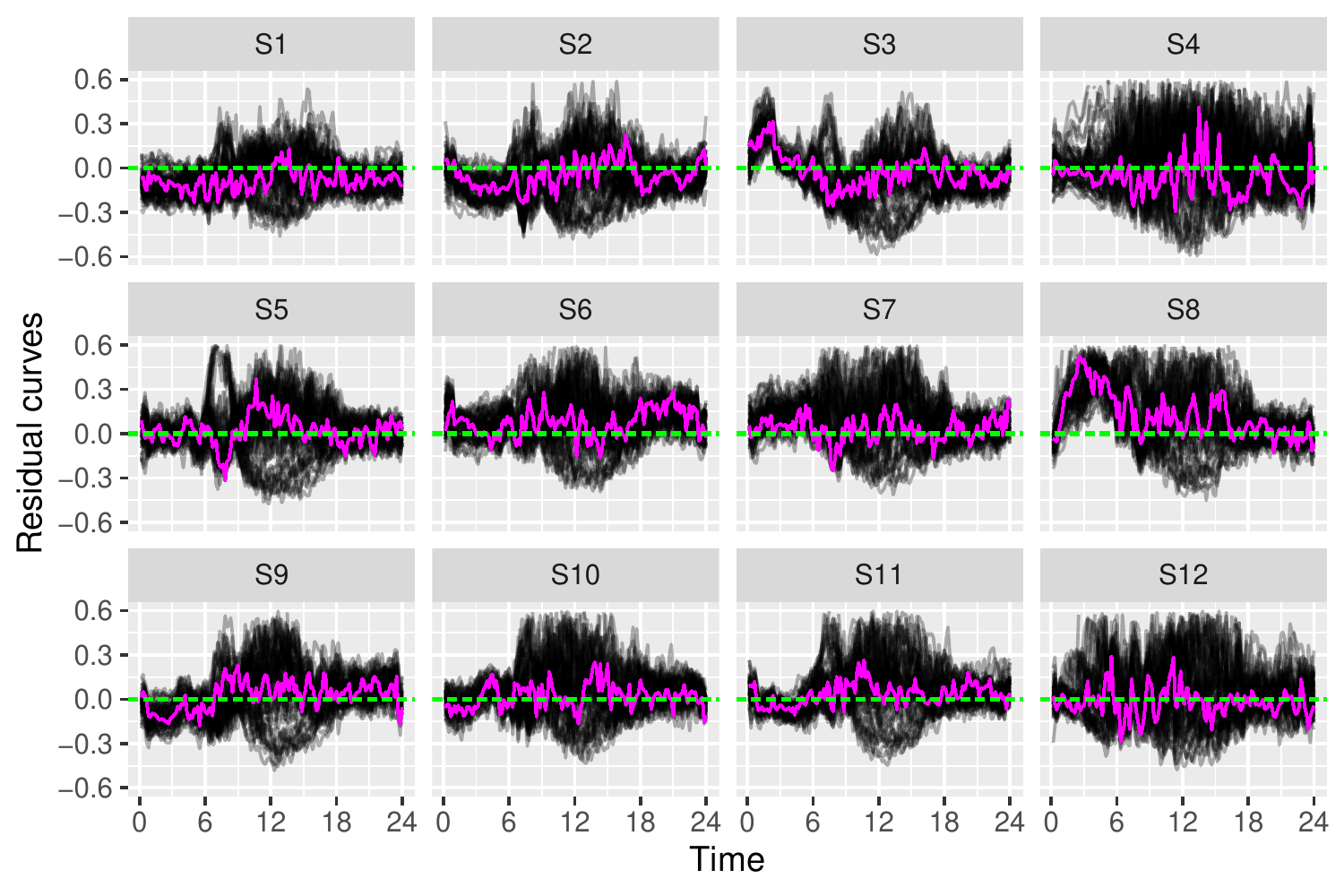}(b)

  \caption{Results of the clustering aggregated data model with three clusters: (a) Observed aggregated load data (in gray) in kWh over estimated aggregated curves (in magenta) and (b) relative error curves (in gray), median residual curves (in green) and zero reference line (in magenta) for the 12 substations.}
  \label{fig:gavin-cl3-res}
\end{figure}

The fitted curves over the observed aggregated data are displayed in Figure~\ref{fig:gavin-cl3-res}a. There are no apparent differences in fitted values  compared to the two-cluster approach in Figure~\ref{fig:gavin-cl2-res}b. Recall that substation markets are mostly dominated by C1 customers, the ones with similar estimated typical curves in all clusters, with substations with more C2 customers like S4 and S12 remaining in the same cluster. Hence, the impact of different C2 typical curves for Llantarnam, for example, might not be evident in the estimated aggregated load. Therefore, the residual curves in Figure~\ref{fig:gavin-cl3-res} yield the same characteristics as the two-cluster residual plot in Figure~\ref{fig:gavin-cl2-res}b. This figure suggests a good model fit represented by the median residual curves around the zero-reference line in most substations, except for a slight overestimation in substation S6. Comparisons between the two- and three-cluster models will be detailed in Section~\ref{sec:gavin-compare}.

\begin{table}
\caption{\label{tab:gavin-pars-cl3}Estimated dispersion ($\sigma_{cb}$) and decay ($\omega_{cb}$) parameters for customer type $c$ in cluster $b$ of the aggregated three-cluster model.}
\centering
\fbox{
\begin{tabular}{lrr}
Parameter & Value & 95\% Confidence Interval\\
\midrule
$\sigma_{11}$ & 1.5367 & (1.4798, 1.5936)\\
$\sigma_{21}$ & 1.5269 & (1.495, 1.5588)\\
$\omega_{11}$ & 0.1584 & (0.091, 0.2258)\\
$\omega_{21}$ & 0.0272 & (-0.1216, 0.1761)\\
\addlinespace
$\sigma_{12}$ & 1.0743 & (1.0732, 1.0754)\\
$\sigma_{22}$ & 1.2794 & (1.2762, 1.2825)\\
$\omega_{12}$ & 0.1197 & (0.1085, 0.1308)\\
$\omega_{22}$ & 0.0905 & (0.0159, 0.1650)\\
\addlinespace
$\sigma_{13}$ & 0.4278 & (0.4151, 0.4405)\\
$\sigma_{23}$ & 5.1783 & (5.1706, 5.1859)\\
$\omega_{13}$ & 0.0202 & (0.0096, 0.0307)\\
$\omega_{23}$ & 0.3743 & (0.3487, 0.3998)\\
\end{tabular}}
\end{table}

Table~\ref{tab:gavin-pars-cl3} displays the estimated covariance parameters for each combination of cluster and customer type. Cluster 1 probably has a homogeneous dispersion because the two values are close, and their 95\% confidence intervals overlap. However, there is high uncertainty in the decay parameters, especially for C2, where its confidence interval is large enough to contain zero, although we know this is not possible due to the parameter positive restriction. Cluster 2, the one with the lowest estimated C2 typical curves, has distinct dispersion and decay parameters for both customers and narrow confidence intervals. Lastly, Cluster 3 presents the largest distinction between dispersion parameters, which is visible in the confidence bands of the estimated typical curve of type C2 in Figure~\ref{fig:gavin-cl3-mc}.

In summary, the clustering aggregated data model with three clusters divided the large cluster in the two-cluster approach into two groups represented mostly by their primaries. The estimated typical curves for customers of type C1 still show similarities between clusters, but now enable the estimated typical curves of C2 customers to accommodate three different electrical energy consumption profiles.



\subsubsection{Model comparison}
\label{sec:gavin-compare}

 The two- and three-cluster aggregated data models resulted in good model fits according to the fitted values (Figures \ref{fig:gavin-cl2-res}a and \ref{fig:gavin-cl3-res}a) and the residual curves (Figures~\ref{fig:gavin-cl2-res}b and \ref{fig:gavin-cl3-res}b). The difference between them can be observed in the large cluster with 10 substations for the two-cluster model, which is split into two clusters in the three-cluster model (Tables \ref{tab:gavin-cl-pi} and \ref{tab:gavin-cl3-pi}).To decide which model is best suited to the observed aggregated data, the model comparison tools described in  Section~\ref{sec:modelcheck} were used.

Table \ref{tab:cl-res-comparison} shows the functional mean squared relative error (fMSRE) of the fitted and observed data under the two- and three-cluster models at each substation. In a comparison of substation fMSREs, S4 and S10 are highlighted because they have the largest differences between models. Observing both the fitted over observed values and the relative residual curves in Figures~\ref{fig:gavin-cl2-res}a and \ref{fig:gavin-cl2-res}b under the two-cluster model, it is clear how far their median curves are from the zero-reference line. On the other hand, observing the same substations S4 and S10 in Figures~\ref{fig:gavin-cl3-res}a and \ref{fig:gavin-cl3-res}b under the three-cluster model, it is apparent that their medians are closer to the zero line. In other words, the three-cluster model improves the model fit for these substations and consequently reduces their fMSRE. The other substations have minor differences between models in terms of fMSRE.

\begin{table}
\caption{\label{tab:cl-res-comparison}Functional mean squared relative error (fMSRE) of fitted and observed data under the two- and three-cluster models at each substation.}
\centering
\fbox{
\begin{tabular}{lrr}
Substation & Clusters & fMSRE\\
\midrule
 & 2 & 0.1330\\

\multirow{-2}{*}{\raggedright\arraybackslash S1} & 3 & 0.1403\\
\cmidrule{1-3}
 & 2 & 0.1552\\

\multirow{-2}{*}{\raggedright\arraybackslash S2} & 3 & 0.1569\\
\cmidrule{1-3}
 & 2 & 0.1676\\

\multirow{-2}{*}{\raggedright\arraybackslash S3} & 3 & 0.1663\\
\cmidrule{1-3}
 & 2 & 0.3009\\

\multirow{-2}{*}{\raggedright\arraybackslash S4} & 3 & 0.2354\\
\cmidrule{1-3}
 & 2 & 0.1603\\

\multirow{-2}{*}{\raggedright\arraybackslash S5} & 3 & 0.1603\\
\cmidrule{1-3}
 & 2 & 0.1818\\

\multirow{-2}{*}{\raggedright\arraybackslash S6} & 3 & 0.1678\\
\cmidrule{1-3}
 & 2 & 0.1971\\

\multirow{-2}{*}{\raggedright\arraybackslash S7} & 3 & 0.1851\\
\cmidrule{1-3}
 & 2 & 0.2284\\

\multirow{-2}{*}{\raggedright\arraybackslash S8} & 3 & 0.2636\\
\cmidrule{1-3}
 & 2 & 0.1750\\

\multirow{-2}{*}{\raggedright\arraybackslash S9} & 3 & 0.1584\\
\cmidrule{1-3}
 & 2 & 0.2689\\

\multirow{-2}{*}{\raggedright\arraybackslash S10} & 3 & 0.1739\\
\cmidrule{1-3}
 & 2 & 0.1590\\

\multirow{-2}{*}{\raggedright\arraybackslash S11} & 3 & 0.1633\\
\cmidrule{1-3}
 & 2 & 0.1977\\

\multirow{-2}{*}{\raggedright\arraybackslash S12} & 3 & 0.1977\\
\end{tabular}}
\end{table}

Another complementary tool is to compare the two models by their approximated BIC values. The two-cluster BIC is \ensuremath{707,308.3} and the three-cluster BIC is \ensuremath{704,571.4}.
Because the selection is favourable to models with the smallest BIC values, it again favours the aggregated three-cluster data model, although its BIC value is only 0.3\% smaller than the BIC for the two-cluster model. 

Therefore, the aggregated three-cluster model performed better in terms of fMSRE and BIC. Moreover, substations are grouped in a meaningful way, related to their primaries and avoiding the large cluster in the two-cluster approach. Hence, the three-cluster model seems to be a reasonable choice to group the electrical energy substations in the UK electrical load data.

\section{Conclusion}

The proposed aggregated data model has proved to be a useful tool to separate substation aggregated electrical load data into typical curves for each type of supplied customer and to comprehend their covariance structure. Our methodology includes novel approaches such as typical surface estimation as a function of time and temperature and explanatory variables and substation clustering based on the similarity of their estimated typical curves. By assuming a Gaussian process and using basis function expansions, our methodology becomes part of a family of functional models with favourable mathematical properties and well-established inference techniques.

Some estimation methods were crucial to the success of the proposed model, such as the least-squares estimator for the typical curves and the proposed initial value evaluation in the clustering approach in order to drastically reduce computing time and improve estimation performance.  

The estimated typical curves demonstrated robustness to wrong covariance structure assumptions when analyzing the UK dataset in Section \ref{chap:results} as well as in simulated studies presented in the Supplementary Material. Furthermore, this work has assessed the results of misspecified scenarios and how they relate to the true parameters, for example, when scalar dispersion parameters are assumed instead of functional variances.



The full aggregated model with explanatory variables and additional functional component demonstrated sophistication and flexibility with both real and simulated data. Suggestions on how to use the additional component properly were provided to avoid poor decisions in temperature ranges with little information. In any case, when working with real data, the confidence intervals of the estimated typical curves and surfaces will indicate ranges of uncertainty.





\section*{Code availability}
The methodology proposed in this work is implemented as an R package called \texttt{aggrmodel}, which is currently available online at the GitHub repository \href{https://www.github.com/gabrielfranco89/aggrmodel}{github.com/gabrielfranco89/aggrmodel}. The repository contains the functions used to perform all the analyses conducted in the paper as well as examples to illustrate package usability, which can be easily explored by the reader.

\section*{Supplementary Material}

The file supplementary\_material.pdf contains results from simulation studies conducted to show the performance of our proposed methods under various controlled scenarios.

\section*{Acknowledgements}
This study was financed in part by the Coordenação de Aperfeiçoamento de Pessoal de Nível Superior – Brasil (CAPES) – Finance Code 001, by the National Science and Engineering Research Council of Canada and FAPESP grant 019/00787-7. We thank Professor Gavin Shaddick for providing the data.   


\bibliographystyle{rss}
\bibliography{src/refs}

\end{document}


\maketitle
\tableofcontents












\section{Simulation studies}

This section evaluates the proposed aggregated data model in simulated scenarios. This approach provides control over the true model parameters that generate the data and the possibility of assessing the performance of estimated parameters in multiple simulation runs. All parameters used in our simulation studies are based on the estimated typical curves and estimated covariance parameters obtained from the analysis of the UK electrical energy substation data in Section 4 of the main paper. 

Two independent simulation studies were performed: one for the full aggregated data model, and the other for the clustering aggregated data model. Section~\ref{sec:simu-diag} describes the performance measures used to assess the quality of the estimated typical curves.  Section~\ref{sec:scenarios} introduces the  simulated scenarios for both studies. Section~\ref{sec:simu-fm} presents the first study with its typical surface, explanatory variables and functional variance; focusing on the precision of the estimated parameters under two  model fits: one considering a homogeneous covariance structure and the other a complete structure as in data generation. Section~\ref{sec:simu-cluster} describes the second study involving the clustering aggregated data model and its results. Section~\ref{chap:supl-tabs} contains two additional tables with results.  

\subsection{Simulation performance measures}
\label{sec:simu-diag}

To assess the performance of the estimated typical curves, the relative residual curve $R_c(t)$ of the customer of type $c$ is defined as

\begin{equation}
    R_c(t)
    =
    \frac{\hat{\alpha}_c(t) - \alpha_c(t)}{\alpha_c(t)}.
    \label{eq:resid-curve}
\end{equation}

\noindent Analogously, the relative residual curve of the estimated variance functionals is also defined as

\begin{equation}
    R_c(t)
    =
    \frac{\hat{\eta}_c(t) - \eta_c(t)}{\eta_c(t)}.
    \label{eq:resid-curve2}
\end{equation}

\noindent Division by the true value in (\ref{eq:resid-curve}) and (\ref{eq:resid-curve2}) is desirable to make the residual curves comparable under different magnitudes.

Let $R_{rc}$ be the relative residual curve of the customer of type $c$ in the $r$-th simulation run. Define the functional Mean Squared Relative Error (fMSRE)  as the mean of the integrals of the squared relative residual curves over time $t$. That is,

\begin{equation}
\label{eq:fMSRE}
    fMSRE_c
    =
    \frac{1}{R}
    \sum_{r=1}^R
    \int_0^T 
    R_{rc}^2(t) dt
       \approx
    \frac{1}{R}
    \sum_{r=1}^R
    \left\{
    \frac{T}{N}
    \sum_{t=t_1}^{t_N}
    R_{rc}^2(t)
    \right\},
\end{equation}

\noindent where $N$ is the number of observed points in time in the data set and $T$ the upper limit of the time domain. Because in this thesis the time frequencies are equally distanced, the fraction $T/N$ is the equally spaced time difference band that approximates the $dt$ of the integral on the left-hand side.



\subsection{Simulated scenario setup}
\label{sec:scenarios}

The simulated scenarios are different combinations of number of observed days, representing the amount of information available, and market balance, which is detailed below.

In real substation data, it is sometimes observed that a particular customer type may be overrepresented, with more than 95\% of the market. If this dominance occurs in all observed substations, this situation is called an unbalanced market scenario, and a balanced market scenario otherwise. To study this phenomenon, the markets were generated as follows:
\begin{itemize}
    \item Unbalanced: all substations have markets with more customers of Type 1 than Type 2 with percentage varying between 70\% and 95\%.
    \item Balanced: six substations have most of their customers of Type 1 and six substations have most of their customers of Type 2, with the majority percentages varying between 70\% and 95\%.
\end{itemize}

\noindent The percentages are relative to the number of customers for each substation, which is displayed in Table~\ref{tab:simu-mkttotal}. 

\begin{table}[htb]
\caption{\label{tab:simu-mkttotal}Fixed number of customers for each substation in the simulation study.}
\centering
\begin{tabular}{l|rrrrrrrrrrrr}
\toprule
Substation & 1 &2 &3 &4 &5 &6 &7 &8 &9 &10 &11 &12 \\
\midrule
Total & 231 &151 &156 &109 &225 &172 &206 &182 &175 &160 &254 &69\\
\bottomrule
\end{tabular}
\end{table}

The combinations of market balance and number of observed days compose the eight different simulated scenarios presented in Table~\ref{tab:scenarios}. Scenarios 1 to 4 are related to the full aggregated data model study and Scenarios 5 to 8 to the clustering aggregated data model study. Each scenario is composed of two types of customers observed at 30 minutes time frequency at 12 substations and replicated 15 times. In other words, 15 datasets were generated with these configurations and studied in detail, as described in Sections \ref{sec:simu-fm} and \ref{sec:simu-cluster}.

\begin{table}[htb]
    \centering
    \caption{Covariance structure, number of clusters, number of observed days, market balance and number of generated datasets (replicates). Eight simulated scenarios were proposed: Scenarios 1 to 4 for the full aggregated data models and Scenarios 5 to 8 for the clustering aggregated data model.}
    \begin{tabular}{crrrrr}
    \toprule
       Scenario  &  Covariance & Clusters & Days & Market & Replications\\
       \midrule
        1 & Complete & 1 & 5 & Unbalanced & 15 \\
        2 & Complete & 1 & 5 & Balanced & 15 \\
        3 & Complete & 1 & 30 & Unbalanced & 15 \\
        4 & Complete & 1 & 30 & Balanced & 15 \\
        \addlinespace
        5 & Homogeneous & 3 & 5 & Unbalanced & 15 \\
        6 & Homogeneous & 3 & 5 & Balanced & 15 \\
        7 & Homogeneous & 3 & 30 & Unbalanced & 15 \\
        8 & Homogeneous & 3 & 30 & Balanced & 15 \\
         \bottomrule
    \end{tabular}
    \label{tab:scenarios}
\end{table}

\subsection{Full aggregated data model}
\label{sec:simu-fm}

The full aggregated data model studies the typical surface together with explanatory variables related to substations. In this case, the surface is a function of time and daily air temperature, as presented in Section 3.2 of the main paper. Section~\ref{sec:simu-true-fm1} describes the air temperature functional and the explanatory variables used in this simulation, Section \ref{sec:simu-fm-results} presents the main results and Section~\ref{sec:simu-fm-results} contains a discussion and the conclusions of this study.


\subsubsection{True parameters}\label{sec:simu-true-fm1}

Recall the typical surface in Equation (3) introduced in Section 3.2 of the main manuscript:
$$
  \tc_{ic}(t) = 
  \tc_{ic} \bigl( u(t), v_i(t) \bigr).
$$
\noindent In this section, $u(t)=t$  and $v_i(t) = T_i(t)$ are used as the temperature at day $i$. Then, the typical surface is given by
\begin{equation}
    \alpha_{ic}
    \Big(
    t, T_i(t)
    \Bigr)
    =
    b_c(t) 
    \times 
    \left(
      1 - 
      \frac{1}{2}
      \Phi\bigl(T_i(t)-1\bigr)
    \right),
    \label{eq:fm-suf}
\end{equation}

\noindent  with $b_c(t)$ as the baseline curve for customer of type $c$ and $\Phi(\cdot)$ as the cumulative density function of the standard normal distribution. Hence when the temperature drops below 1\textdegree C the typical surface area increases considerably. 

\begin{figure}[htp]
  \begin{subfigure}{\textwidth}
  \centering
\includegraphics[width=.65\linewidth]{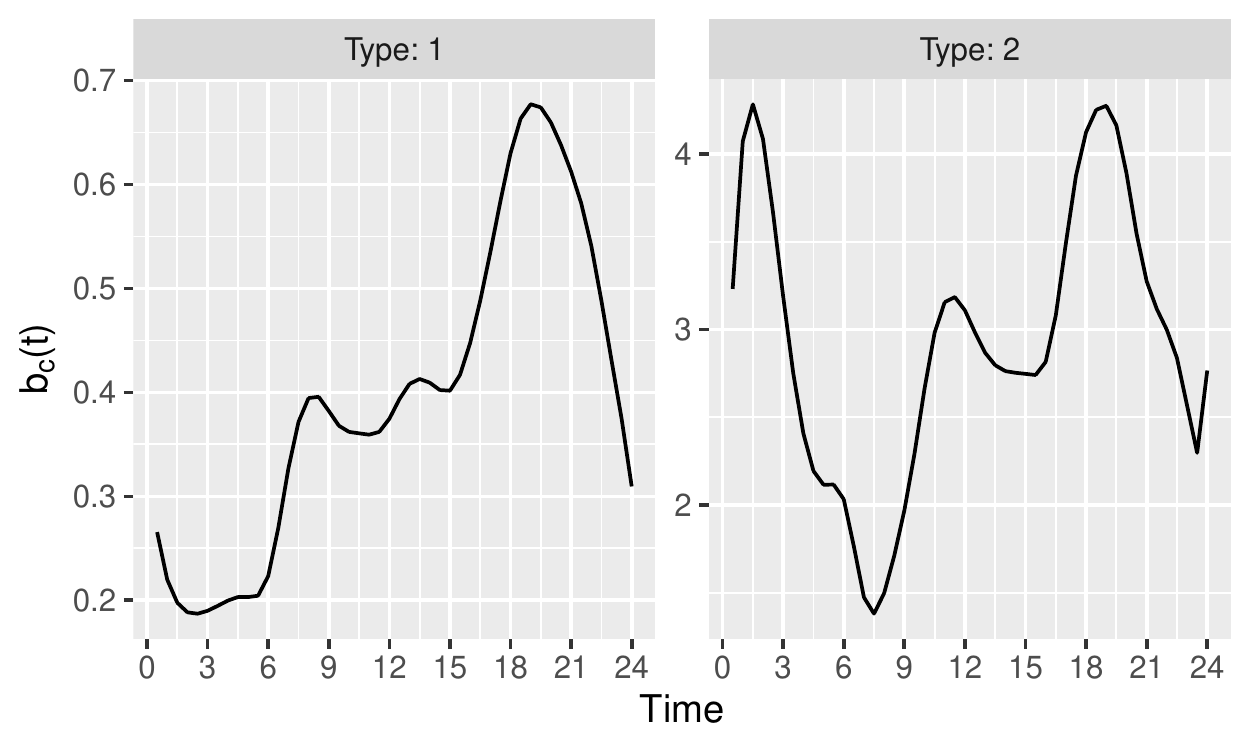} 

  \caption{Baseline curves}
\end{subfigure}
\begin{subfigure}{\textwidth}
  \centering
\includegraphics[width=.65\linewidth]{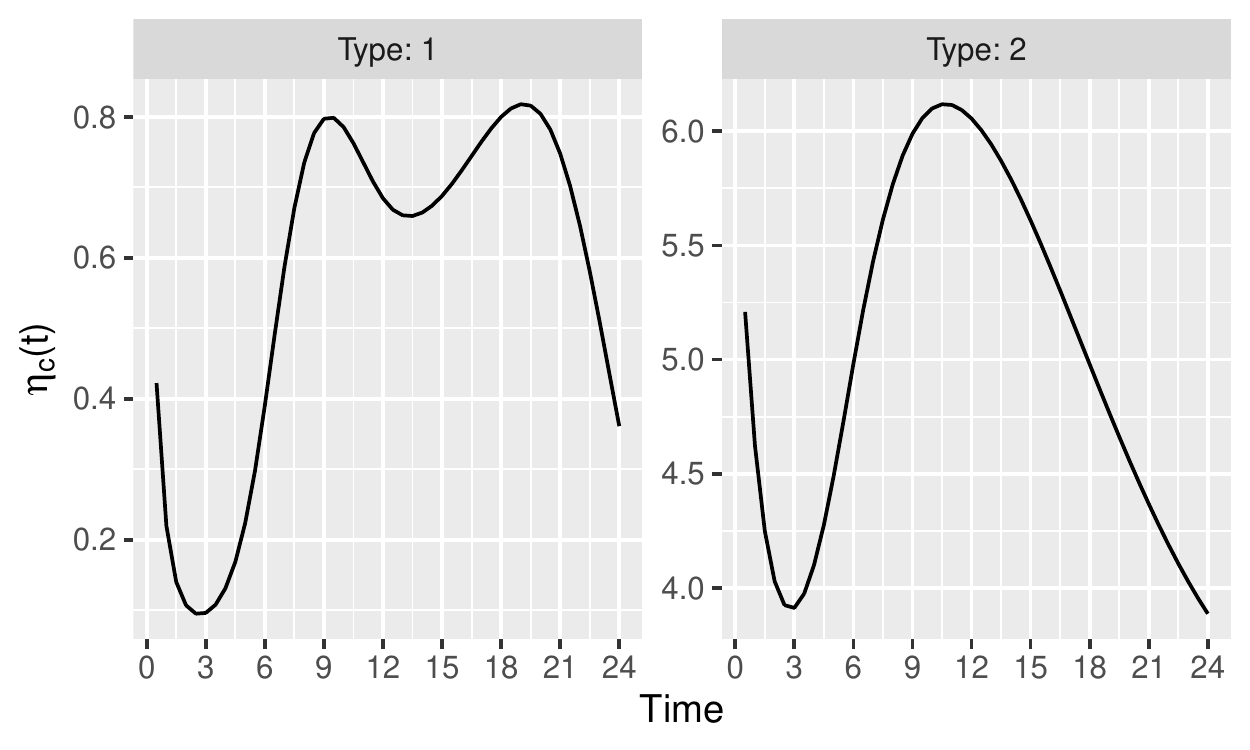} 

  \caption{Variance functionals}
  \end{subfigure}
\begin{subfigure}{\textwidth}
  \centering
\includegraphics[width=.65\linewidth]{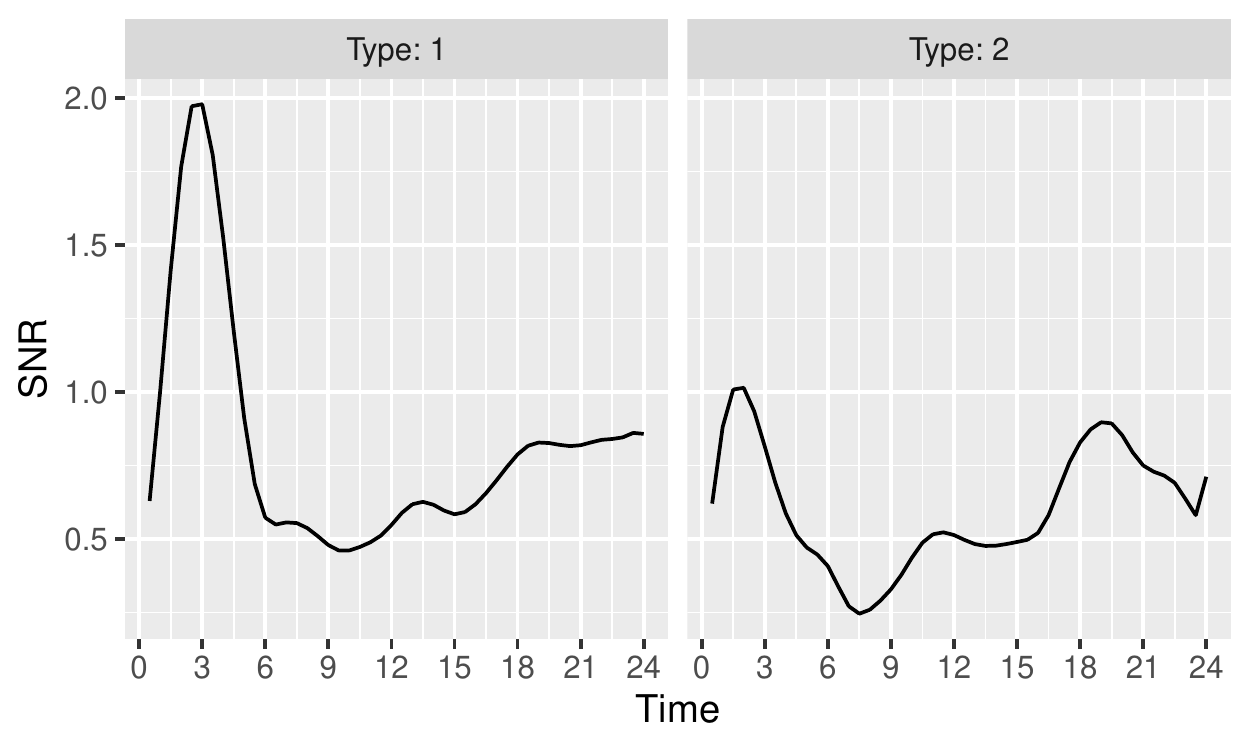} 

  \caption{Signal-to-noise ratio (SNR)}
  \end{subfigure}  
  \caption{Baseline curves, true variance functionals and signal-to-noise ratio at time $t$ of the simulation study.}
  \label{fig:fm-true-typical}
\end{figure}

\begin{figure}[p]
  \begin{subfigure}{\textwidth}
    \centering
\includegraphics[width=.8\linewidth]{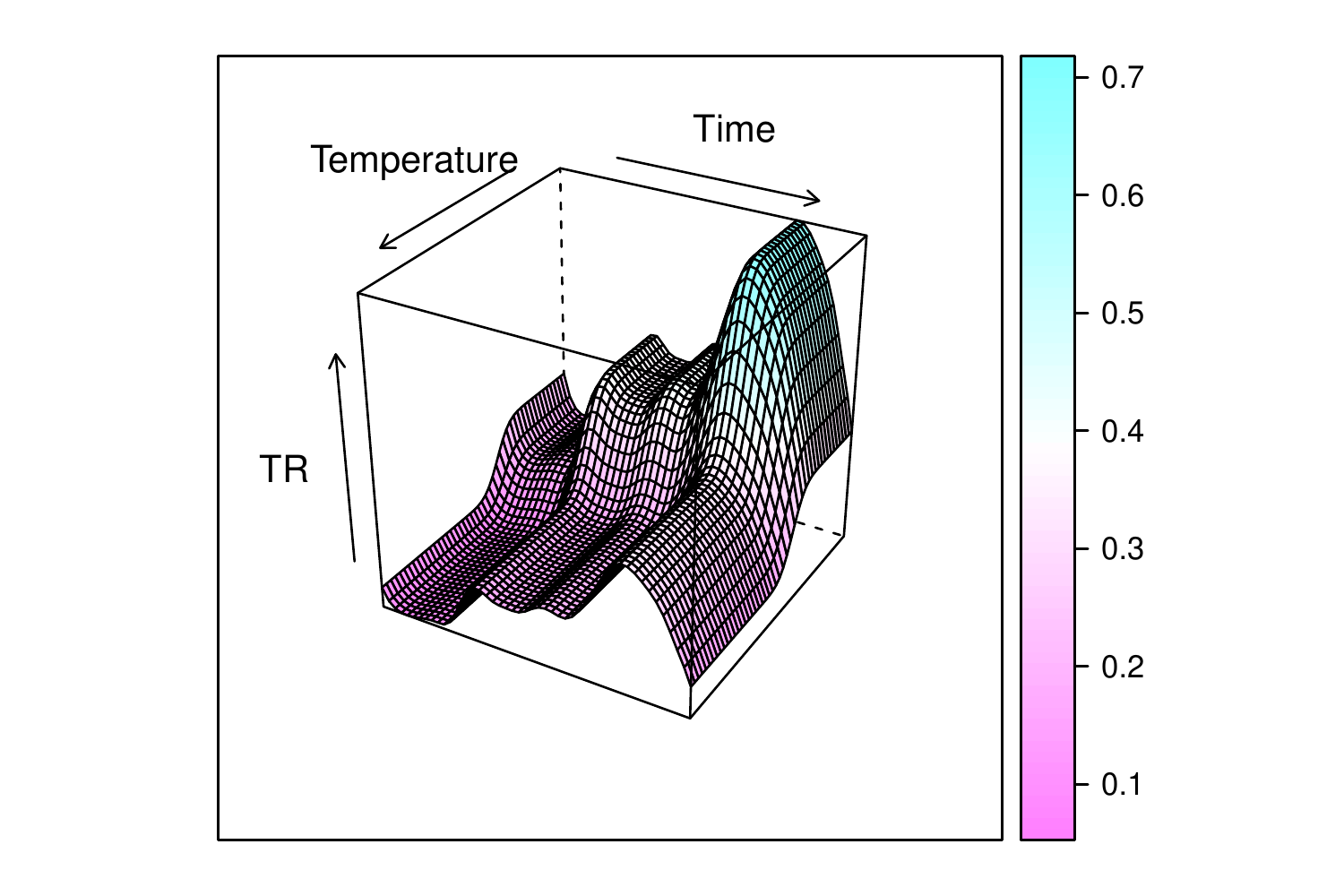} 

    \caption{Typical surface for Type 1}
  \end{subfigure}
  \begin{subfigure}{\textwidth}
    \centering
\includegraphics[width=.8\linewidth]{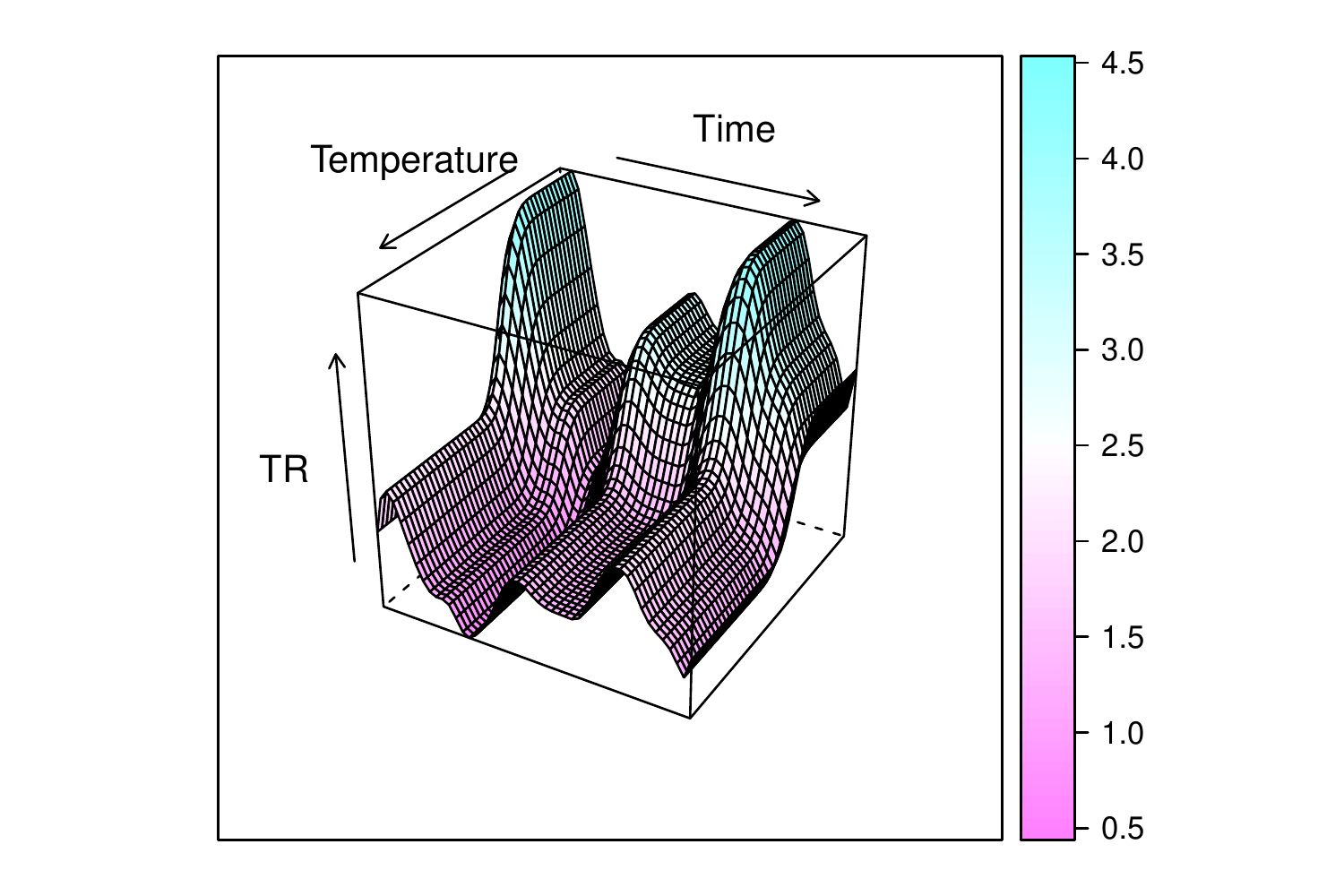} 

    \caption{Typical surface for Type 2}
  \end{subfigure}

  \caption{Typical surfaces (TR) for each customer type as functions of time and the full range of observed temperatures in the three primaries.}
  \label{fig:fm-simu-surf}
\end{figure}

Figure~\ref{fig:fm-true-typical} shows the baseline curves, the variance functionals and the signal-to-noise ratio (SNR) for each customer type. The SNR is simply the ratio of the typical curve to the variance functional at time $t$. The baseline curves and variance functionals were based on the estimated typical curves obtained from the real data analysis in Section 4 of the main text. The type 1 baseline curve mimics the unrestricted domestic customer with lower consumption in early morning, increasing after 8 AM and reaching its peak at 8 PM. The Type 2 curve mimics the ``Economy 7'' customer with peaks around 2am and 8pm but with considerably larger electrical load values than Type 1. Customers variance functionals have higher values around the work period between 9am and 5pm, although Type 1 has two peaks that possibly represent when people leave from and arrive at their homes.  The typical surfaces are shown in Figure~\ref{fig:fm-simu-surf}.

\begin{figure}[ht]
  \centering
\includegraphics[width=.8\linewidth]{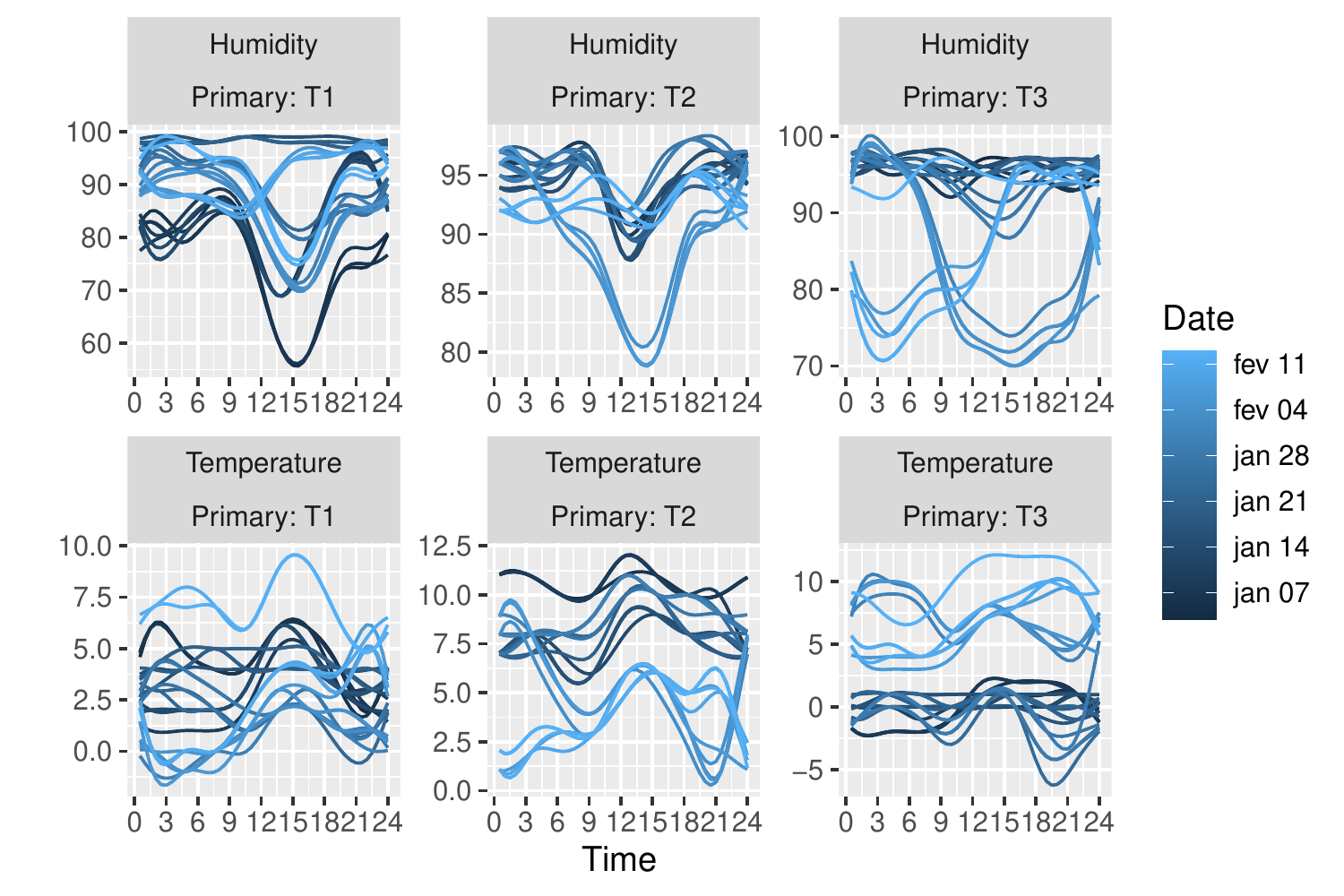} 

  \caption{Air temperature and air humidity for each primary T1, T2 and T3 used in the simulation study.}
  \label{fig:fm_simu_temp}
\end{figure}

The weather data containing temperature and air humidity were also based on real measurements for winter 2013 in Wales, United Kingdom. For this study, three sets of data were generated, representing three locations labelled T1, T2, and T3. Substations 1 to 4 were assigned to location T1, substations 5 to 8 to location T2, and the remaining substations 9 to 12 to location T3. Figure~\ref{fig:fm_simu_temp} shows the temperature and air humidity profiles for each location observed over 30 days. In fact, only data for scenarios 3, 4, 7, and 8 were generated in this manner. For scenarios 1, 2, 5, and 6, only the first five days were considered. In this study, temperature was used as the second component of the typical surface, and air humidity was used as an explanatory variable of the full aggregated data model with constant coefficient.

 Furthermore, two explanatory variables were considered: air humidity as a functional variable, and a binary variable with value 1 for substations 1 and 2 and 0 otherwise, with associated coefficients  $1/90=0.0111$ and $13$, respectively. Therefore, from Section 3.2 of the main paper, the full aggregated complete model can be written as
\begin{equation}
     \Yijt =
            \Biggl(
            \sum_{c=1}^C
            \mjc
            \tc_c \bigl( t, T_i(t) \bigr)
            \Biggr)
            +
          13\, \covar_{j1}  +
          0.0111 \covar_{ij2}(t)  +
          \erroijt, 
\end{equation}

\noindent where $\covar_{j1}$ is the dummy variable for substations 1 and 2 and $\covar_{j2} = \covar_{ij2}(t)$ the air humidity of substation $j$ at time $t$ of day $i$.
Finally, the true covariance decay parameters for each customer type were defined as $\omega_1 = 0.03$ and $\omega_2=0.7$.

\subsubsection{Results} 
\label{sec:simu-fm-results}

In this study, for each of Scenarios 1 to 4 in Table~\ref{tab:scenarios}, two models were fitted: one homogeneous and one complete aggregated data model. The homogeneous fit tests the performance of typical surface estimation under an under-parameterized covariance structure and the behaviour of the dispersion parameters by reducing the variance functional to a scalar. On the other hand, the complete model tests check whether, under the correct scenario, the proposed model performs well in terms of typical surface and covariance parameter estimation.

Throughout this section, the number of observed days and the market balance are explicitly shown to avoid consulting Table~\ref{tab:scenarios} to remember the scenario setup.

\begin{figure}[htp!]
  \centering
  \begin{subfigure}{.48\textwidth}
    \centering
\includegraphics[width=\linewidth]{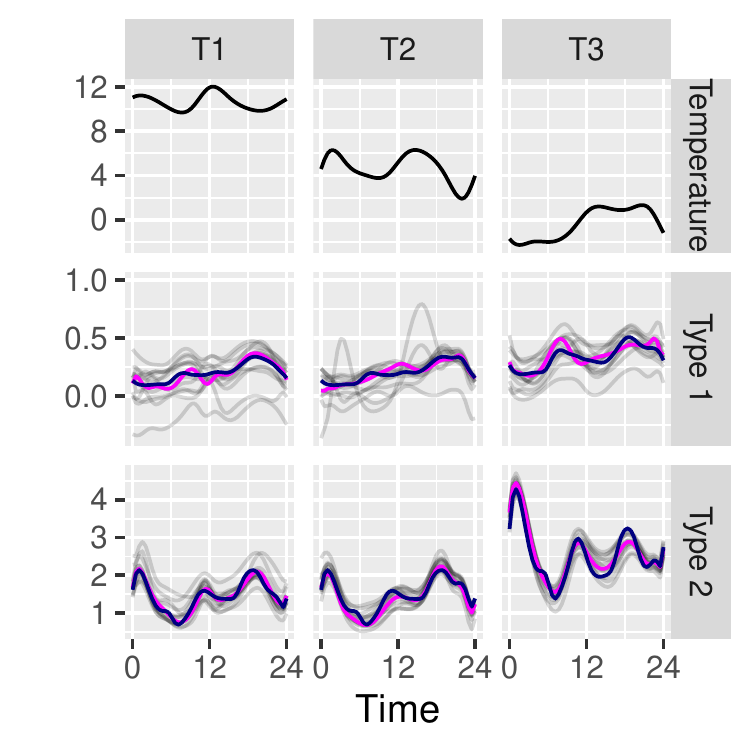} 

    \caption{Balanced market with 5-day data}
  \end{subfigure}
  \begin{subfigure}{.48\textwidth}
    \centering
\includegraphics[width=\linewidth]{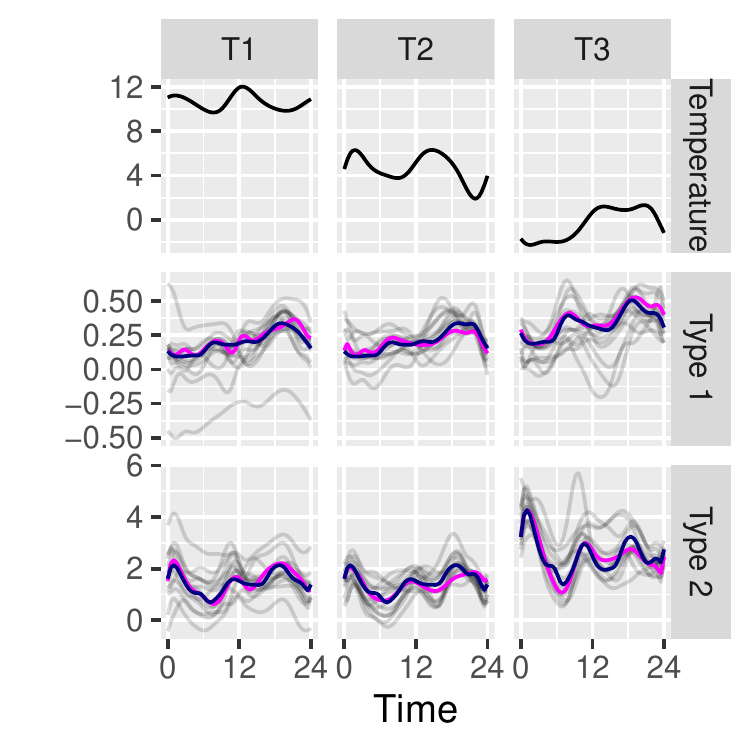} 

    \caption{UnBalanced market with 5-day data}
  \end{subfigure}
    \begin{subfigure}{.48\textwidth}
      \centering
\includegraphics[width=\linewidth]{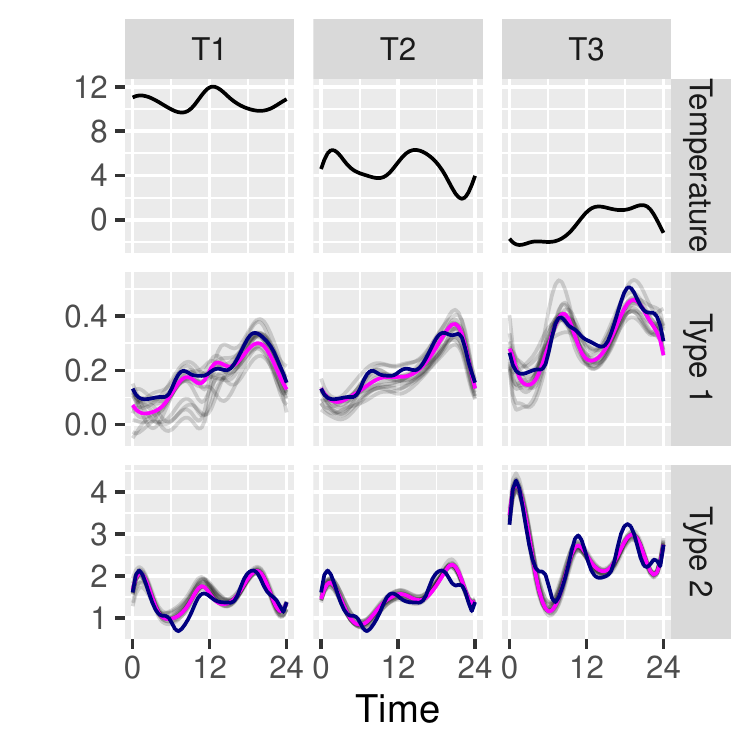} 

    \caption{Balanced market with 30-day data}
  \end{subfigure}
    \begin{subfigure}{.48\textwidth}
      \centering
\includegraphics[width=\linewidth]{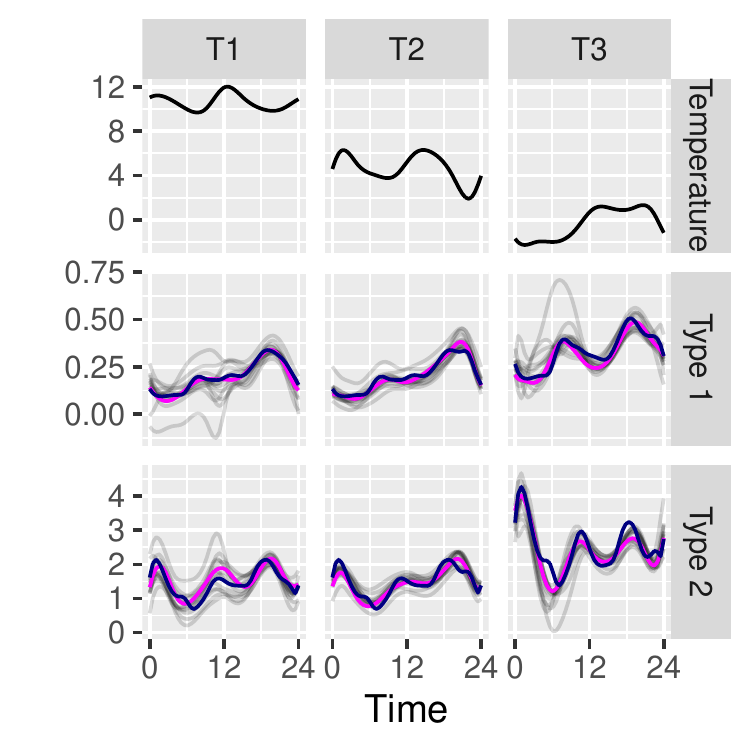} 

    \caption{Unbalanced market with 30-day data}
  \end{subfigure}
  \caption{At every panel, the first row represents the temperatures $T(t)$ for each temperature set T1, T2 and T3; the following rows represent the estimated typical curves of $\alpha_c(t,T(t))$ for customers of Type 1 and 2 in Scenarios 1 to 4 under the homogeneous model fit. Median depth lines are represented in magenta,  true typical curves in blue and estimated typical curves for each simulation run in gray.}
  \label{fig:fm-hm-typical}
\end{figure}

\begin{figure}[htp!]
  \centering
  \begin{subfigure}{.48\textwidth}
    \centering
\includegraphics[width=\linewidth]{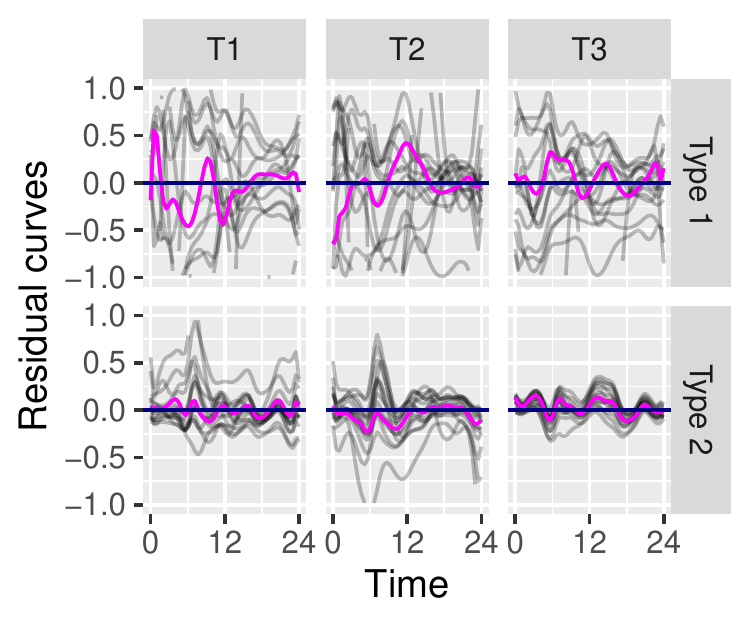}

    \caption{Balanced market with 5-day data}
  \end{subfigure}
  \begin{subfigure}{.48\textwidth}
    \centering
\includegraphics[width=\linewidth]{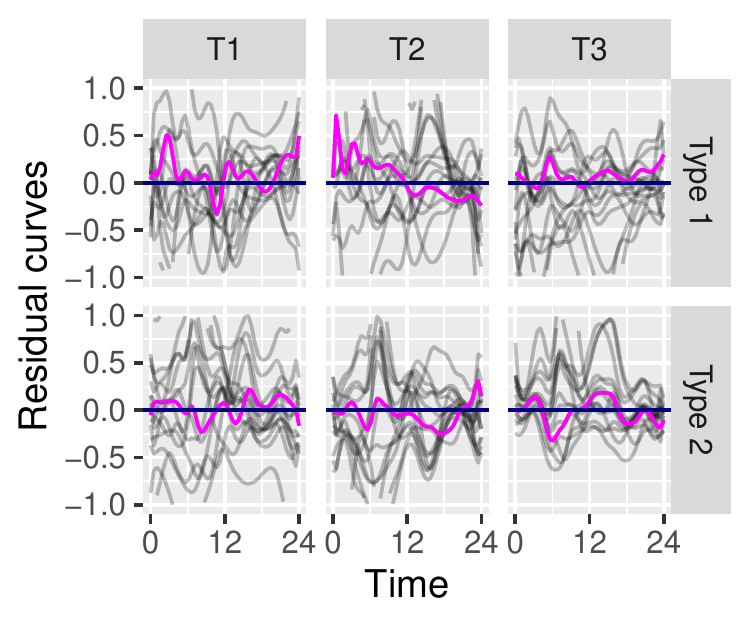}

    \caption{Unbalanced market with 5-day data}
  \end{subfigure}
    \begin{subfigure}{.48\textwidth}
      \centering
\includegraphics[width=\linewidth]{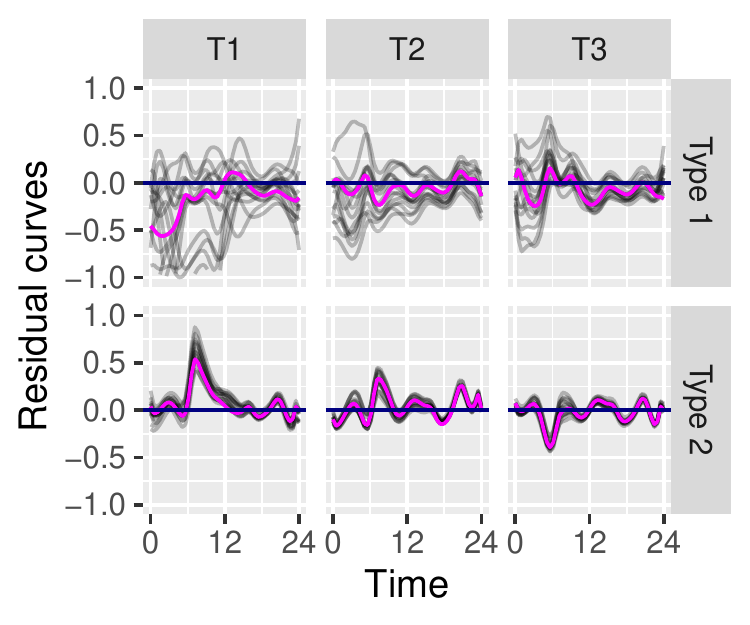} 

    \caption{Balanced market with 30-day data}
  \end{subfigure}
    \begin{subfigure}{.48\textwidth}
      \centering
\includegraphics[width=\linewidth]{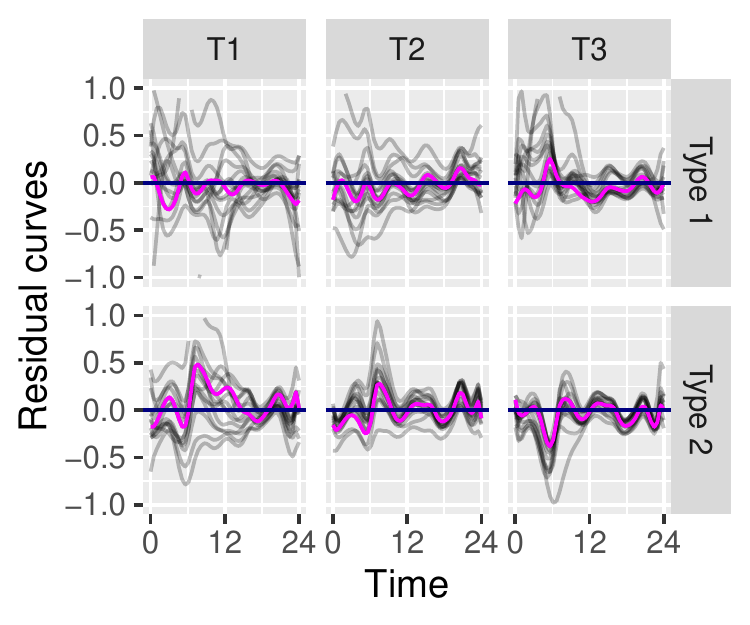} 

    \caption{Unbalanced market with 30-day data}
  \end{subfigure}
  \caption{Residual curves of the estimated typical curves for Scenarios 1 to 4 in grey and their median depth in magenta under the homogeneous model fit in Figure~\ref{fig:fm-hm-typical}.}
  \label{fig:fm-hm-typical-mse}
\end{figure}

Starting with the homogeneous fit study, Figure~\ref{fig:fm-hm-typical} shows the estimated typical surfaces $\alpha_{ic}(t,T(t))$ for some temperature curves $T(t)$ for every combination of observed days and market balance.  The first row in the panels represents a single instance of the temperature $T(t)$ on the first observed day in the simulated data and in its respective primary group T1, T2 or T3. Observe that Figures~\ref{fig:fm-hm-typical}a and \ref{fig:fm-hm-typical}b show estimated typical surfaces with noticeable variability, where some curves assume negative values. However, the balanced scenario in Figure~\ref{fig:fm-hm-typical}a presents estimated curves for Type 2 with lower variability than those in Figure~\ref{fig:fm-hm-typical}b. On the lower panels, Figures~\ref{fig:fm-hm-typical}c and \ref{fig:fm-hm-typical}d show lower variability than the five-day scenarios. Furthermore, observe that the estimated curves for Type 2 in Figure~\ref{fig:fm-hm-typical}c have even lower variability. In general, the median curves in the four scenarios show that the estimated curves are concentrated around their true values. This proximity to the true curve is better visualized in the residual curves shown in Figure~\ref{fig:fm-hm-typical-mse}. As presented in Section \ref{sec:simu-diag}, these curves are standardized so that the scenario performance can be compared. Note that the residual curves for Type 2 in the five-day scenarios have lower variability in Figure~\ref{fig:fm-hm-typical-mse}a than their respective ones in Figure~\ref{fig:fm-hm-typical-mse}b, as mentioned earlier. The same event occurs in the 30-day scenarios, but with lower variability than the five-day scenarios. The four panels of Figure~\ref{fig:fm-hm-typical-mse} show median curves oscillating around the horizontal zero-reference line, with no major differences among scenarios. To summarize the precision of the estimated typical surfaces shown in Figure~\ref{fig:fm-hm-typical}, Table~\ref{tab:simu-fmse} shows the functional Mean Squared Relative Error for Scenarios 1 to 4 fitted by the homogeneous model. Clearly, the fMSRE for the estimated Type 1 typical curves is considerably higher in the five-day scenarios. It seems that the magnitude of the curves influences the variability of the estimates because the curves with greater magnitude in Type 2 have lower fMSRE than those with lower magnitude in Type 1. Moreover, all fMSRE for the 30-day scenarios are lower than the respective ones in the five-day scenarios.

\begin{figure}[htp]
  \centering
  \begin{subfigure}{\textwidth}
  \centering
\includegraphics[width=.8\linewidth]{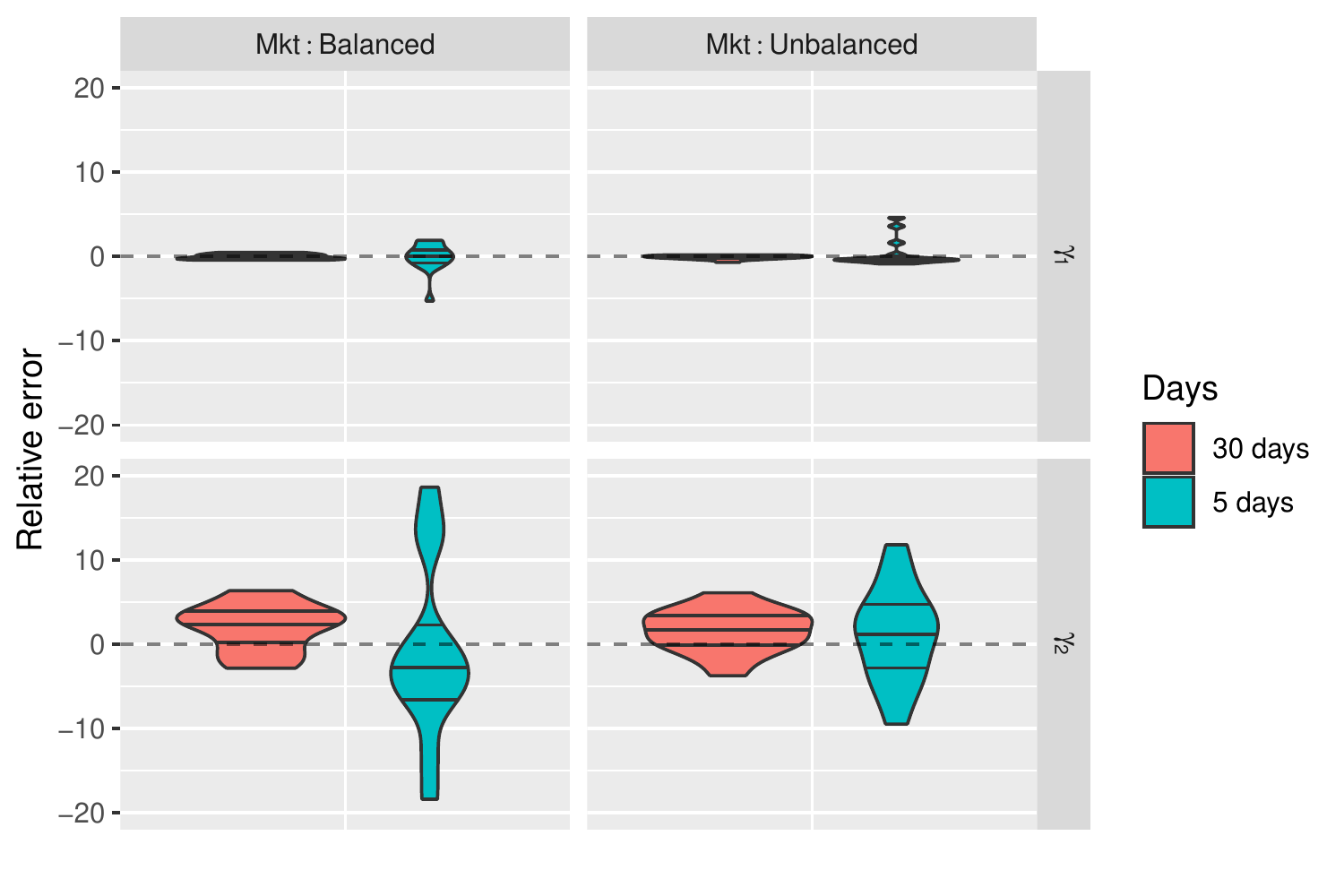} 
\caption{Homogeneous covariance structure}
\label{fig:fm-gamma-hm}
\end{subfigure}
  \begin{subfigure}{\textwidth}
  \centering
\includegraphics[width=.8\linewidth]{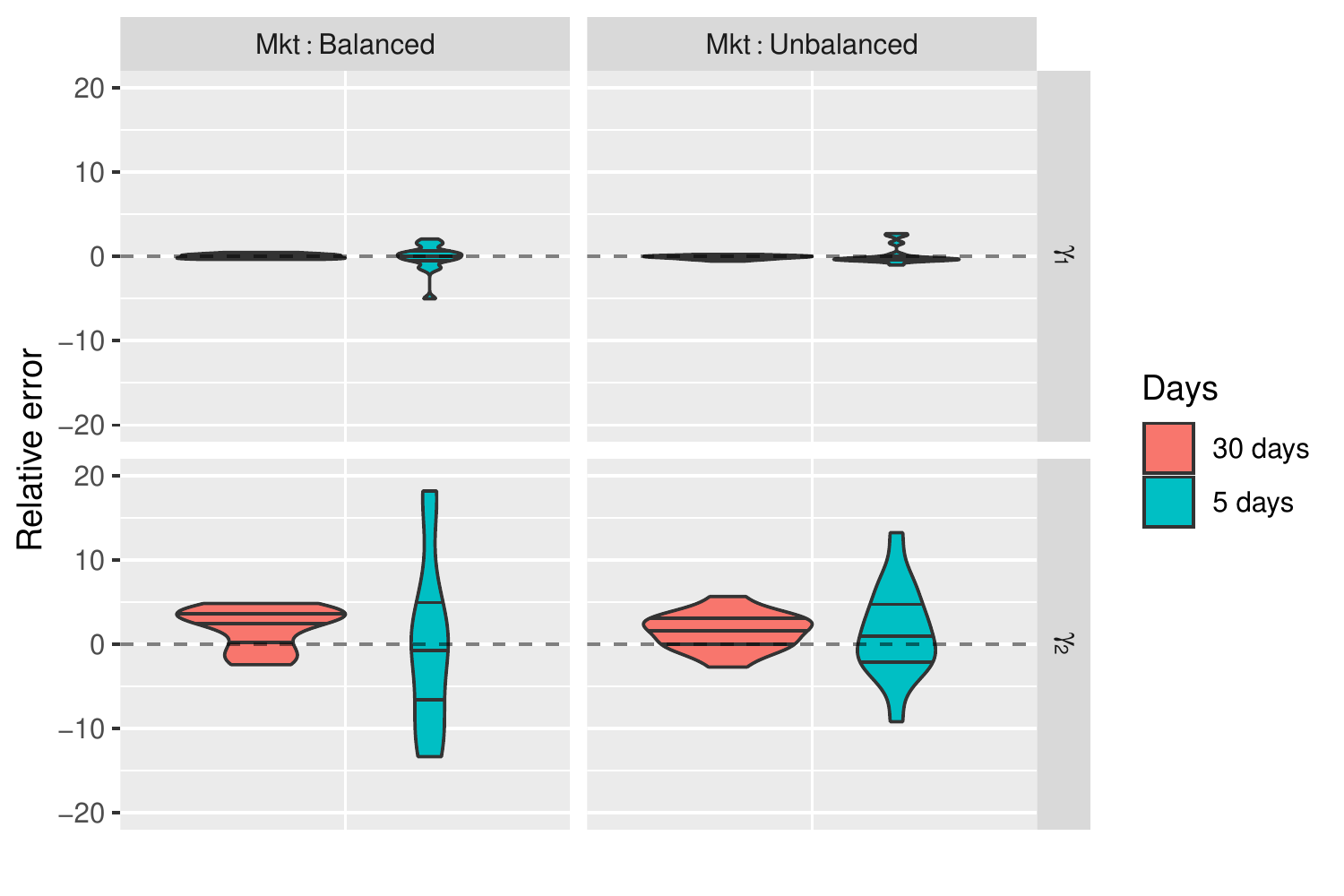} 
  \caption{Complete covariance structure}
  \label{fig:fm-gamma-comp}
  \end{subfigure}
  \caption{Relative errors of the estimated explanatory variables coefficients, $\gamma_1 = 13$ and $\gamma_2 = 0.0011$, and their relative error distribution under a) the homogeneous model fit and b) under the complete model fit for Scenarios 1 to 4.}
  \label{fig:fm-gamma}
  \end{figure}

Figure~\ref{fig:fm-gamma}a shows  violin plots of the relative error of the estimated coefficients associated with the explanatory variables $\gamma_1=13$ and $\gamma_2=0.0011$. One run was excluded from the plot in the balanced scenario because it showed an absolute relative error greater than 38. In all scenarios, the estimates with $\gamma_2$ have larger violins than those with $\gamma_1$. The 30-day scenarios have lower expected variability than the five-day scenario estimates, but their median reference lines above the zero line show visible underestimation of the parameter $\gamma_2$. Furthermore, Table~\ref{tab:gamma} shows the mean, median and square root of the Mean Squared Relative Error (srMSRE) of the estimated parameters. Observe that parameter $\gamma_1$ has estimates with considerably lower srMSRE than $\gamma_2$. The  underestimation is notable in the mean and median values of $\gamma_2$. Nevertheless, the statistics of parameter $\gamma_2$ show slight overestimation of the mean and larger srMSREs in all scenarios, especially the balanced five-day scenario, the one that presented a run with relative error greater than 38.

\begin{figure}[htp]
  \centering
  \begin{subfigure}{\textwidth}
    \centering
\includegraphics[width=.8\linewidth]{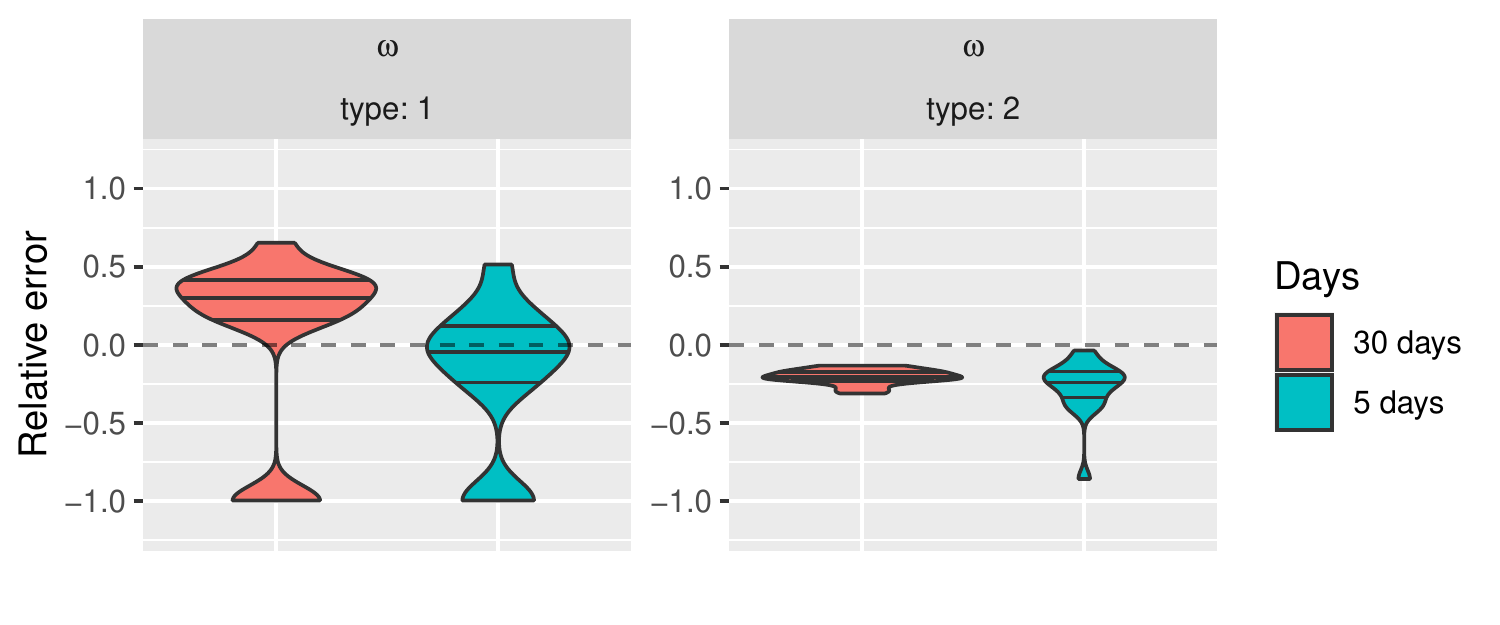} 

  \caption{Unbalanced market}
  \end{subfigure}
  \begin{subfigure}{\textwidth}
    \centering
\includegraphics[width=.8\linewidth]{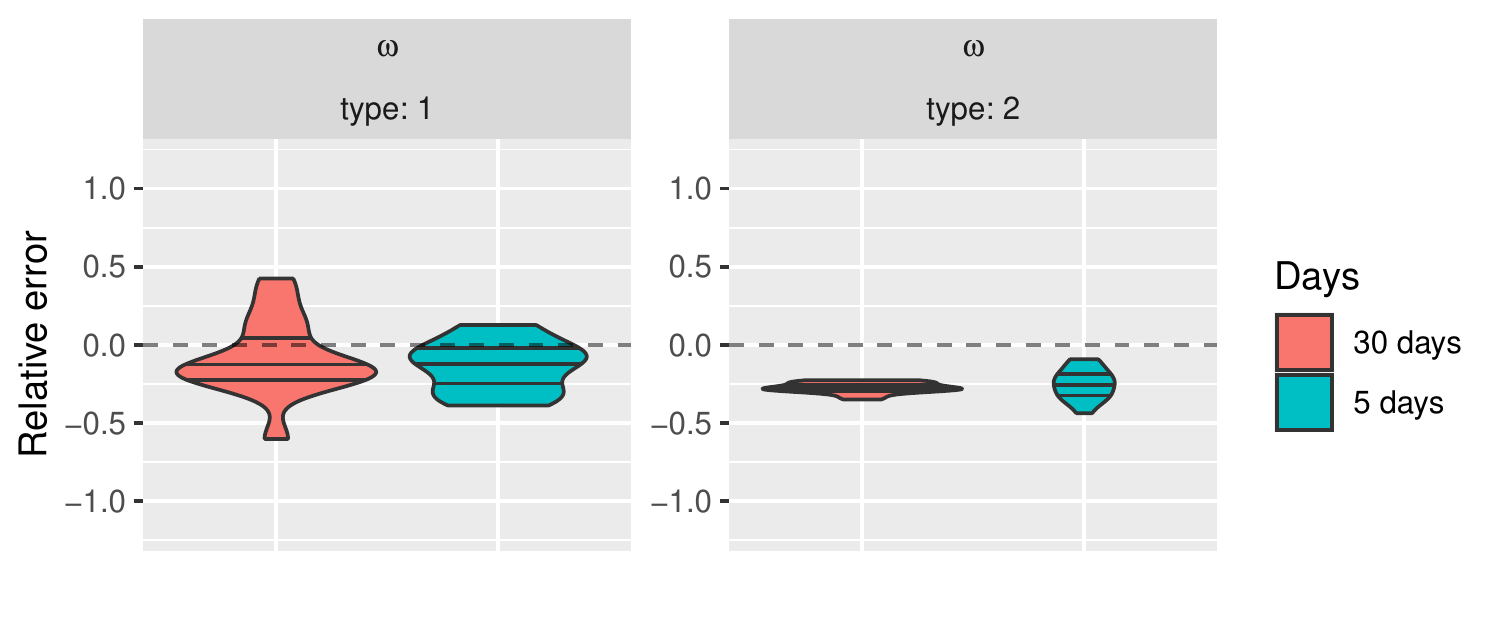} 

  \caption{Balanced market}
  \end{subfigure}
  \caption{Relative errors of the estimated covariance parameters $\omega_1 = 0.03$ and $\omega_2 = 0.70$ under the homogeneous model fit for Scenarios 1 to 4.}
  \label{fig:fm-covar}
\end{figure}

\begin{figure}[htp]
  \begin{subfigure}{.45\textwidth}
    \centering
\includegraphics[width=.8\linewidth]{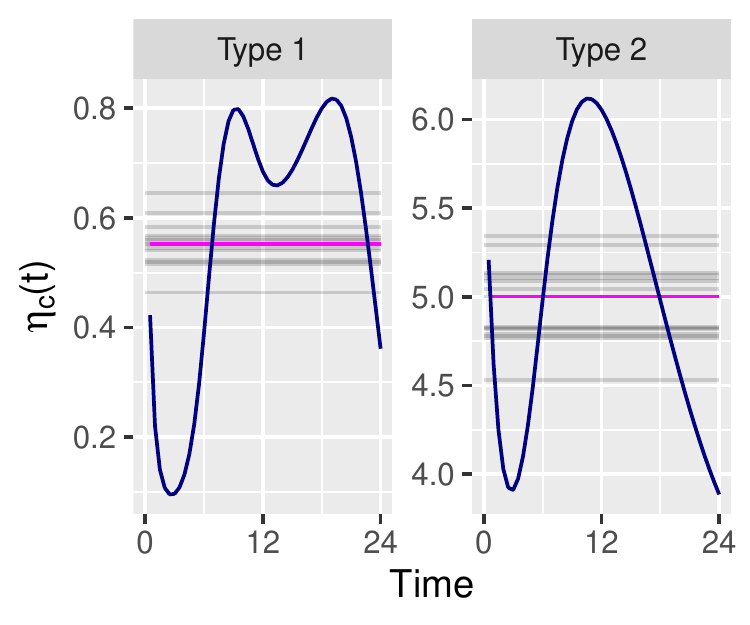} 

    \caption{Balanced market with 5-day data}
  \end{subfigure}
  \centering
  \begin{subfigure}{.48\textwidth}
    \centering
\includegraphics[width=.8\linewidth]{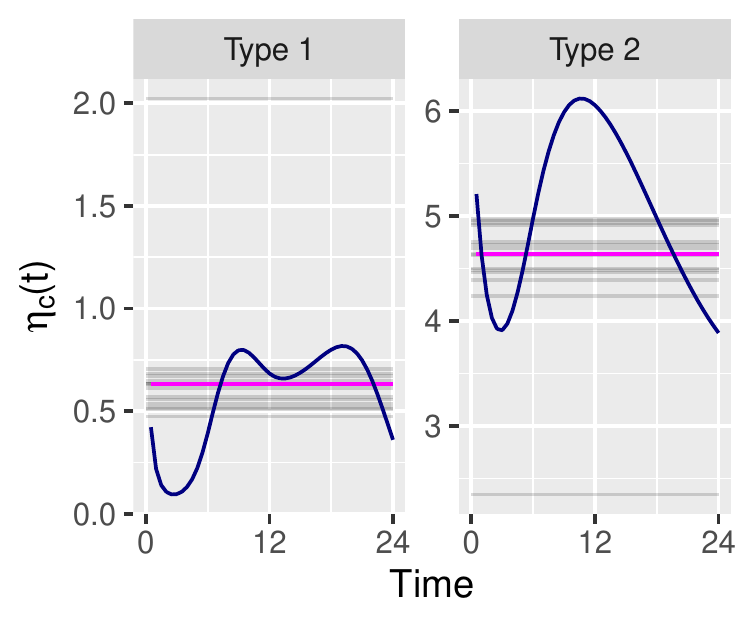} 

    \caption{Unbalanced market with 5-day data}
  \end{subfigure}
  \begin{subfigure}{.48\textwidth}
    \centering
\includegraphics[width=.8\linewidth]{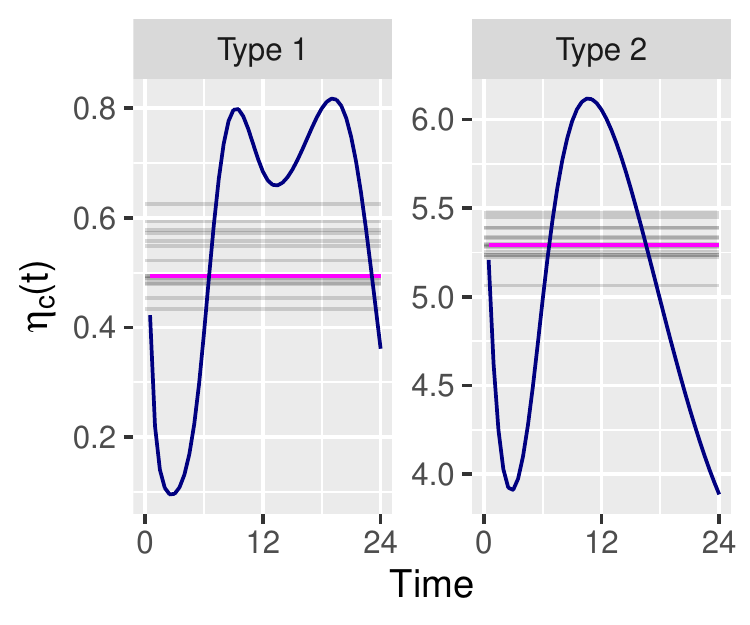} 

    \caption{Balanced market with 30-day data}
  \end{subfigure}
  \begin{subfigure}{.48\textwidth}
    \centering
\includegraphics[width=.8\linewidth]{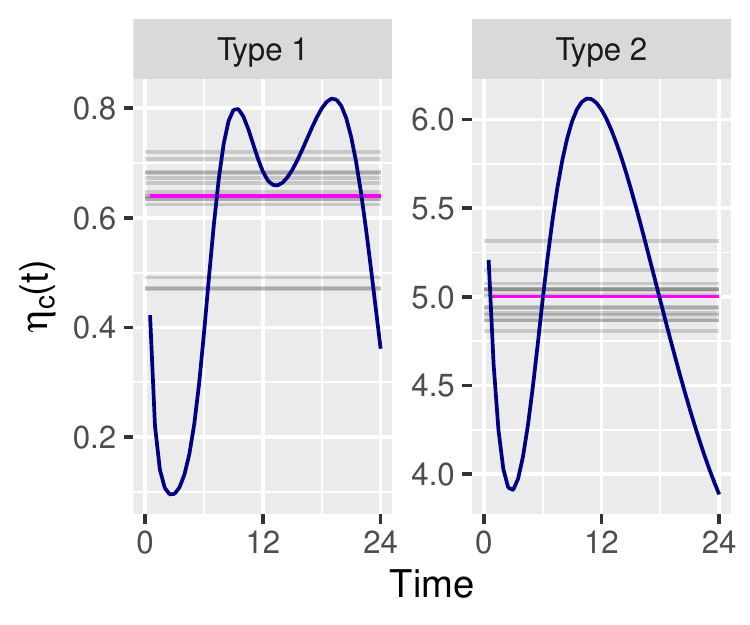} 

    \caption{Unbalanced market with 30-day data}
  \end{subfigure}  
  \caption{Estimated dispersion parameters for Scenarios 1 to 4 under the homogeneous model fit represented by the horizontal gray lines. Median lines are represented in magenta and the true variance functionals in blue.}
  \label{fig:simu-hm-funcvar}
  \end{figure}

The estimated covariance parameters for Scenarios 1 to 4 are displayed in Figures~\ref{fig:fm-covar} and \ref{fig:simu-hm-funcvar} and Table~\ref{tab:simu-fm-tab}. Figure~\ref{fig:simu-hm-funcvar} shows the estimated dispersion parameters represented over the true variance functionals, Figure~\ref{fig:fm-covar} the violin plots of the estimated decay parameters  and Table~\ref{tab:simu-fm-tab} the mean, median and square root of the Mean Squared Relative Error (MSRE)of the estimated decay parameters. Because the homogeneous model estimates a scalar as the dispersion parameter, the estimated values in Figure~\ref{fig:simu-hm-funcvar} are represented as constant lines over time. It seems that the horizontal lines are trying to capture an average of the variance functionals over time. In fact, taking the average of the variance functionals in Figure~\ref{fig:simu-hm-funcvar} over $t \in T$ yields $0.572$ for Type 1 and $5.03$ for Type 2, which are close to the respective median lines at $0.6324$ and $4.6375$. Moreover, the visibly overestimated valuefor Type 1 and the underestimated one for Type 2 in the unbalanced five-day scenario belong to the same run. On the other hand, the estimated decay parameters show systematic underestimation for Type 2 in all scenarios, as shown in Figure~\ref{fig:fm-covar}. The reduced estimate variability for the 30-day scenarios is observed only for estimated values of $\omega_2$. Again, the difference in magnitude of the parameters seems to have an influence on their performance, because $\omega_2>\omega_1$.   Furthermore, Table~\ref{tab:simu-fm-tab} shows the underestimation of $\omega_2$ in the median and mean values and smaller srMSREs in favour of balanced markets in the five-day scenarios.

\begin{figure}[htp]
  \centering
  \begin{subfigure}{.48\textwidth}
\includegraphics[width=\linewidth]{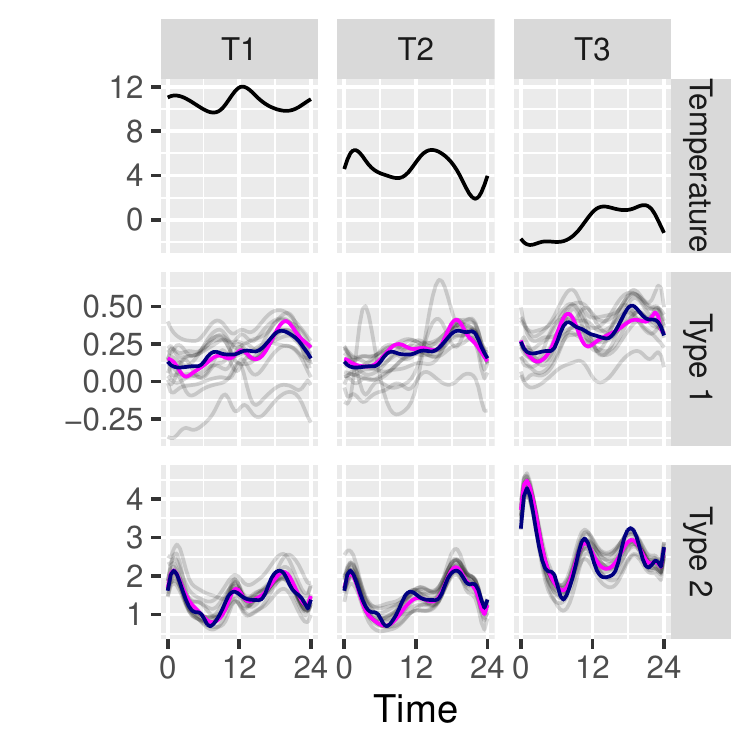} 

    \caption{Balanced market with 5-day data}
  \end{subfigure}
  \begin{subfigure}{.48\textwidth}
\includegraphics[width=\linewidth]{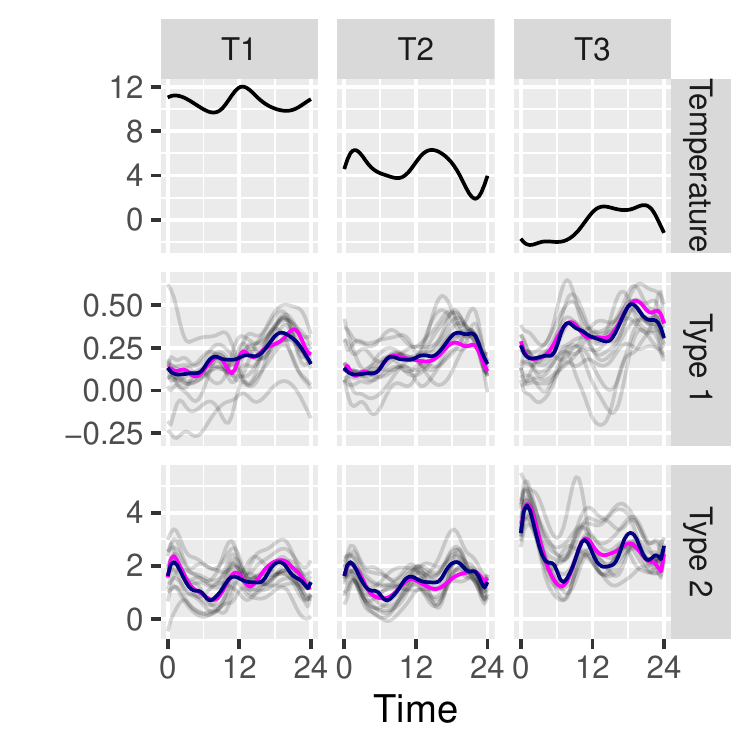} 

    \caption{Unbalanced market with 5-day data}
  \end{subfigure}
    \begin{subfigure}{.48\textwidth}
\includegraphics[width=\linewidth]{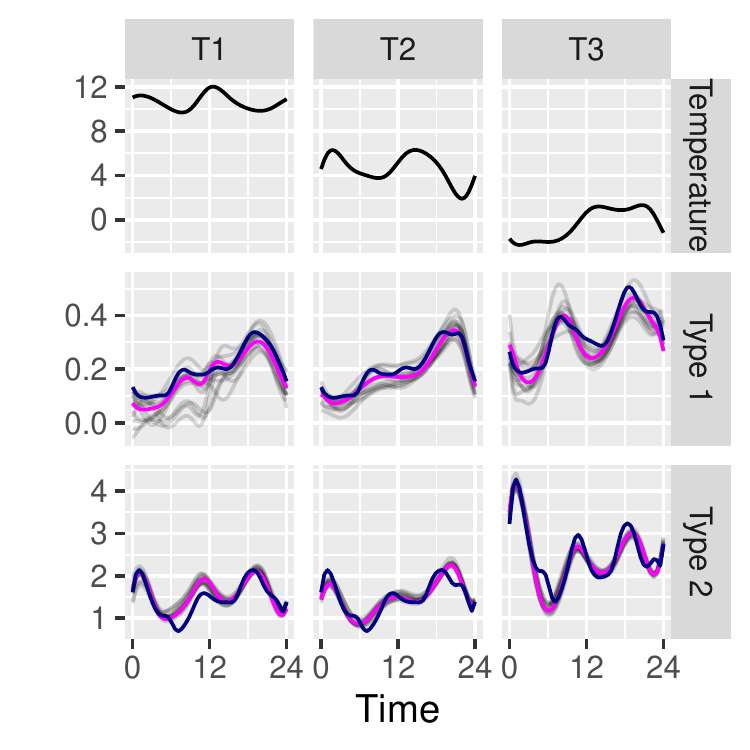} 

    \caption{Balanced market with 30-day data}
  \end{subfigure}
    \begin{subfigure}{.48\textwidth}
\includegraphics[width=\linewidth]{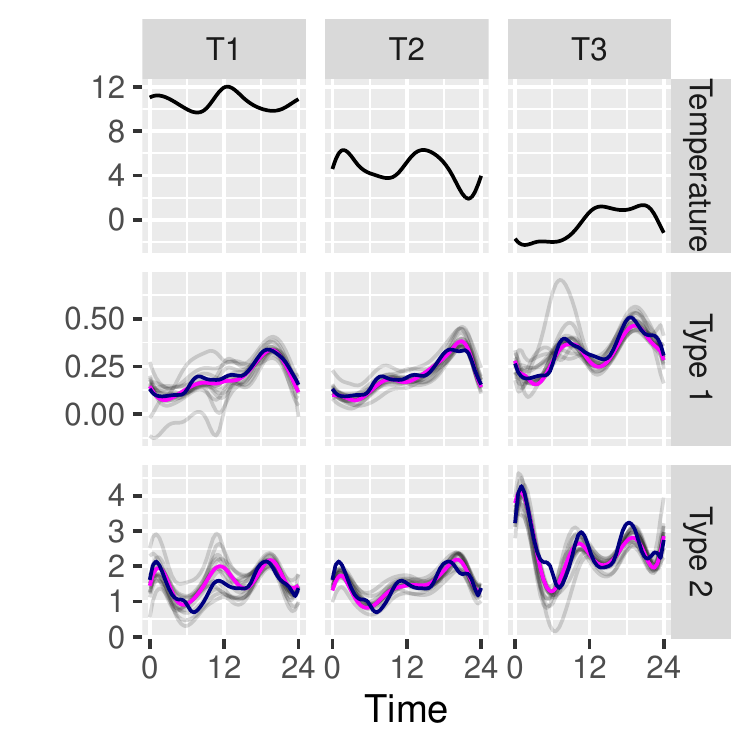} 

    \caption{Unbalanced market with 30-day data}
  \end{subfigure}
  \caption{In every panel, the first row represents the temperatures $T(t)$ for each temperature set T1, T2 and T3; the following rows represent the estimated typical curves of $\alpha_c(t,T(t))$ for customers of Type 1 and 2 in Scenarios 1 to 4 under the complete model fit. Median depth lines are represented in magenta, true typical curves in blue and estimated typical curves for each simulation run in gray.}
  \label{fig:fm-typical-comp}
\end{figure}

\begin{figure}[htp!]
  \centering
  \begin{subfigure}{.48\textwidth}
    \centering
\includegraphics[width=\linewidth]{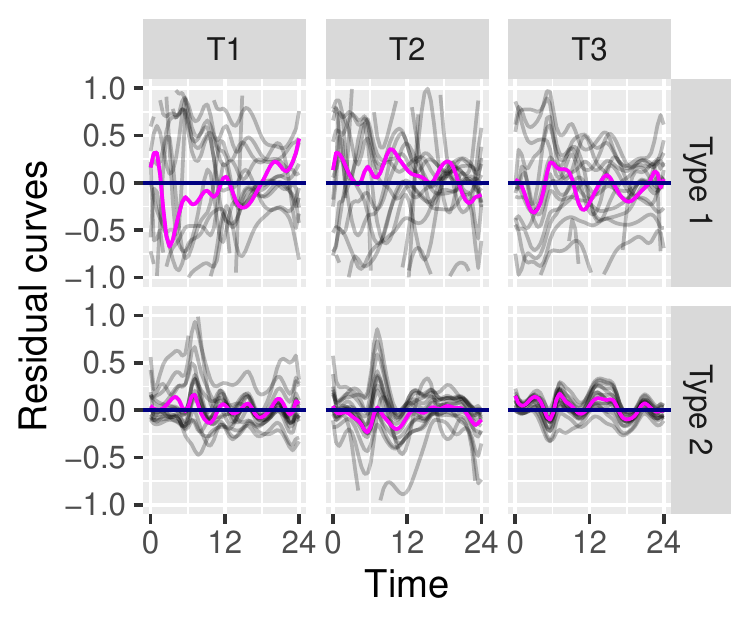}
    \caption{Balanced market with 5-day data}
  \end{subfigure}
  \begin{subfigure}{.48\textwidth}
    \centering
\includegraphics[width=\linewidth]{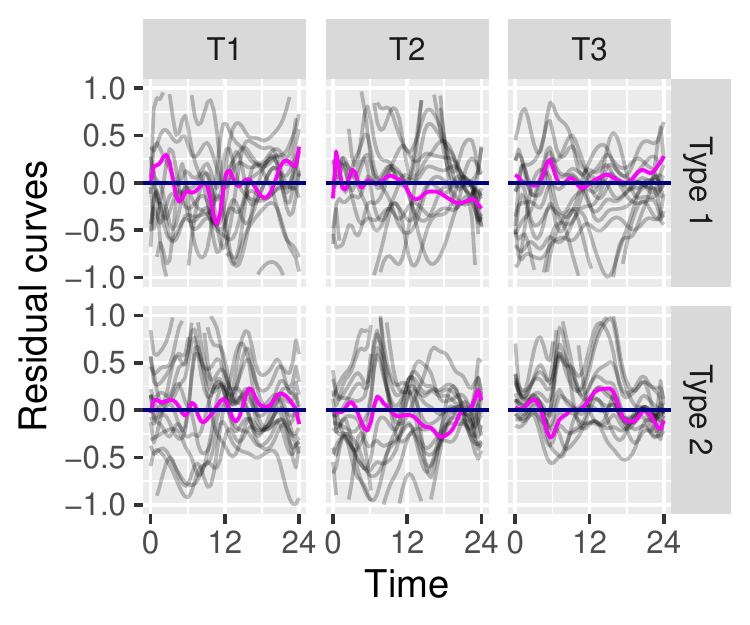}

    \caption{Unbalanced market with 5-day data}
  \end{subfigure}
    \begin{subfigure}{.48\textwidth}
      \centering
\includegraphics[width=\linewidth]{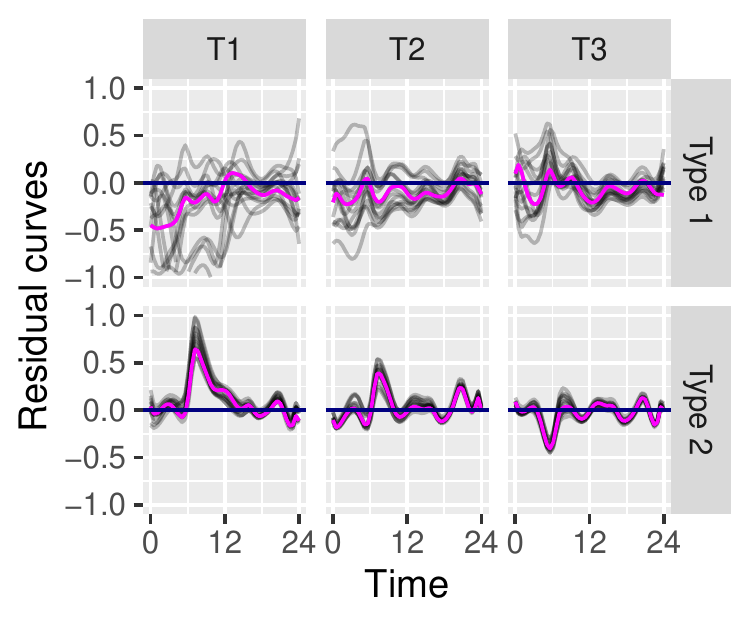} 

    \caption{Balanced market with 30-day data}
  \end{subfigure}
    \begin{subfigure}{.48\textwidth}
      \centering
\includegraphics[width=\linewidth]{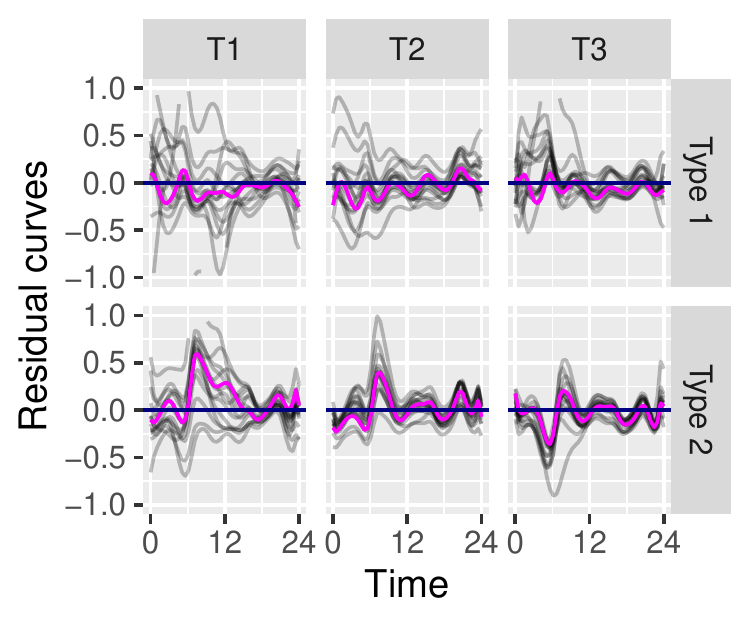} 

    \caption{Unbalanced market with 30-day data}
  \end{subfigure}
  \caption{Residual curves of the estimated typical curves in grey and their median depth in magenta under the complete model fit in Figure~\ref{fig:fm-typical-comp}.  }
  \label{fig:fm-comp-typical-mse}
\end{figure}

Figure~\ref{fig:fm-typical-comp} analogously shows the estimated typical surfaces for Scenarios 1 to 4 under the complete model fit. Again, observe that the estimated curve variability is reduced in the 30-day scenarios, especially for Type 2 under balanced markets. In addition, the advantage of balanced markets under the five-day scenarios can be seen from the lower variability of the residual curves in Figure~\ref{fig:fm-comp-typical-mse} and  the lower fMSRE in Table~\ref{tab:simu-fmse}. The complete model does not present clear superiority in terms of fMSRE  compared with the homogeneous model study.

Figure~\ref{fig:fm-gamma}b displays the relative errors of the estimated coefficients $\gamma_1$ and $\gamma_2$ associated with the explanatory variables. The characteristics of the violins are much like the respective ones in the homogeneous model. In fact, note that the srMSREs in Table~\ref{tab:gamma} of both studies have similar values. Consequently, the complete model case shares the aspect of smaller srMSREs for estimates of $\gamma_1$, especially in the 30-day scenarios.

\begin{figure}[htp]
  \begin{subfigure}{.45\textwidth}
    \centering
\includegraphics[width=.8\linewidth]{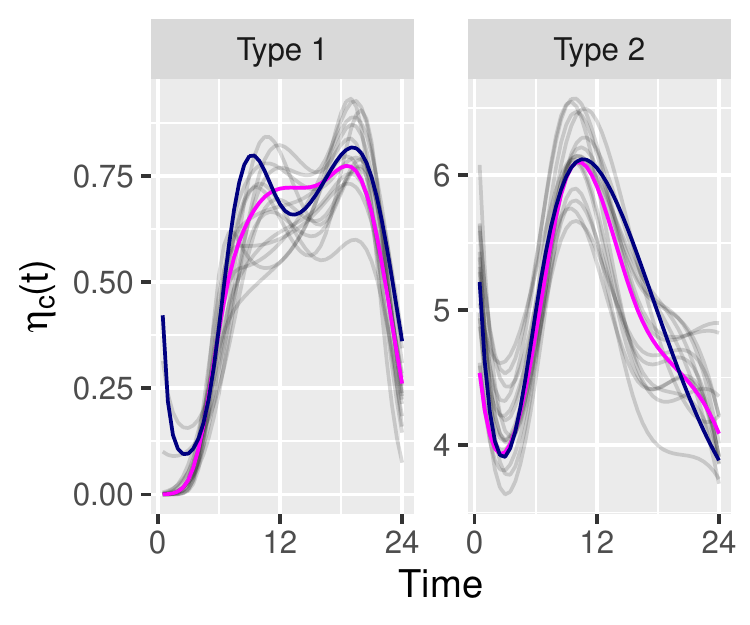} 

    \caption{Balanced market with 5-day data}
  \end{subfigure}
  \centering
  \begin{subfigure}{.48\textwidth}
    \centering
\includegraphics[width=.8\linewidth]{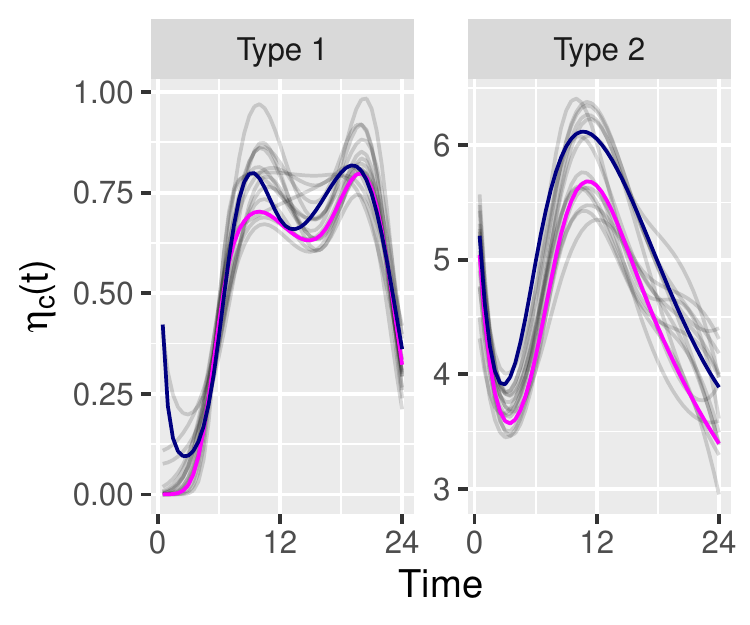} 

    \caption{Unbalanced market with 5-day data}
  \end{subfigure}
  \begin{subfigure}{.48\textwidth}
    \centering
\includegraphics[width=.8\linewidth]{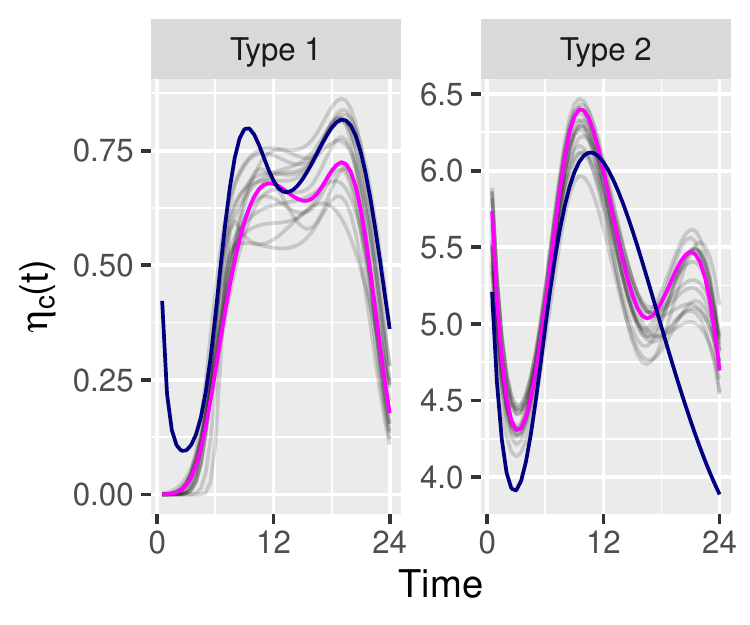} 

    \caption{Balanced market with 30-day data}
  \end{subfigure}
  \begin{subfigure}{.48\textwidth}
    \centering
\includegraphics[width=.8\linewidth]{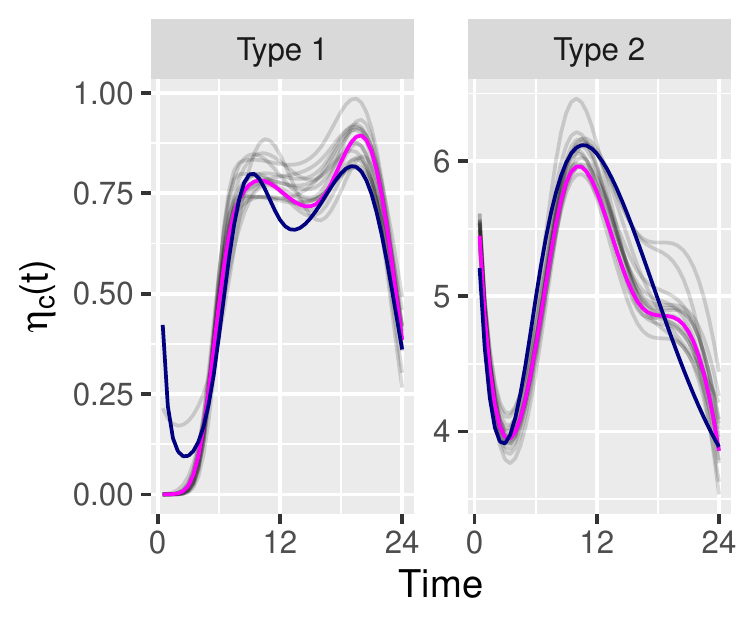} 

    \caption{Unbalanced market with 30-day data}
  \end{subfigure}  
  \caption{Estimated variance functionals for Scenarios 1 to 4 under the complete model fit. Median depth lines are represented in magenta,  true variance functionals in blue and estimated curves in gray.}
  \label{fig:simu-funcvar}
  \end{figure}

\begin{figure}[htp]
 \centering
  \begin{subfigure}{.45\textwidth}
    \centering
\includegraphics[width=.96\linewidth]{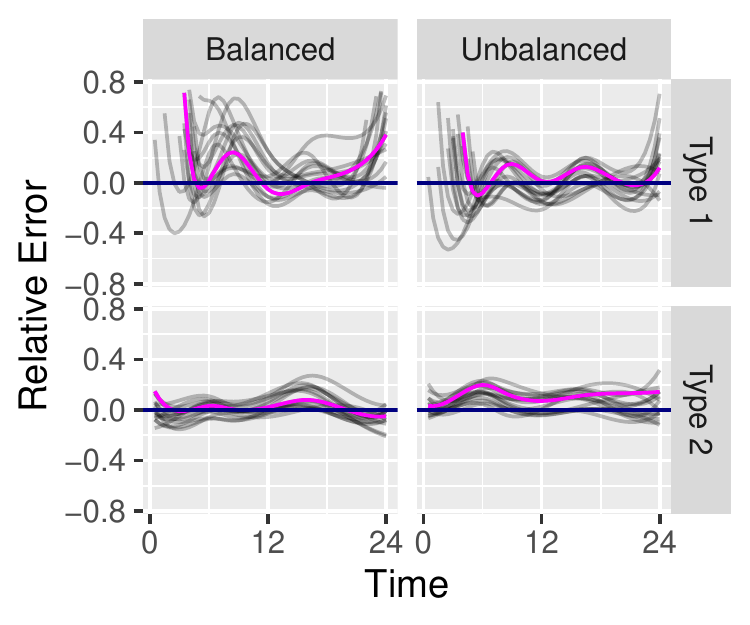} 
    \caption{5 days}
  \end{subfigure}
\begin{subfigure}{.45\textwidth}
    \centering
\includegraphics[width=.96\linewidth]{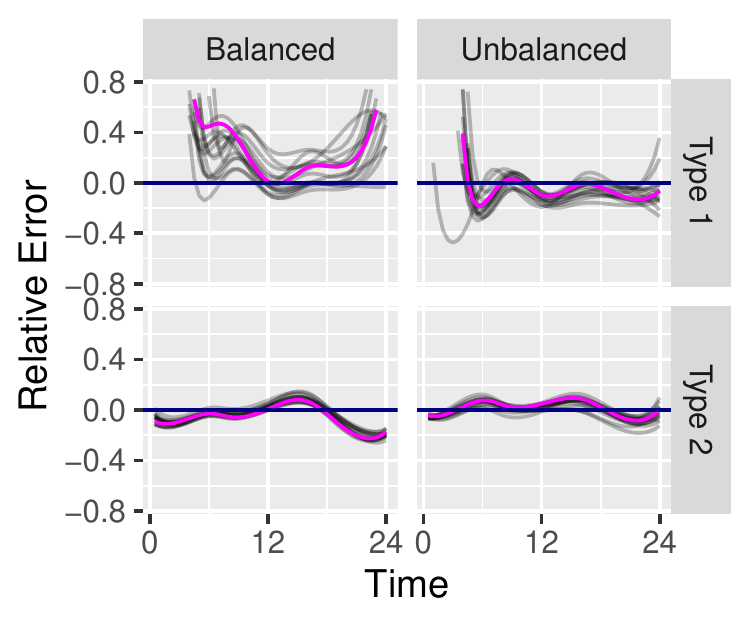} 

    \caption{30 days}
  \end{subfigure}
  \caption{Residual curves of the estimated variance functionals for Scenarios 1 to 4 under the complete model fit. Median lines are represented in magenta  and residual  curves in gray.}
  \label{fig:simu-funcvar-mse}
  \end{figure}

\begin{figure}[htp]
  \centering
  \begin{subfigure}{\textwidth}
    \centering
\includegraphics[width=.8\linewidth]{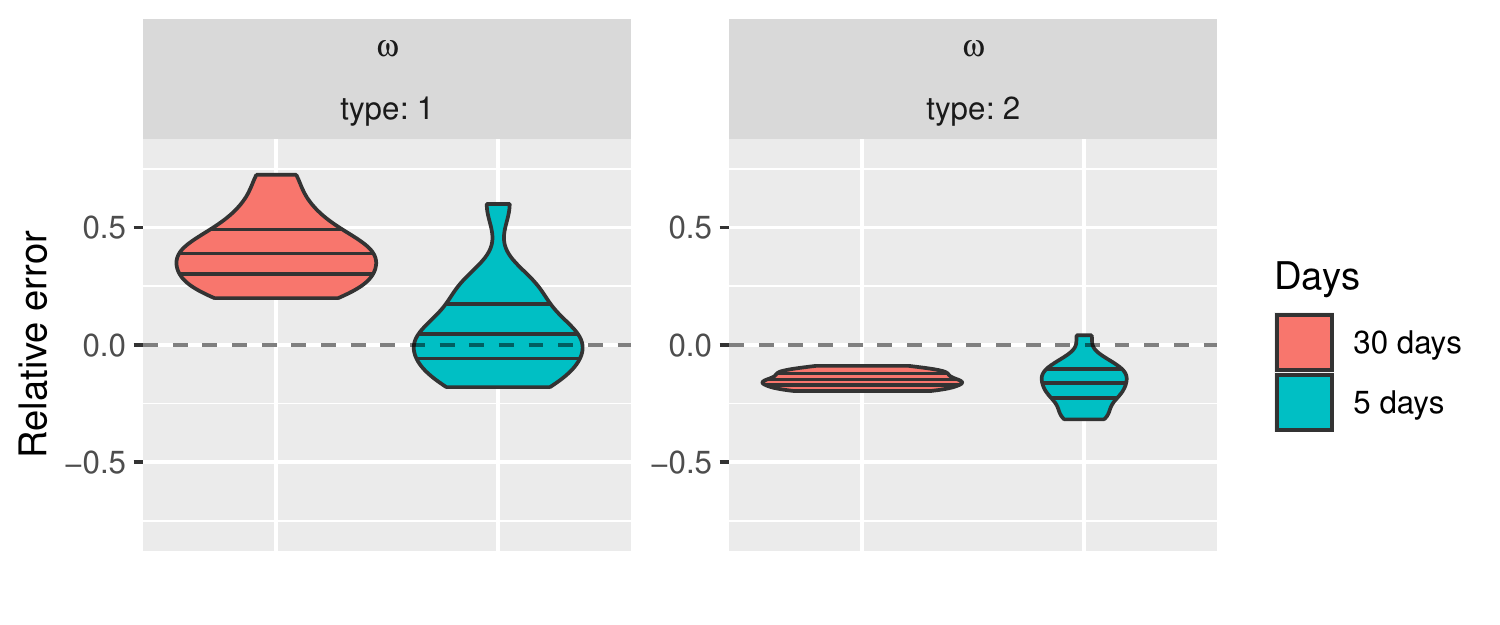} 

  \caption{Unbalanced market}
  \end{subfigure}
  \begin{subfigure}{\textwidth}
    \centering    
\includegraphics[width=.8\linewidth]{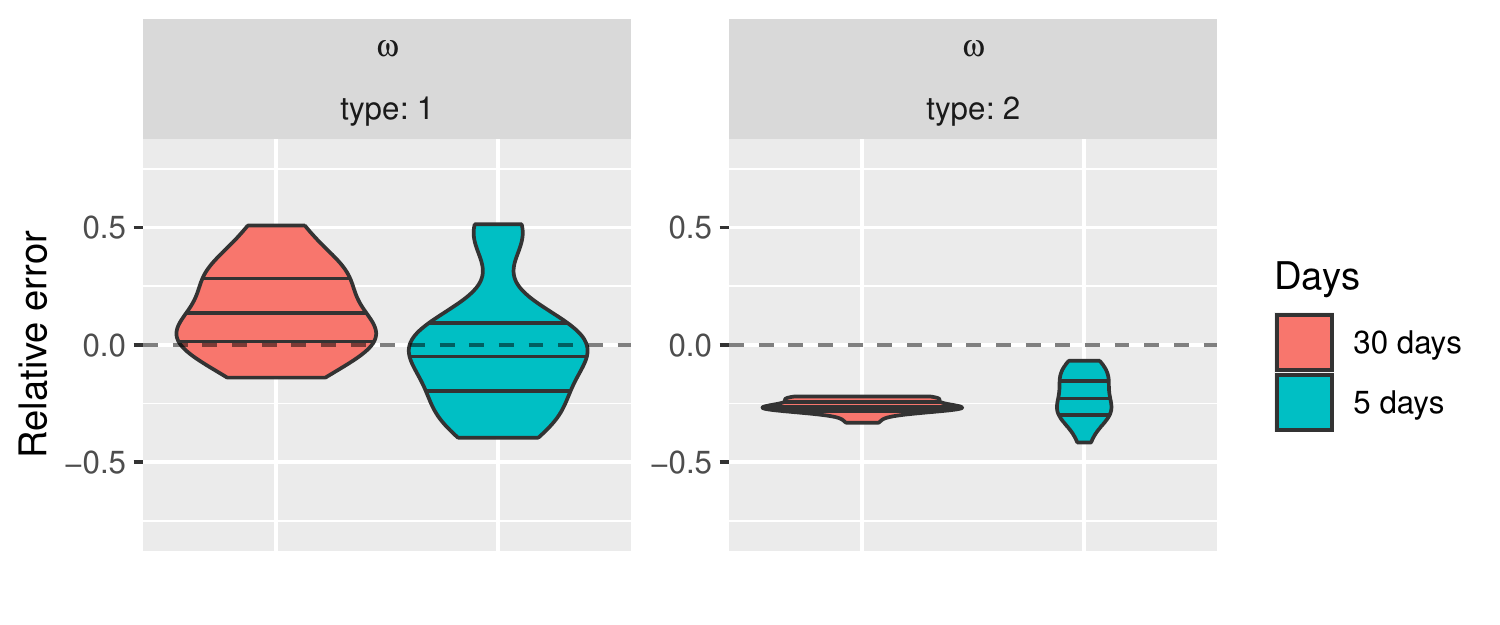} 

  \caption{Balanced market}
  \end{subfigure}
  \caption{Relative errors of the estimated covariance parameters $\omega_1=0.03$ and $\omega_2=0.70$ for Scenarios 1 to 4, under the complete aggregated data model fit.}
  \label{fig:fm-covar-comp}
\end{figure}

Finally, Figure~\ref{fig:simu-funcvar} shows the estimated variance functionals of the complete model and Figure~\ref{fig:simu-funcvar-mse} their respective residual curves. As observed in the typical curves, the 30-day scenarios present lower estimate variability than the five-day scenarios. In general, the estimates capture the main features of the true curves, such as the prolonged higher values for customers of Type 1 and the decreasing values after 12 AM for Type 2. However, in some regions, the estimated curves present behaviour different from the true curve. In all scenarios, observe that the Type 1 curves begin almost at zero, whereas the true curve has a small peak with rapid decay. Moreover, in the balanced 30-day scenario, the estimated variance functionals for Type 2 customers present a nonexistent local peak at the end of the day. The violin plots of the relative errors of the estimated decay parameters are displayed in Figure~\ref{fig:fm-covar-comp} and their summary statistics in Table~\ref{tab:simu-fm-tab}. Essentially, the complete model offers estimates with smaller srMSRE  compared with the homogeneous model, but the underestimation of $\omega_2$ persists. 

In addition, because the homogeneous model is nested in the complete model, Table~\ref{tab:simu-rvm} in  Section~\ref{chap:supl-tabs} shows the likelihood ratio test for all runs in every scenario. In all cases the test favours the complete model fit with p-values smaller than $0.0001$.

\begin{table}[bhpt]
        \caption{Functional mean squared relative errors of the estimated typical curves under the homogeneous (Figure~\ref{fig:fm-hm-typical}) and complete (Figure~\ref{fig:fm-typical-comp}) model fit for Scenarios 1 to 4.}
    \label{tab:simu-fmse}
\centering    
\begin{tabular}{llllr}
\toprule
Model & Days & Type & Market balance & fMSRE\\
\midrule
 &  &  & Balanced & 17.6706\\
\cmidrule{4-5}
 &  & \multirow{-2}{*}{\raggedright\arraybackslash Type 1} & Unbalanced & 18.4121\\
\cmidrule{3-5}
 &  &  & Balanced & 0.8899\\
\cmidrule{4-5}
 & \multirow{-4}{*}{\raggedright\arraybackslash 5 days} & \multirow{-2}{*}{\raggedright\arraybackslash Type 2} & Unbalanced & 4.6557\\
\cmidrule{2-5}
 &  &  & Balanced & 1.9525\\
\cmidrule{4-5}
 &  & \multirow{-2}{*}{\raggedright\arraybackslash Type 1} & Unbalanced & 2.1867\\
\cmidrule{3-5}
 &  &  & Balanced & 0.5814\\
\cmidrule{4-5}
\multirow{-8}{*}{\raggedright\arraybackslash Homogeneous} & \multirow{-4}{*}{\raggedright\arraybackslash 30 days} & \multirow{-2}{*}{\raggedright\arraybackslash Type 2} & Unbalanced & 1.0923\\
\cmidrule{1-5}
 &  &  & Balanced & 19.5322\\
\cmidrule{4-5}
 &  & \multirow{-2}{*}{\raggedright\arraybackslash Type 1} & Unbalanced & 14.9073\\
\cmidrule{3-5}
 &  &  & Balanced & 0.9747\\
\cmidrule{4-5}
 & \multirow{-4}{*}{\raggedright\arraybackslash 5 days} & \multirow{-2}{*}{\raggedright\arraybackslash Type 2} & Unbalanced & 3.9616\\
\cmidrule{2-5}
 &  &  & Balanced & 1.9122\\
\cmidrule{4-5}
 &  & \multirow{-2}{*}{\raggedright\arraybackslash Type 1} & Unbalanced & 1.9651\\
\cmidrule{3-5}
 &  &  & Balanced & 0.7720\\
\cmidrule{4-5}
\multirow{-8}{*}{\raggedright\arraybackslash Complete} & \multirow{-4}{*}{\raggedright\arraybackslash 30 days} & \multirow{-2}{*}{\raggedright\arraybackslash Type 2} & Unbalanced & 1.1850\\
\bottomrule
\end{tabular}
\end{table}

\begin{table}[b]

\caption{\label{tab:gamma}Mean, median and square root of the Mean Squared Relative Error (MSRE) of the estimated explanatory variables parameters under the homogeneous and complete model fit for Scenarios 1 to 4.}
\centering
\begin{tabular}{llrrrrr}
\toprule
Model & Parameter & Days & Market & Mean & Median & $\sqrt{MSRE}$\\
\midrule
 &  &  & Balanced & 12.7193 & 12.0985 & 0.3054\\

 &  & \multirow{-2}{*}{\centering\arraybackslash 30 days} & Unbalanced & 11.8776 & 12.1702 & 0.2516\\
\cmidrule{3-7}
 &  &  & Balanced & 12.0501 & 12.1832 & 1.7045\\

 & \multirow{-4}{*}{\centering\arraybackslash $\gamma_1=13$} & \multirow{-2}{*}{\centering\arraybackslash 5 days} & Unbalanced & 17.2909 & 8.3669 & 1.6263\\
\cmidrule{2-7}
 &  &  & Balanced & 0.0330 & 0.0421 & 3.4016\\

 &  & \multirow{-2}{*}{\centering\arraybackslash 30 days} & Unbalanced & 0.0289 & 0.0328 & 2.9574\\
\cmidrule{3-7}
 &  &  & Balanced & 0.0156 & -0.0159 & 14.6551\\

\multirow{-8}{*}{\centering\arraybackslash Homogeneous} & \multirow{-4}{*}{\centering\arraybackslash $\gamma_2=0.0011$} & \multirow{-2}{*}{\centering\arraybackslash 5 days} & Unbalanced & 0.0231 & 0.0334 & 5.6259\\
\cmidrule{1-7}
 &  &  & Balanced & 13.3503 & 12.9929 & 0.2578\\

 &  & \multirow{-2}{*}{\centering\arraybackslash 30 days} & Unbalanced & 11.8177 & 12.1773 & 0.2295\\
\cmidrule{3-7}
 &  &  & Balanced & 11.3529 & 15.1660 & 1.6096\\

 & \multirow{-4}{*}{\centering\arraybackslash $\gamma_1=13$} & \multirow{-2}{*}{\centering\arraybackslash 5 days} & Unbalanced & 15.3926 & 9.1294 & 1.1091\\
\cmidrule{2-7}
 &  &  & Balanced & 0.0330 & 0.0490 & 3.0979\\

 &  & \multirow{-2}{*}{\centering\arraybackslash 30 days} & Unbalanced & 0.0283 & 0.0337 & 2.5957\\
\cmidrule{3-7}
 &  &  & Balanced & 0.0254 & 0.0104 & 15.3070\\

\multirow{-8}{*}{\centering\arraybackslash Complete} & \multirow{-4}{*}{\centering\arraybackslash $\gamma_2=0.0011$} & \multirow{-2}{*}{\centering\arraybackslash 5 days} & Unbalanced & 0.0274 & 0.0264 & 5.4715\\
\bottomrule
\end{tabular}
\end{table}

\begin{table}[b]\centering
\caption{Mean, median and square root of the Mean Squared Relative Error (MSRE) of the estimated decay parameters for Scenarios 1 to 4, under the homogeneous model fit.}  

\begin{tabular}{llrrrrr}
\toprule
Model & Parameter & Days & Market  & Median & Mean & $\sqrt{MSRE}$\\
\midrule
  &  &  & Balanced  & 0.0257 & 0.0275 & 0.2549\\

 &  & \multirow{-2}{*}{\centering\arraybackslash 30 days} & Unbalanced  & 0.0382 & 0.0322 & 0.5544\\
\cmidrule{3-7}
 &  &  & Balanced  & 0.0270 & 0.0259 & 0.2107\\

 & \multirow{-4}{*}{\centering\arraybackslash $\omega_1=0.03$} & \multirow{-2}{*}{\centering\arraybackslash 5 days} & Unbalanced & 0.0287 & 0.0764 & 6.6934\\
\cmidrule{2-7}
 &  &  & Balanced  & 0.5028 & 0.5079 & 0.2764\\

 &  & \multirow{-2}{*}{\centering\arraybackslash 30 days} & Unbalanced & 0.5520 & 0.5562 & 0.2111\\
\cmidrule{3-7}
 &  &  & Balanced & 0.5226 & 0.5251 & 0.2675\\

\multirow{-7}{*}{\centering\arraybackslash Homogeneous} & \multirow{-4}{*}{\centering\arraybackslash $\omega_2=0.70$} & \multirow{-2}{*}{\centering\arraybackslash 5 days} & Unbalanced  & 0.5427 & 0.5073 & 0.3318\\
\cmidrule{1-7}
 &  &  & Balanced & 0.0316 & 0.0298 & 0.4243\\

 &  & \multirow{-2}{*}{\centering\arraybackslash 30 days} & Unbalanced & 0.0419 & 0.0418 & 0.4208\\
\cmidrule{3-7}
 &  &  & Balanced  & 0.0288 & 0.0291 & 0.2516\\

 & \multirow{-4}{*}{\centering\arraybackslash $\omega_1=0.03$} & \multirow{-2}{*}{\centering\arraybackslash 5 days} & Unbalanced  & 0.0312 & 0.0318 & 0.2065\\
\cmidrule{2-7}
 &  &  & Balanced  & 0.5169 & 0.5178 & 0.2621\\

 &  & \multirow{-2}{*}{\centering\arraybackslash 30 days} & Unbalanced  & 0.5899 & 0.5962 & 0.1521\\
\cmidrule{3-7}
 &  &  & Balanced & 0.5323 & 0.5450 & 0.2418\\

\multirow{-7}{*}{\centering\arraybackslash Complete} & \multirow{-4}{*}{\centering\arraybackslash $\omega_2=0.70$} & \multirow{-2}{*}{\centering\arraybackslash 5 days} & Unbalanced & 0.5819 & 0.5831 & 0.1917\\
\bottomrule
\end{tabular}
\label{tab:simu-fm-tab}
\end{table}

\subsubsection{Discussion and conclusion}
\label{sec:simu-fm-conclustion}

In all scenarios, the estimated typical surfaces of the homogeneous model are robust to the misspecification of the covariance structure, as shown in Figure~\ref{fig:fm-hm-typical}. Both studies show an expected better performance for the 30-day scenarios in terms of estimation variability and fMSRE, which is also true for balanced scenarios compared to unbalanced ones.

Interestingly, it seems that the magnitude of the parameters may influence the quality of the estimate. The estimated typical curves, for example, show better estimates for customers of Type 2, the ones with higher consumption curves compared to Type 1. The same characteristic is observed in the relative errors of the estimates of $\gamma_1$, a parameter much greater than $\gamma_2$. Nonetheless, this is not as evident in the decay parameter estimation. The latter seems to be especially difficult to estimate because its performance in terms of precision and srMSRE (square root of the mean squared relative error) is not improved under 30-day scenarios or balanced markets. In fact, the estimates present a systematic underestimation of $\omega_2=0.70$. Still on the covariance structure, the estimated variance functionals in the complete study can capture the main features of the true ones, despite an unexpected local peak in the 30-day scenario with balanced market.

In general, the advantages of balanced markets and 30-day scenarios is evident. The complete model provides a functional variance structure that can capture different dispersions over time. However, in terms of typical surface estimation, there is no clear difference between the homogeneous and the complete model fit, which could be attributed to the fact that the least-squares estimators are unbiased independently of the covariance structure, as noted in Section 3.6 of the main text.

\subsection{Clustering the aggregated model}
\label{sec:simu-cluster}

This section studies the clustering approach of the aggregated data model presented in Section 3.5 of the main manuscript. The method was tested in Scenarios 5 to 8 and is presented in Table~\ref{tab:scenarios}, where data from three clusters were simulated under the homogeneous covariance structure. In contrast to Section~\ref{sec:simu-fm}, typical surfaces were not considered here for the clustering model.

Section~\ref{sec:simu-cluster-setup} details the clustering configuration and the true parameters. Sections \ref{sec:simu-cluster-results} describes the main results and \ref{sec:simu-cluster-conclusion} contains the  discussion and conclusions of this simulation study.


\subsubsection{Clustering setup and true parameters}
\label{sec:simu-cluster-setup}

Scenarios 5 to 8 in Table~\ref{tab:scenarios} are made up of variations of the number of days (5 and 30) and the market (balanced and unbalanced). The remaining simulation parameters were fixed to three clusters and two types of customers observed in 12 substations every 30 minutes.  The true cluster allocation is displayed in Table~\ref{tab:simu-cl-true}, where substations  1 to 6 are assigned to Cluster 1, substations 7 to 10 to Cluster 2 and finally substations 11 and 12  to Cluster 3.  The chosen covariance structure is the homogeneous one, where each customer type has its own dispersion and decay parameters. 

\begin{table}[hb]
    \centering
       \caption{True cluster assignment for each substation in the simulation study.}
    \begin{tabular}{l|cccccccccccc}
    \toprule
       Substation  & 1 & 2 &3 &4 &5 &6 &7 &8 &9 &10 &11 &12\\
       \midrule
        True cluster &1 &1 &1 &1 &1 &1 &2 &2 &2 &2 &3 &3\\
        \bottomrule
    \end{tabular}
     \label{tab:simu-cl-true}
\end{table}

\begin{figure}[thp]
    \centering
\includegraphics[width=.8\linewidth]{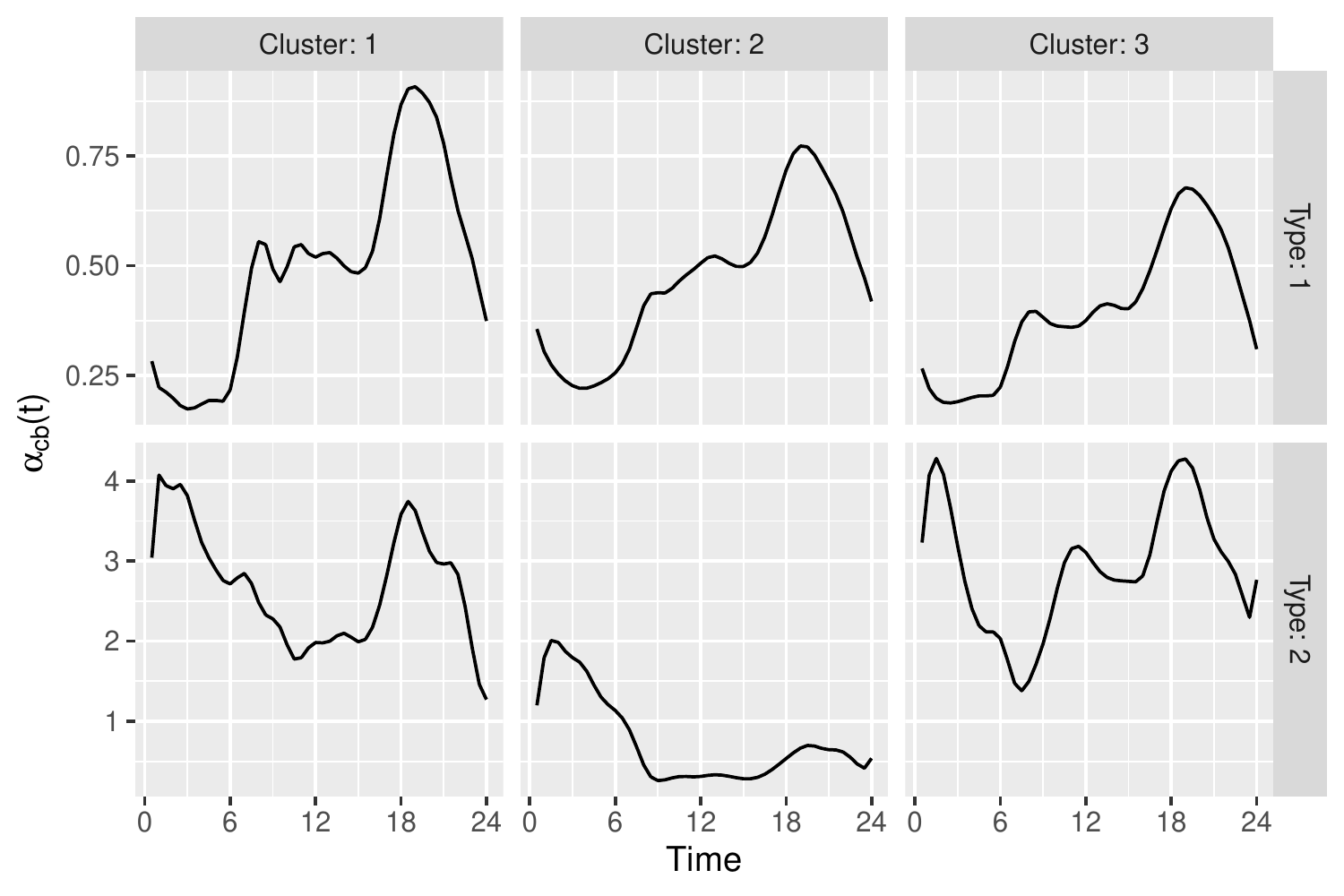} 
    \caption{True typical curves for the clustering simulation considering three clusters and two customers types.}
    \label{fig:true_mc_cl}
\end{figure}

Figure~\ref{fig:true_mc_cl} shows the six typical curves divided into the two customer types for each of the three clusters. Their shapes were based on the estimated typical curves for the UK energy grid dataset in Section 4.3 of the main paper. Hence, Type 1 mimics the unrestricted domestic customer, with similar shapes among clusters, whereas Type 2 mimics the ``Economy 7'' customer.

The covariance parameters that compose the homogeneous covariance structure of the simulated scenarios are presented in Table~\ref{tab:true_cp_cl} divided by cluster, parameter and customer type, where, again, their values were based on the estimated covariance parameters for the UK energy grid dataset described in Section 4.3 of the main manuscript.

\begin{table}[hb]
    \centering
        \caption{True covariance parameters for clustering simulation considering three clusters and two customer types.}
    \label{tab:true_cp_cl}
\begin{tabular}{llrr}
\toprule
Cluster & Parameter & Type & Value\\
\midrule
 &  & $c=1$ & 1.54\\

 & \multirow{-2}{*}{\centering\arraybackslash $\sigma_{cb}$} & $c=2$ & 1.53\\
\cmidrule{2-4}
 &  & $c=1$ & 0.16\\

\multirow{-4}{*}{\centering\arraybackslash $b=1$} & \multirow{-2}{*}{\centering\arraybackslash $\omega_{cb}$} & $c=2$ & 0.03\\
\cmidrule{1-4}
 &  & $c=1$ & 1.07\\

 & \multirow{-2}{*}{\centering\arraybackslash $\sigma_{cb}$} & $c=2$ & 1.28\\
\cmidrule{2-4}
 &  & $c=1$ & 0.12\\

\multirow{-4}{*}{\centering\arraybackslash $b=2$} & \multirow{-2}{*}{\centering\arraybackslash $\omega_{cb}$} & $c=2$ & 0.09\\
\cmidrule{1-4}
 &  & $c=1$ & 0.43\\

 & \multirow{-2}{*}{\centering\arraybackslash $\sigma_{cb}$} & $c=2$ & 5.18\\
\cmidrule{2-4}
 &  & $c=1$ & 0.02\\

\multirow{-4}{*}{\centering\arraybackslash $b=3$} & \multirow{-2}{*}{\centering\arraybackslash $\omega_{cb}$} & $c=2$ & 0.37\\
\bottomrule
\end{tabular}


\end{table}

\subsubsection{Results}
\label{sec:simu-cluster-results}

Scenarios 5 to 8 were subjected to two model fits: clustering homogeneous aggregated data models with two and three clusters. The two-cluster fit tested how the model would perform if the number of clusters were underdetermined, that is, how the method groups substations and consequently what are the characteristics of the estimated typical curves and covariance parameters. On the other hand, the three-cluster fit evaluates model performance under correct scenarios.

Before presenting the results, it is necessary to note the number of runs that did not converge or converged to a local maximum in this simulation at each model fit. The non-convergent runs in the two-cluster fit were the following: two runs in the unbalanced five-day market scenario, one run in the balanced five-day market scenario, and two runs in the balanced 30-day market scenario. Moreover, the three-cluster fit had three non-convergent runs in both balanced and unbalanced five-day market scenarios, four in the unbalanced 30-day scenario, and two in the balanced 30-day scenarios. The runs that converged to local maxima presented anomalous estimated typical curves with negative and discrepant values.

Let us begin with the two-cluster fit and its respective substation clustering as shown in Table~\ref{tab:simu-cl-assign}. In all runs, substations are assigned with high probability to the same cluster configuration, and therefore Substations 1 to 6 were assigned to Cluster 1, Substations 7 to 10 to Cluster 2, and Substations 11 and 12 to Cluster 3. Note that the substations of true Cluster 3 were assigned to the larger Cluster 1 in the two-cluster model. Recall also that the true Clusters 1 and 3 in Figure~\ref{fig:true_mc_cl} have similar typical curves for Type 1 and Type 2 and that both have approximately the same magnitude and characteristics over time, and hence it is reasonable that they merge into a single cluster in the two-cluster model.

\begin{figure}[p]
  \centering
  \begin{subfigure}{.49\textwidth}
    \centering
\includegraphics[width=.8\linewidth]{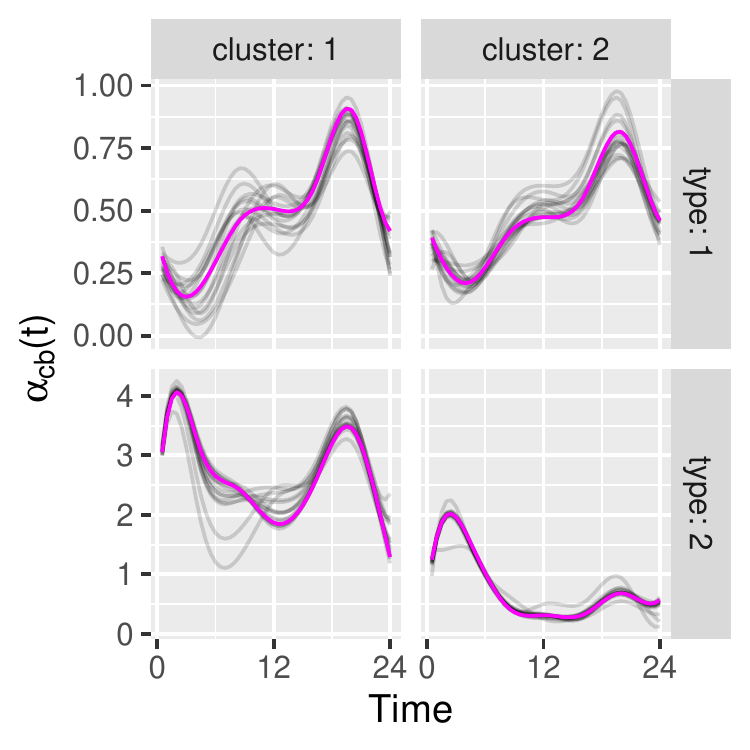} 

    \caption{Balanced market with 5-day data}
  \end{subfigure}
  \begin{subfigure}{.49\textwidth}
    \centering
\includegraphics[width=.8\linewidth]{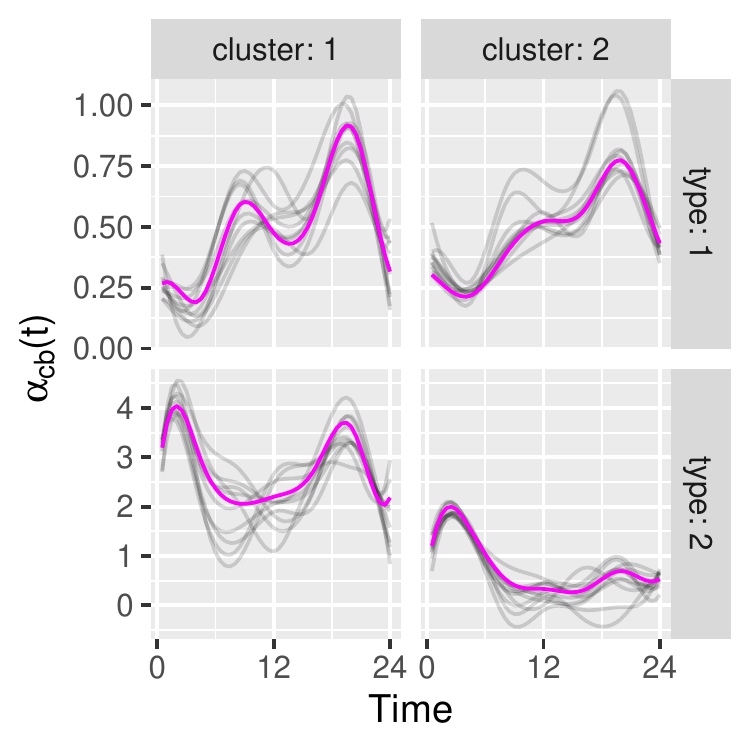} 

    \caption{Unbalanced market with 5-day data}
  \end{subfigure}
    \begin{subfigure}{.49\textwidth}
    \centering
\includegraphics[width=.8\linewidth]{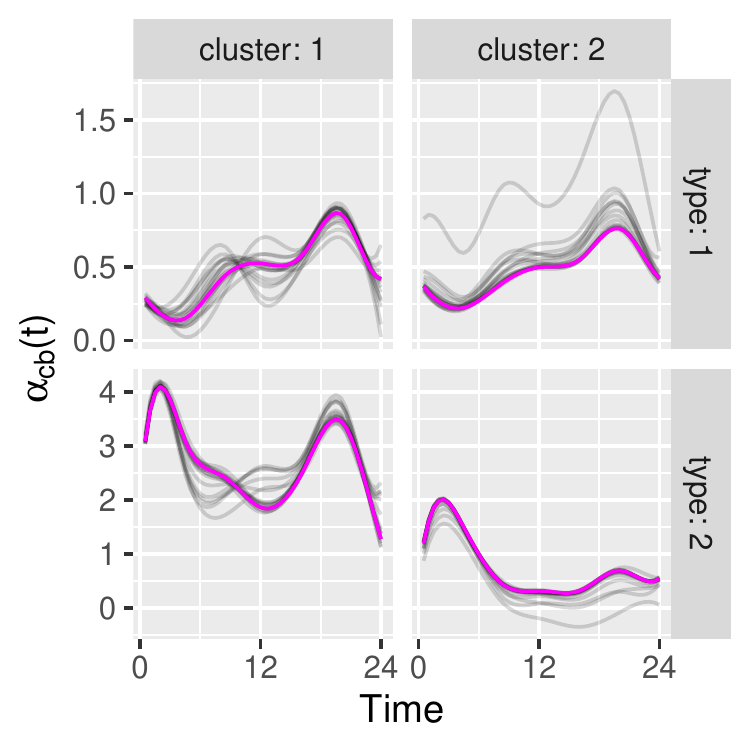} 

    \caption{Balanced market with 30-day data}
  \end{subfigure}
    \begin{subfigure}{.49\textwidth}
    \centering
\includegraphics[width=.8\linewidth]{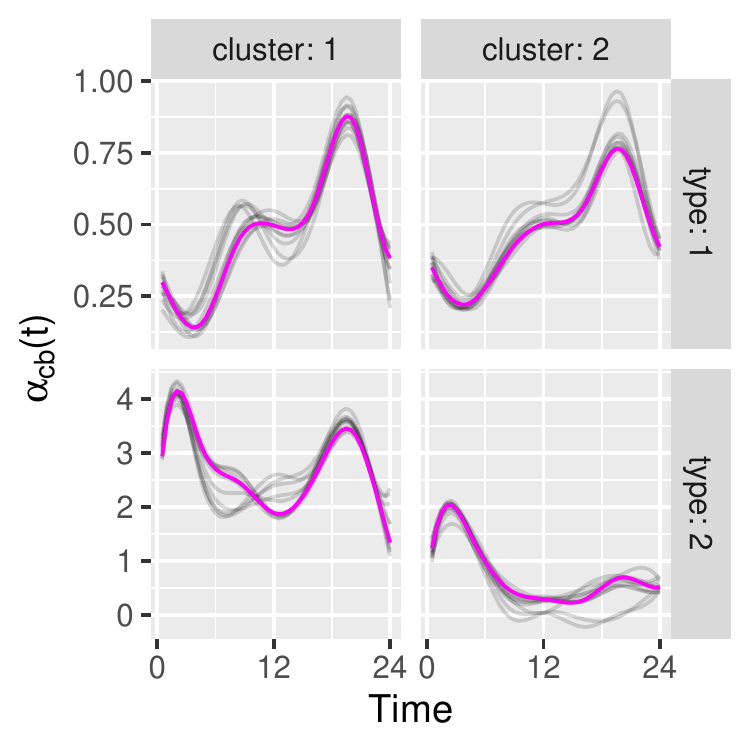} 

    \caption{Unbalanced market with 30-day data}
  \end{subfigure}
  \caption{Estimated typical curves for Scenarios 5 to 8, represented by the combination of market balance and number of days, under the two-cluster model. Median curves are represented in magenta and estimated typical curves in gray.}
  \label{fig:simu-cl2-typical}
\end{figure}

Figure~\ref{fig:simu-cl2-typical} shows the estimated typical curves for Scenarios 5 to 8 under the two- cluster model fit. In general, observe that Cluster 1 curves capture the main characteristics of Clusters 1 and 3: the work period stability, the 8 PM peak of Type 1 curves, and the 2 AM and 8 PM peaks of Type 2 curves. The 30-day scenarios have slightly lower estimate variability than the five-day scenarios, but note that the Type 2 curves in Cluster 1 have runs with different estimated characteristics of the work period, as shown in Figures~\ref{fig:simu-cl2-typical}c and \ref{fig:simu-cl2-typical}d. In fact, the main difference between true Clusters 1 and 3 is the work period characterization of Type 2, and therefore it is to be expected that some runs could estimate typical curves in favour of the true Cluster 1 or Cluster 3.

Table~\ref{tab:simu-cl2-tab} shows the summary statistics of the estimated covariance parameters of the two-cluster model fit. Because the estimated Cluster 2 substations coincide with the substations in the true Cluster 2, it is to be expected that their estimated covariance parameters are close to their true values.  Observe in Table~\ref{tab:simu-cl2-tab}that the median and mean of the estimated parameters for Cluster 2 are close to their true values in the Reference column, especially for 30-day scenarios. Under five-day scenarios, balanced markets have better estimates in terms of precision. On the other hand, Cluster 1 estimates are located between the true values of true Clusters 1 and 3, and therefore the Reference column for Cluster 1 in Table~\ref{tab:simu-cl2-tab} represents the mean of the covariance parameters of the true Clusters 1 and 3. The proximity of estimated and true covariance parameters is greater for customers of Type 2. In contrast, the Type 2 true dispersion parameters of Clusters 1 and 3 present the largest difference in Table~\ref{tab:true_cp_cl}, 1.54 and 5.18 respectively. Nonetheless, there is no clear evidence that the estimated covariance parameters in Cluster 1 are close to the average of the true parameters of Clusters 1 and 3.

\begin{figure}[thp]
  \centering
  \begin{subfigure}{.49\textwidth}
    \centering
\includegraphics[width=.8\linewidth]{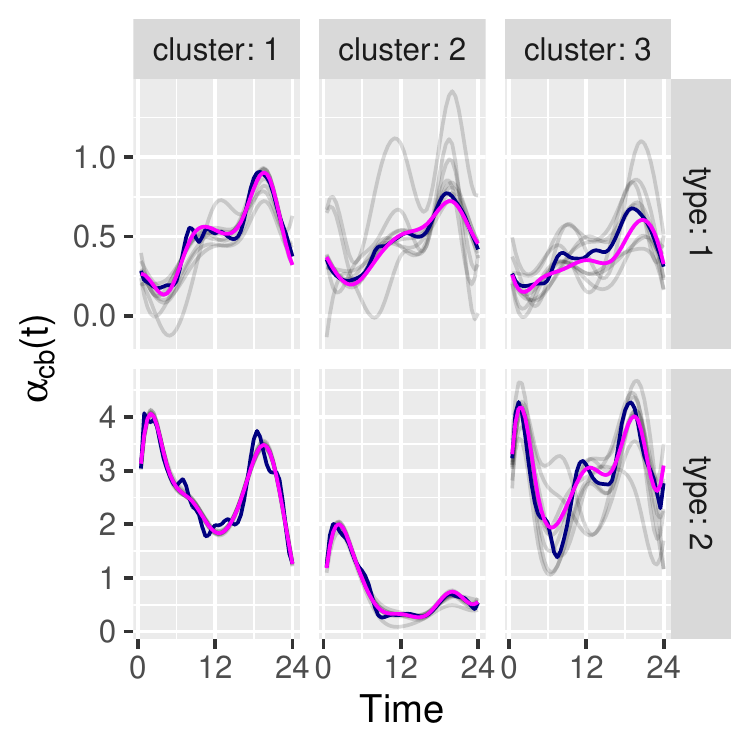} 

    \caption{Balanced market with 5-day data}
  \end{subfigure}
  \begin{subfigure}{.49\textwidth}
    \centering
\includegraphics[width=.8\linewidth]{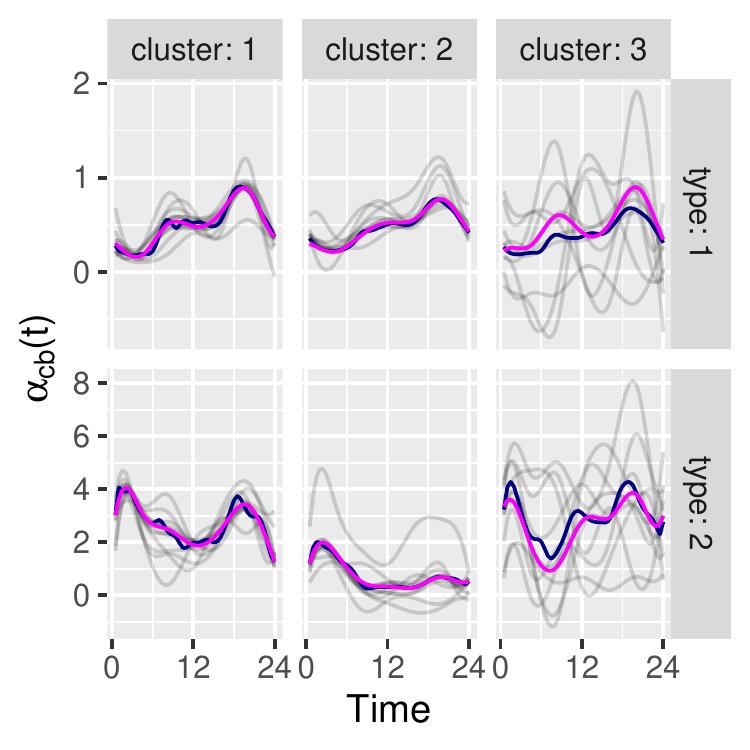} 

    \caption{Unbalanced market with 5-day data}
  \end{subfigure}
    \begin{subfigure}{.49\textwidth}
    \centering
\includegraphics[width=.8\linewidth]{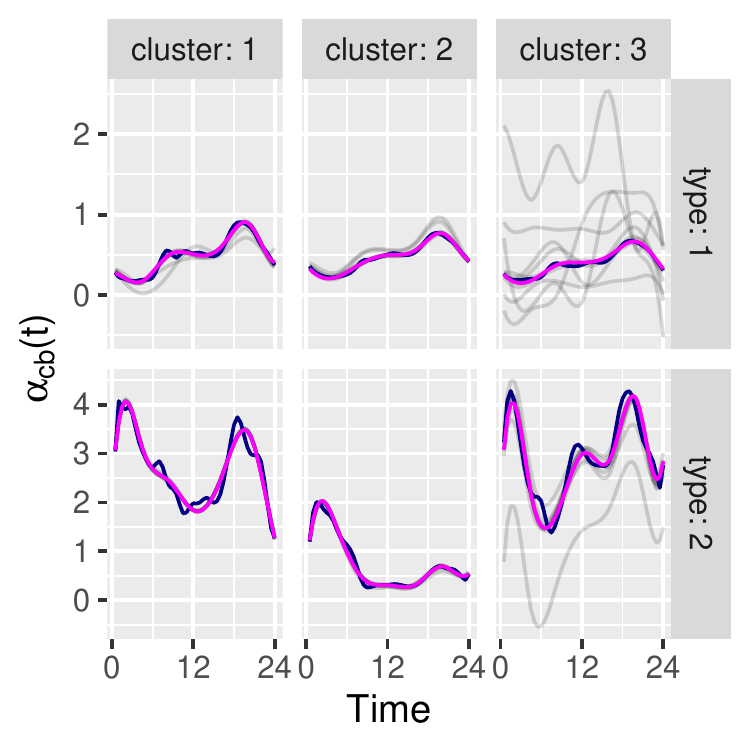} 

    \caption{Balanced market with 30-day data}
  \end{subfigure}
    \begin{subfigure}{.49\textwidth}
    \centering
\includegraphics[width=.8\linewidth]{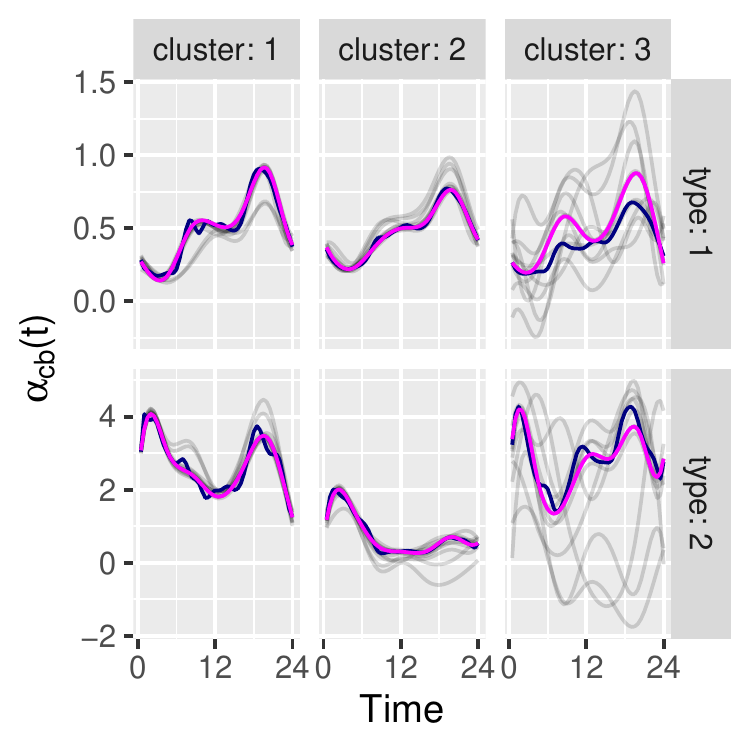} 

    \caption{Unbalanced market with 30-day data}
  \end{subfigure}
  \caption{Estimated typical curves for Scenarios 5 to 8, represented by the combination of market balance and number of days, under the three-cluster model. Median curves are represented in magenta and estimated typical curves in gray.}
  \label{fig:simu-cl3-typical}
\end{figure}

\begin{figure}[thp]
  \centering
  \begin{subfigure}{.49\textwidth}
    \centering
\includegraphics[width=.8\linewidth]{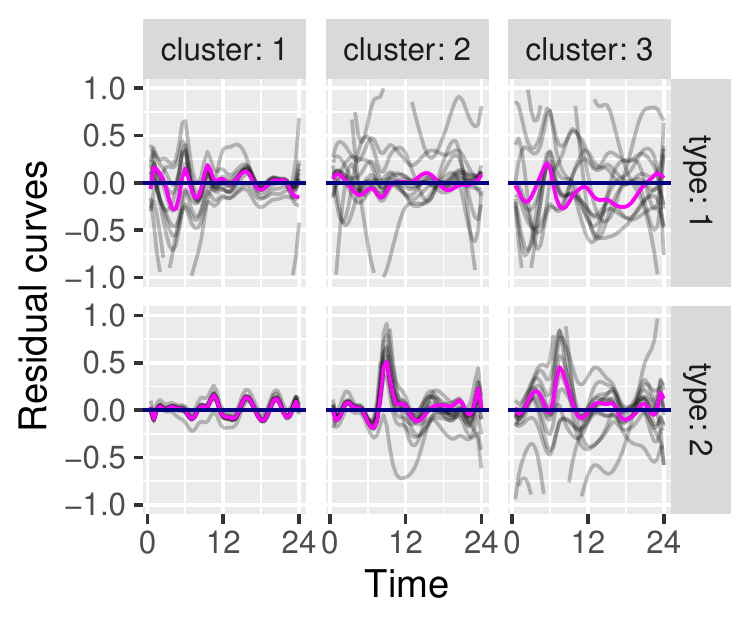} 

    \caption{Balanced market with 5-day data}
  \end{subfigure}
  \begin{subfigure}{.49\textwidth}
    \centering
\includegraphics[width=.8\linewidth]{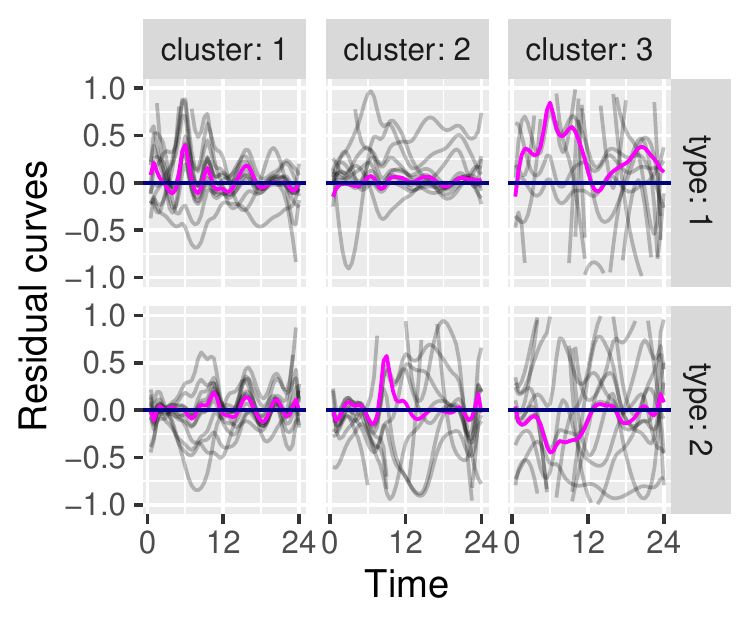} 

    \caption{Unbalanced market with 5-day data}
  \end{subfigure}
    \begin{subfigure}{.49\textwidth}
    \centering
\includegraphics[width=.8\linewidth]{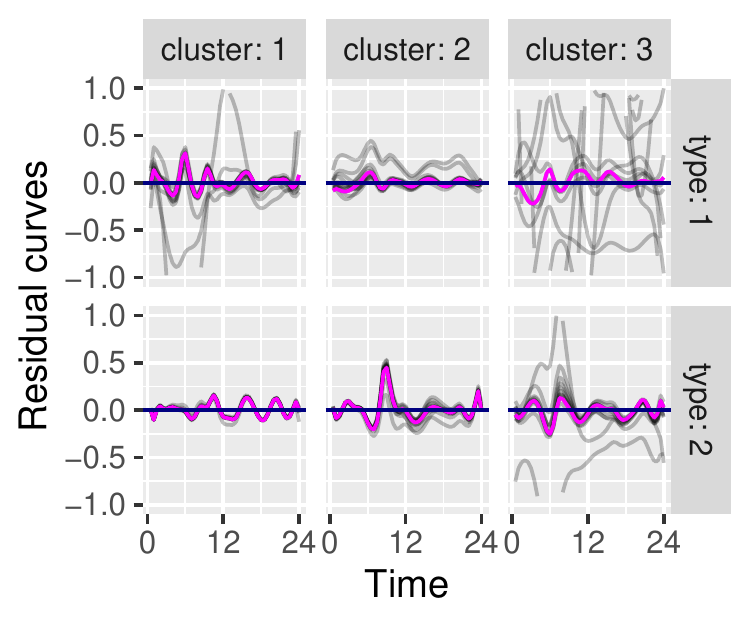} 

    \caption{Balanced market with 30-day data}
  \end{subfigure}
    \begin{subfigure}{.49\textwidth}
    \centering
\includegraphics[width=.8\linewidth]{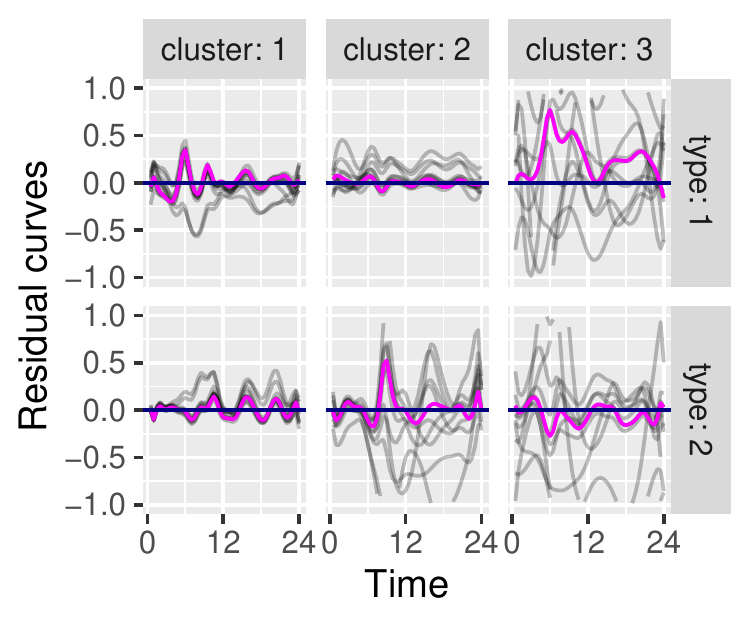} 

    \caption{Unbalanced market with 30-day data}
  \end{subfigure}
  \caption{Residual curves for Scenarios 5 to 8, represented by the combination of market balance and number of days, under the three-cluster model. Median curves are represented in magenta and estimated typical curves in gray..}
  \label{fig:simu-cl3-mse}
\end{figure}

The estimated typical curves of the three-cluster model are displayed in Figure~\ref{fig:simu-cl3-typical} and their associated residual curves in Figure~\ref{fig:simu-cl3-mse}. In general, the median curves show that the estimated curves capture the main characteristics of the true typical curves, although there are visible discrepant examples, more frequently seen in the unbalanced scenarios, as shown in Figures \ref{fig:simu-cl3-typical}b and \ref{fig:simu-cl3-typical}c. Negative values could be avoided by restricting the typical curve estimation, but to avoid overextending the computational burden of this simulation, it was decided to retain the least-squares estimators in exchange for some negative values, and also to show that in general the estimator is robust for different scenarios because the median curves are positive along the entire time axis. Once more, the residual curves show that the 30-day scenarios are more concentrated around the zero-reference line than the five-day scenarios, with even better fits for balanced scenarios. In this clustering approach, the relative residual curves for Cluster 3 in the unbalanced scenario do not concentrate around the zero-reference line, as shown in Figures~\ref{fig:simu-cl3-mse}b and \ref{fig:simu-cl3-mse}d. Furthermore, Cluster 1 has residual curves with lower variability than Cluster 3 in all scenarios. In fact, Cluster 1 contains six substations, whereas Cluster 3 contains two substations, the minimum number required for model identifiability. It seems that the larger the number of substations in the cluster, the better is the precision, and consequently this might be the reason for the Cluster 3 overestimation of the typical curves, particularly under unbalanced scenarios.

The estimated covariance parameters of the three-cluster models are represented by
their mean, median and srMSRE in Table~\ref{tab:simu-cl3-tab}. As observed in previous results throughout this section, smallest srMSRE are associated with parameters with larger magnitudes, for example,  $\sigma_{22}=1.28$ and $\sigma_{23} = 5.18$. n contrast, the largest srMSRE are associated with parameters with smaller magnitudes, for example, $\omega_{13}=0.02$. In the latter case, the unbalanced scenarios presented smaller srMSRE than the balanced markets. In cases like $\omega_{22}=0.09$, increasing the number of observation days from 5 to 30 improved srMSRE. The same behaviour was observed in most parameters, particularly those with small magnitude. On the other hand, parameters $\omega_{21}=0.03$ and $\omega_{13}=0.02$ were the smallest parameters in the simulation, but the srMSREs of $\omega_{21}$ were mostly around 1.5, whereas the srMSREs of $\omega_{13}$ had three values greater than 10. Recall that Cluster 3 contained only two substations. Therefore, as mentioned earlier for typical curve estimation, both the number of substations and the number of observation days are important to improve parameter estimation in each cluster.

To avoid an overextended table in this section, the comparison of BIC values between the two- and three-cluster models is presented in  Table~\ref{tab:simu-bic} in Section~\ref{chap:supl-tabs}. In all cases, the BIC is favourable to the three-cluster model with differences mostly of order $10^2$.

\begin{table}[bh]
\caption{\label{tab:simu-cl-assign} Cluster allocation of the 12 substations under the clustering models with two and three clusters. The proportion of runs assigned to that cluster are 100\% in all runs in both model fit.}
\centering
\begin{tabular}{lrrr}
\toprule
Substation & True & two-cluster fit & three-cluster fit \\
\midrule
1 & 1 & 1  & 1 \\
2 & 1 & 1  & 1  \\
3 & 1 & 1  & 1  \\
4 & 1 & 1  & 1  \\
5 & 1 & 1  & 1  \\
6 & 1 & 1  & 1  \\
\addlinespace
7 & 2 & 2  & 2  \\
8 & 2 & 2  & 2  \\
9 & 2 & 2  & 2  \\
10 & 2 & 2  & 2  \\
\addlinespace
11 & 3 & 1  & 3  \\
12 & 3 & 1  & 3  \\
\bottomrule
\end{tabular}
\end{table}

\begin{table}[b]\centering
\caption{Summary statistics for the estimated covariance parameters for Scenarios 5 to 8 under the two-cluster model fit. The reference column for Cluster 1 (parameter subindex ending in 1) is the mean value between covariance parameters of the true Clusters 1 and 3 and for Cluster 2 (parameter subindex ending in 2) is the true covariance parameters for the true Cluster 2.}
\begin{tabular}{ccrrrrr}
\toprule
Parameter & Days & Market & Ref & Median & Mean & Std Dev\\
\midrule
 &  & Balanced & 0.985 & 1.3690 & 1.1717 & 0.6952\\

 & \multirow{-2}{*}{\centering\arraybackslash 30 days} & Unbalanced & 0.985 & 1.4175 & 1.4324 & 0.1123\\
\cmidrule{2-7}
 &  & Balanced & 0.985 & 1.2360 & 1.0226 & 0.5916\\

\multirow{-4}{*}{\centering\arraybackslash $\sigma_{11}$} & \multirow{-2}{*}{\centering\arraybackslash 5 days} & Unbalanced & 0.985 & 1.3903 & 0.9552 & 0.7257\\
\cmidrule{1-7}
 &  & Balanced & 0.090 & 0.1863 & 1.7907 & 3.2170\\

 & \multirow{-2}{*}{\centering\arraybackslash 30 days} & Unbalanced & 0.090 & 0.1303 & 0.1457 & 0.0323\\
\cmidrule{2-7}
 &  & Balanced & 0.090 & 0.2262 & 2.4892 & 3.6509\\

\multirow{-4}{*}{\centering\arraybackslash $\omega_{11}$} & \multirow{-2}{*}{\centering\arraybackslash 5 days} & Unbalanced & 0.090 & 0.4885 & 1.7235 & 2.2490\\
\cmidrule{1-7}
 &  & Balanced & 3.355 & 3.3149 & 3.7087 & 1.0352\\

 & \multirow{-2}{*}{\centering\arraybackslash 30 days} & Unbalanced & 3.355 & 3.8129 & 3.3212 & 0.7789\\
\cmidrule{2-7}
 &  & Balanced & 3.355 & 4.0515 & 4.0523 & 0.7751\\

\multirow{-4}{*}{\centering\arraybackslash $\sigma_{21}$} & \multirow{-2}{*}{\centering\arraybackslash 5 days} & Unbalanced & 3.355 & 3.7831 & 4.1194 & 0.9537\\
\cmidrule{1-7}
 &  & Balanced & 0.200 & 0.1247 & 0.1559 & 0.0653\\

 & \multirow{-2}{*}{\centering\arraybackslash 30 days} & Unbalanced & 0.200 & 0.1859 & 0.1728 & 0.0768\\
\cmidrule{2-7}
 &  & Balanced & 0.200 & 0.1983 & 0.1848 & 0.0644\\

\multirow{-4}{*}{\centering\arraybackslash $\omega_{21}$} & \multirow{-2}{*}{\centering\arraybackslash 5 days} & Unbalanced & 0.200 & 0.2595 & 0.2048 & 0.0891\\
\cmidrule{1-7}
 &  & Balanced & 1.070 & 1.1663 & 1.2281 & 0.3476\\

 & \multirow{-2}{*}{\centering\arraybackslash 30 days} & Unbalanced & 1.070 & 1.0870 & 1.1621 & 0.1524\\
\cmidrule{2-7}
 &  & Balanced & 1.070 & 1.0906 & 1.1274 & 0.3569\\

\multirow{-4}{*}{\centering\arraybackslash $\sigma_{12}$} & \multirow{-2}{*}{\centering\arraybackslash 5 days} & Unbalanced & 1.070 & 1.2591 & 1.1776 & 0.3398\\
\cmidrule{1-7}
 &  & Balanced & 0.120 & 0.1290 & 0.3108 & 0.7708\\

 & \multirow{-2}{*}{\centering\arraybackslash 30 days} & Unbalanced & 0.120 & 0.1143 & 0.1152 & 0.0071\\
\cmidrule{2-7}
 &  & Balanced & 0.120 & 0.1080 & 0.3592 & 1.0368\\

\multirow{-4}{*}{\centering\arraybackslash $\omega_{12}$} & \multirow{-2}{*}{\centering\arraybackslash 5 days} & Unbalanced & 0.120 & 0.1134 & 1.0093 & 2.3378\\
\cmidrule{1-7}
 &  & Balanced & 1.280 & 1.4533 & 1.3147 & 0.3727\\

 & \multirow{-2}{*}{\centering\arraybackslash 30 days} & Unbalanced & 1.280 & 1.3541 & 1.2999 & 0.8166\\
\cmidrule{2-7}
 &  & Balanced & 1.280 & 1.4583 & 0.9738 & 0.7445\\

\multirow{-4}{*}{\centering\arraybackslash $\sigma_{22}$} & \multirow{-2}{*}{\centering\arraybackslash 5 days} & Unbalanced & 1.280 & 0.3949 & 0.8713 & 1.0385\\
\cmidrule{1-7}
 &  & Balanced & 0.090 & 0.0958 & 0.3894 & 0.8949\\

 & \multirow{-2}{*}{\centering\arraybackslash 30 days} & Unbalanced & 0.090 & 0.1233 & 1.4038 & 2.7188\\
\cmidrule{2-7}
 &  & Balanced & 0.090 & 0.0998 & 2.2790 & 3.1786\\

\multirow{-4}{*}{\centering\arraybackslash $\omega_{22}$} & \multirow{-2}{*}{\centering\arraybackslash 5 days} & Unbalanced & 0.090 & 3.6099 & 4.7933 & 4.9760\\
\bottomrule
\end{tabular}
\label{tab:simu-cl2-tab}
\end{table}




\begin{longtable}[t]{cccccc}
\caption{\label{tab:simu-cl3-tab}Mean, median and square root of the Mean Squared Relative Error (srMSRE) of the estimated covariance parameters for Scenarios 5 to 8, under the three-cluster model fit.}\\
\toprule
Parameter & Days & Market & Median & Mean & $\sqrt{MSRE}$\\
\midrule
\endfirsthead
\multicolumn{6}{c}{\tablename~\thetable~(continued)}\\
\addlinespace
\toprule
Parameter & Days & Market & Median & Mean & $\sqrt{MSRE}$\\
\midrule
\endhead
\midrule
 &  & Balanced & 1.0862 & 1.0114 & 0.6470\\

 & \multirow{-2}{*}{\centering\arraybackslash 30 days} & Unbalanced & 1.3929 & 1.2079 & 0.4677\\
\cmidrule{2-6}
 &  & Balanced & 0.9430 & 0.8281 & 0.6991\\

\multirow{-4}{*}{\centering\arraybackslash $\sigma_{11}=1.54$} & \multirow{-2}{*}{\centering\arraybackslash 5 days} & Unbalanced & 1.3859 & 1.2022 & 0.4923\\
\cmidrule{1-6}
 &  & Balanced & 0.3220 & 2.2018 & 3.5878\\

 & \multirow{-2}{*}{\centering\arraybackslash 30 days} & Unbalanced & 0.1295 & 0.3441 & 1.1971\\
\cmidrule{2-6}
 &  & Balanced & 3.9317 & 6.5328 & 6.3132\\

\multirow{-4}{*}{\centering\arraybackslash $\omega_{11}=0.16$} & \multirow{-2}{*}{\centering\arraybackslash 5 days} & Unbalanced & 0.1274 & 0.9780 & 2.3491\\
\cmidrule{1-6}
 &  & Balanced & 2.8842 & 2.9041 & 0.9477\\

 & \multirow{-2}{*}{\centering\arraybackslash 30 days} & Unbalanced & 2.7891 & 2.8831 & 0.9404\\
\cmidrule{2-6}
 &  & Balanced & 2.8640 & 2.9372 & 0.9590\\

\multirow{-4}{*}{\centering\arraybackslash $\sigma_{21}=1.53$} & \multirow{-2}{*}{\centering\arraybackslash 5 days} & Unbalanced & 2.7464 & 2.7632 & 0.9386\\
\cmidrule{1-6}
 &  & Balanced & 0.0759 & 0.0777 & 1.2608\\

 & \multirow{-2}{*}{\centering\arraybackslash 30 days} & Unbalanced & 0.0960 & 0.0985 & 1.5111\\
\cmidrule{2-6}
 &  & Balanced & 0.0736 & 0.0807 & 1.3000\\

\multirow{-4}{*}{\centering\arraybackslash $\omega_{21}=0.03$} & \multirow{-2}{*}{\centering\arraybackslash 5 days} & Unbalanced & 0.1057 & 0.4043 & 3.5322\\
\cmidrule{1-6}
 &  & Balanced & 1.0274 & 1.0346 & 0.4578\\

 & \multirow{-2}{*}{\centering\arraybackslash 30 days} & Unbalanced & 1.0810 & 1.1472 & 0.3197\\
\cmidrule{2-6}
 &  & Balanced & 0.6081 & 0.6888 & 0.7774\\

\multirow{-4}{*}{\centering\arraybackslash $\sigma_{12}=1.07$} & \multirow{-2}{*}{\centering\arraybackslash 5 days} & Unbalanced & 1.2345 & 1.2259 & 0.4379\\
\cmidrule{1-6}
 &  & Balanced & 0.1208 & 0.6763 & 2.1698\\

 & \multirow{-2}{*}{\centering\arraybackslash 30 days} & Unbalanced & 0.1226 & 0.1344 & 0.4184\\
\cmidrule{2-6}
 &  & Balanced & 1.4476 & 2.4465 & 4.4133\\

\multirow{-4}{*}{\centering\arraybackslash $\omega_{12}=0.12$} & \multirow{-2}{*}{\centering\arraybackslash 5 days} & Unbalanced & 0.1083 & 0.5807 & 2.0123\\
\cmidrule{1-6}
 &  & Balanced & 1.5417 & 1.5858 & 0.4888\\

 & \multirow{-2}{*}{\centering\arraybackslash 30 days} & Unbalanced & 1.2823 & 1.2391 & 0.4312\\
\cmidrule{2-6}
 &  & Balanced & 1.5092 & 1.5814 & 0.4852\\

\multirow{-4}{*}{\centering\arraybackslash $\sigma_{22}=1.28$} & \multirow{-2}{*}{\centering\arraybackslash 5 days} & Unbalanced & 0.0363 & 0.5207 & 0.9565\\
\cmidrule{1-6}
 &  & Balanced & 0.0960 & 0.0965 & 0.2692\\

 & \multirow{-2}{*}{\centering\arraybackslash 30 days} & Unbalanced & 0.1057 & 0.4156 & 1.9330\\
\cmidrule{2-6}
 &  & Balanced & 0.0877 & 0.0884 & 0.2444\\

\multirow{-4}{*}{\centering\arraybackslash $\omega_{22}=0.09$} & \multirow{-2}{*}{\centering\arraybackslash 5 days} & Unbalanced & 5.2492 & 5.6967 & 7.8929\\
\cmidrule{1-6}
 &  & Balanced & 0.3390 & 1.4096 & 1.7326\\

 & \multirow{-2}{*}{\centering\arraybackslash 30 days} & Unbalanced & 1.3438 & 1.4685 & 1.5540\\
\cmidrule{2-6}
 &  & Balanced & 0.0282 & 0.4760 & 1.1353\\

\multirow{-4}{*}{\centering\arraybackslash $\sigma_{13}=0.43$} & \multirow{-2}{*}{\centering\arraybackslash 5 days} & Unbalanced & 1.0871 & 1.0075 & 1.3569\\
\cmidrule{1-6}
 &  & Balanced & 3.9775 & 3.2759 & 12.7591\\

 & \multirow{-2}{*}{\centering\arraybackslash 30 days} & Unbalanced & 0.1446 & 0.3754 & 4.2157\\
\cmidrule{2-6}
 &  & Balanced & 9.9767 & 10.3302 & 22.7049\\

\multirow{-4}{*}{\centering\arraybackslash $\omega_{13}=0.02$} & \multirow{-2}{*}{\centering\arraybackslash 5 days} & Unbalanced & 0.9252 & 2.7126 & 11.6030\\
\cmidrule{1-6}
 &  & Balanced & 5.2455 & 5.0097 & 0.3249\\

 & \multirow{-2}{*}{\centering\arraybackslash 30 days} & Unbalanced & 3.9596 & 3.3274 & 0.5980\\
\cmidrule{2-6}
 &  & Balanced & 4.7792 & 4.6787 & 0.3502\\

\multirow{-4}{*}{\centering\arraybackslash $\sigma_{23}=5.18$} & \multirow{-2}{*}{\centering\arraybackslash 5 days} & Unbalanced & 4.0859 & 3.9616 & 0.4917\\
\cmidrule{1-6}
 &  & Balanced & 0.2624 & 0.2492 & 0.5713\\

 & \multirow{-2}{*}{\centering\arraybackslash 30 days} & Unbalanced & 0.2968 & 0.6871 & 1.1099\\
\cmidrule{2-6}
 &  & Balanced & 0.1930 & 0.1960 & 0.6859\\

\multirow{-4}{*}{\centering\arraybackslash $\omega_{23}=0.37$} & \multirow{-2}{*}{\centering\arraybackslash 5 days} & Unbalanced & 0.2759 & 0.8170 & 1.3016\\
\bottomrule
\end{longtable}

\subsubsection{Discussion and conclusion}
\label{sec:simu-cluster-conclusion}

In both fitted models, substations are allocated to the same clusters throughout the series of runs. In the two-cluster model, Substations 11 and 12, which belong to the true Cluster 3, are always assigned to Cluster 1 together with Substations 1 to 6. In this case, with an underdetermined number of clusters, the method groups the clusters with more similarity. Hence the estimated typical curves for Cluster 1 still capture the main features of the true curves for Clusters 1 and 3. Similarly, the estimated covariance parameters for Cluster 1 present values between the true covariance parameters of the true Clusters 1 and 3. In the three-cluster model, substations are assigned to the correct cluster. Consequently, except for some cases under unbalanced scenarios, estimated typical curves for this model are well located around their true curves. In general, 30-day scenarios have less dispersed estimates than five-day scenarios, and balanced markets have less dispersed estimates than unbalanced ones.

The differences among scenarios have a similar impact on the estimation of covariance parameters. In addition, there is evidence that the number of substations in a cluster is crucial to estimation performance, particularly for small-magnitude parameters. The positive impact of increasing the number of substations on parameter estimation is shown in  \cite{lenzi2017analysis}. Two parameters with small values for Clusters 1 and 3 have distinct srMSRE probably because the information available for Cluster 3 estimation is less than for Cluster 1.

When comparing both models, the three-cluster model presents lower BIC than the two-cluster model in all cases.In a real-world problem, where the true number of clusters is unknown, the BIC can be a useful tool to decide between models.

Users of the clustering aggregated data model are encouraged to be careful with the estimated covariance parameters and to try multiple models with different numbers of clusters using these estimated values as input for their initial values. For example, the estimated values of the aggregated two-cluster data model can be used as an input to fit the aggregated three-cluster data model by repeating one of the results. As shown in Section 2.5, after multiple fits, the user can compare the models using the likelihood test ratio to help decide which model is the most adequate to the data.

\clearpage

\subsection{Additional tables}
\label{chap:supl-tabs}

\begin{longtable}{lrrrrr}
\caption{\label{tab:simu-rvm}Likelihood ratio test comparison table of homogeneous and complete aggregated data models in simulated datasets for each experimental run. The degrees of freedom used to compute the p-value is 10 for all comparisons.}\\
\toprule
& &\multicolumn{2}{c}{Log-likelihood} & &\\
\cmidrule{3-4}
Days & Run & Homogeneous & Complete & Test statistic & p-value\\
\midrule
5 days & 1 & 11356.44 & 11206.11 & 300.6677 & <0.0001\\
5 days & 2 & 11468.83 & 11279.91 & 377.8489 & <0.0001\\
5 days & 3 & 11329.26 & 11129.60 & 399.3154 & <0.0001\\
5 days & 4 & 11393.79 & 11237.21 & 313.1679 & <0.0001\\
5 days & 5 & 11380.05 & 11161.95 & 436.2019 & <0.0001\\
\addlinespace
5 days & 6 & 11412.08 & 11240.27 & 343.6055 & <0.0001\\
5 days & 7 & 11409.96 & 11230.16 & 359.5910 & <0.0001\\
5 days & 8 & 11380.91 & 11221.02 & 319.7964 & <0.0001\\
5 days & 9 & 11296.43 & 11111.48 & 369.8967 & <0.0001\\
5 days & 10 & 11293.50 & 11131.60 & 323.7867 & <0.0001\\
\addlinespace
5 days & 11 & 11297.66 & 11132.56 & 330.1824 & <0.0001\\
5 days & 12 & 11380.86 & 11204.74 & 352.2391 & <0.0001\\
5 days & 13 & 11305.42 & 11135.14 & 340.5573 & <0.0001\\
5 days & 14 & 11357.41 & 11182.27 & 350.2775 & <0.0001\\
5 days & 15 & 11337.32 & 11153.20 & 368.2447 & <0.0001\\
\addlinespace
5 days & 16 & 12252.88 & 12167.35 & 171.0486 & <0.0001\\
5 days & 17 & 12199.07 & 12079.56 & 239.0298 & <0.0001\\
5 days & 18 & 12315.08 & 12222.54 & 185.0706 & <0.0001\\
5 days & 19 & 12394.19 & 12316.81 & 154.7764 & <0.0001\\
5 days & 20 & 12254.24 & 12160.62 & 187.2429 & <0.0001\\
\addlinespace
5 days & 21 & 12286.01 & 12189.36 & 193.3048 & <0.0001\\
5 days & 22 & 12314.49 & 12225.54 & 177.8949 & <0.0001\\
5 days & 23 & 12173.57 & 12073.10 & 200.9313 & <0.0001\\
5 days & 24 & 12248.57 & 12150.72 & 195.6961 & <0.0001\\
5 days & 25 & 12197.71 & 12090.13 & 215.1580 & <0.0001\\
\addlinespace
5 days & 26 & 12304.61 & 12196.86 & 215.4922 & <0.0001\\
5 days & 27 & 12135.35 & 12034.37 & 201.9598 & <0.0001\\
5 days & 28 & 12385.44 & 12302.31 & 166.2589 & <0.0001\\
5 days & 29 & 12359.07 & 12253.33 & 211.4677 & <0.0001\\
5 days & 30 & 12249.15 & 12142.30 & 213.7031 & <0.0001\\
\addlinespace
30 days & 1 & 68412.15 & 67418.28 & 1987.7512 & <0.0001\\
30 days & 2 & 68555.41 & 67692.07 & 1726.6639 & <0.0001\\
30 days & 3 & 68409.85 & 67507.56 & 1804.5693 & <0.0001\\
30 days & 4 & 68848.74 & 67854.20 & 1989.0811 & <0.0001\\
30 days & 5 & 68853.23 & 67912.53 & 1881.3950 & <0.0001\\
\addlinespace
30 days & 6 & 68254.05 & 67181.14 & 2145.8196 & <0.0001\\
30 days & 7 & 68428.22 & 67442.46 & 1971.5111 & <0.0001\\
30 days & 8 & 68720.86 & 67794.68 & 1852.3555 & <0.0001\\
30 days & 9 & 68143.11 & 67241.85 & 1802.5107 & <0.0001\\
30 days & 10 & 68639.51 & 67645.38 & 1988.2693 & <0.0001\\
\addlinespace
30 days & 11 & 68532.07 & 67568.32 & 1927.5025 & <0.0001\\
30 days & 12 & 68452.58 & 67430.66 & 2043.8409 & <0.0001\\
30 days & 13 & 68697.84 & 67687.06 & 2021.5609 & <0.0001\\
30 days & 14 & 68183.63 & 67221.51 & 1924.2479 & <0.0001\\
30 days & 15 & 68828.41 & 67972.03 & 1712.7790 & <0.0001\\
\addlinespace
30 days & 16 & 73834.91 & 73325.45 & 1018.9162 & <0.0001\\
30 days & 17 & 74309.19 & 73894.94 & 828.4990 & <0.0001\\
30 days & 18 & 74396.67 & 73964.25 & 864.8214 & <0.0001\\
30 days & 19 & 74607.51 & 74136.77 & 941.4789 & <0.0001\\
30 days & 20 & 74741.17 & 74284.07 & 914.1883 & <0.0001\\
\addlinespace
30 days & 21 & 74686.06 & 74273.96 & 824.1941 & <0.0001\\
30 days & 22 & 74678.44 & 74242.28 & 872.3236 & <0.0001\\
30 days & 23 & 74295.85 & 73798.20 & 995.2966 & <0.0001\\
30 days & 24 & 73439.84 & 72817.88 & 1243.9196 & <0.0001\\
30 days & 25 & 74510.71 & 74113.48 & 794.4602 & <0.0001\\
\addlinespace
30 days & 26 & 74232.01 & 73790.03 & 883.9603 & <0.0001\\
30 days & 27 & 74783.54 & 74378.31 & 810.4463 & <0.0001\\
30 days & 28 & 74510.23 & 74068.60 & 883.2588 & <0.0001\\
30 days & 29 & 74963.07 & 74588.12 & 749.8957 & <0.0001\\
30 days & 30 & 74913.23 & 74506.08 & 814.3098 & <0.0001\\
\bottomrule
\end{longtable}

\newpage
\begin{longtable}{lrrrr}
\caption{\label{tab:simu-bic}BIC values for the clustering aggregated data models in simulated datasets at each experimental runs. \textsc{NA} values represent the runs that did not converge.}\\
\toprule
& \multicolumn{2}{c}{BIC} & \\
\cmidrule{3-4}
Days & Run & 2 Clusters & 3 Clusters & BIC diff\\
\midrule
5 days & 1 & 22892.98 & 22653.75 & 239.24\\
5 days & 2 & 23037.71 & 22728.32 & 309.38\\
5 days & 3 & 22765.23 & 22522.55 & 242.68\\
5 days & 4 & 23002.98 & 22632.54 & 370.44\\
5 days & 5 & NA & NA & NA\\
\addlinespace
5 days & 6 & 22867.82 & 22589.01 & 278.82\\
5 days & 7 & 22928.48 & 22674.78 & 253.70\\
5 days & 8 & 23057.19 & NA & NA\\
5 days & 9 & 23110.73 & 22797.53 & 313.20\\
5 days & 10 & NA & NA & NA\\
\addlinespace
5 days & 11 & 23023.11 & 22788.12 & 234.99\\
5 days & 12 & 22893.37 & 22633.57 & 259.80\\
5 days & 13 & 23033.60 & 22798.68 & 234.93\\
5 days & 14 & 22996.42 & 22713.55 & 282.87\\
5 days & 15 & 23127.33 & 22832.22 & 295.11\\
\addlinespace
5 days & 16 & 24320.71 & 23984.69 & 336.02\\
5 days & 17 & 24930.08 & 24283.66 & 646.42\\
5 days & 18 & 24508.43 & NA & NA\\
5 days & 19 & 24306.36 & 24053.74 & 252.62\\
5 days & 20 & 24582.05 & 24364.63 & 217.42\\
\addlinespace
5 days & 21 & 24138.96 & 23911.80 & 227.16\\
5 days & 22 & 24580.06 & 24236.24 & 343.83\\
5 days & 23 & 24477.64 & NA & NA\\
5 days & 24 & 24365.16 & 24106.78 & 258.37\\
5 days & 25 & NA & NA & NA\\
\addlinespace
5 days & 26 & 24842.38 & 24266.45 & 575.93\\
5 days & 27 & 23893.08 & 23763.69 & 129.38\\
5 days & 28 & 24568.87 & 24068.67 & 500.20\\
5 days & 29 & 24718.34 & 24275.80 & 442.54\\
5 days & 30 & 24304.65 & 24030.62 & 274.03\\
\addlinespace
30 days & 1 & 140732.75 & 139731.99 & 1000.75\\
30 days & 2 & 137862.55 & 137674.83 & 187.73\\
30 days & 3 & 141129.86 & 140153.32 & 976.53\\
30 days & 4 & 139429.37 & NA & NA\\
30 days & 5 & 138120.81 & 137360.21 & 760.60\\
\addlinespace
30 days & 6 & 138909.99 & NA & NA\\
30 days & 7 & 138244.85 & NA & NA\\
30 days & 8 & 139216.20 & 138601.06 & 615.15\\
30 days & 9 & 138597.82 & 137882.19 & 715.63\\
30 days & 10 & 137915.18 & NA & NA\\
\addlinespace
30 days & 11 & 139329.72 & 138463.56 & 866.16\\
30 days & 12 & 138265.27 & NA & NA\\
30 days & 13 & 139177.05 & 138376.03 & 801.02\\
30 days & 14 & 137716.41 & 136986.04 & 730.38\\
30 days & 15 & 138839.29 & 138544.00 & 295.30\\
\addlinespace
30 days & 16 & 148793.81 & NA & NA\\
30 days & 17 & 151598.28 & 149765.27 & 1833.01\\
30 days & 18 & 148195.67 & 146904.43 & 1291.24\\
30 days & 19 & 146444.55 & 145965.37 & 479.17\\
30 days & 20 & 150741.62 & 148847.58 & 1894.04\\
\addlinespace
30 days & 21 & 145479.22 & 144462.94 & 1016.28\\
30 days & 22 & 147632.69 & 146775.49 & 857.20\\
30 days & 23 & 149746.58 & 148789.10 & 957.48\\
30 days & 24 & NA & 148840.17 & NA\\
30 days & 25 & 151627.47 & 150116.68 & 1510.80\\
\addlinespace
30 days & 26 & NA & 150376.76 & NA\\
30 days & 27 & 150394.95 & NA & NA\\
30 days & 28 & 149657.30 & 147460.43 & 2196.86\\
30 days & 29 & 147228.13 & 146344.21 & 883.91\\
30 days & 30 & 152994.93 & 150468.24 & 2526.69\\
\bottomrule
\end{longtable}

\bibliographystyle{unsrtnat}
\bibliography{src/refs}